\documentclass[12pt,a4paper]{article}
 
\usepackage{jheppub}
\usepackage[usenames,dvipsnames]{xcolor}
\usepackage{amssymb} 
\usepackage{amsmath}
\usepackage{mathtools}
\usepackage{amsfonts}    
\usepackage{dsfont}
\usepackage{pdfpages}
\usepackage{verbatim}
\usepackage{stmaryrd}
\usepackage{graphicx}
\usepackage{bbold}

\usepackage{float}

\usepackage{tensor}
\usepackage{mathrsfs}
\usepackage{textgreek} 
\usepackage{euscript}
\usepackage[smalltableaux]{ytableau}
\ytableausetup{centertableaux}
\usepackage{tikz}
\usetikzlibrary{decorations.pathreplacing, decorations.markings,calc,shapes.misc,decorations.pathmorphing,patterns.meta, math}
\usetikzlibrary{decorations.pathmorphing}
\usetikzlibrary{decorations.markings}
\usepackage{slashed}
\usepackage{datetime}
\usepackage{hyperref}
\usepackage{braket}
\usepackage{bm}
\hypersetup{
  pdfencoding=unicode,
	colorlinks=true,
	urlcolor=Maroon,
	linkcolor=RoyalBlue,
	citecolor=Maroon,
	pdftitle={Minimal String Perturbation Revisited},
	pdfauthor={Victor Rodriguez, Mykhaylo Usatyuk, Zi-Yue Wang},
	pdfdisplaydoctitle=true,
	pdfstartview=FitH,
	linktocpage=true
}

\usetikzlibrary{shapes}
\usetikzlibrary{arrows}
\usetikzlibrary{decorations.pathmorphing}
\usetikzlibrary{decorations.markings}
\usetikzlibrary{shapes.misc}
\tikzset{snake it/.style={decorate, decoration=snake}}
\tikzset{cross/.style={cross out, draw=black, minimum size=2*(#1-\pgflinewidth), inner sep=0pt, outer sep=0pt},
cross/.default={1pt}}
\usetikzlibrary{shapes.geometric}
\tikzset{
    partial ellipse/.style args={#1:#2:#3}{
        insert path={+ (#1:#3) arc (#1:#2:#3)}
    }
}

\usepackage{subcaption}

\newcommand{\ba}{\begin{align}}

\newcommand{\be}{\begin{equation}}
\newcommand{\ee}{\end{equation}}
\def\bd{\begin{tikzpicture}}
\def\ed{\end{tikzpicture}}

\DeclareMathOperator\tr{tr}

\renewcommand\Re{\mathop{\text{Re}}}

\allowdisplaybreaks

\def\XXint#1#2#3{{\setbox0=\hbox{$#1{#2#3}{\int}$}
     \vcenter{\hbox{$#2#3$}}\kern-.5\wd0}}

\definecolor{light-gray}{gray}{0.75}

\newcommand\arccosh{\text{arccosh}}

\newcommand\Tr{\mathrm{Tr}}
\renewcommand\d{\text{d}}

\newcommand{\e}{\mathrm{e}}
\newcommand{\nn}{\nonumber}

\renewcommand{\leq}{\leqslant}
\renewcommand{\geq}{\geqslant}

\newcommand{\cT}{\mathcal{T}}

\newcommand{\gb}{\Gamma_b}
\newcommand{\gbeta}{\Gamma_\beta}

\newcommand{\id}{\mathds{1}}

\newcommand{\rrangle}{\rangle\!\rangle}

\definecolor{bleudefrance}{rgb}{0.19, 0.55, 0.91}
\definecolor{candyapplered}{rgb}{1.0, 0.03, 0.0}

\newcommand{\myred}[1]{\textcolor{BrickRed}{#1}}
\newcommand{\myinf}{\circ}

\def\be#1\ee{\begin{align}#1\end{align}}

\newcommand{\lr}[1]{\left( #1 \right)}
\newcommand{\ol}[1]{\overline{#1}}
\newcommand{\tl}[1]{\tilde{#1}}

\def\b{\beta}
\def\mF{\mathcal{F}}
\def\Gi{\Gamma}

\DeclareMathOperator{\csch}{csch}

\def\t{\text}
\def\rb{\rangle}
\def\lb{\langle}
\def\rb{\rangle}
\def\lb{\langle}
\def\qq{\qquad}
\def\q{\quad}

\def\cN{\mathcal{N}}
\def\cT{\mathcal{T}}

\def \bal#1\eal  {\begin{align} #1 \end{align}}

\def \bal#1\eal  {\begin{align} #1 \end{align}}
\def\({\left(}
\def\){\right)}
\def\[{\left[}
\def\]{\right]}

\newcommand{\pd} {\partial}
\newcommand{\mc} {\mathcal}
\newcommand{\mb} {\mathbb}

\newcommand{\si}{{\sigma}}

\newcommand{\epi}{\epsilon}

\def \bm#1\em {\begin{equation}\begin{matrix*}[l] #1 \end{matrix*}\end{equation}}

\title{ADE Minimal Strings and Multi-Matrix Duals}

\author[1]{Victor A. Rodriguez}\emailAdd{varodriguez@ucsb.edu}
\author[2]{\!\!, Mykhaylo Usatyuk}\emailAdd{musatyuk@kitp.ucsb.edu}
\author[1]{\!\!, Zi-Yue Wang}\emailAdd{zi-yue@ucsb.edu}
\affiliation[1]{Department of Physics, University of California, Santa Barbara, CA 93106, USA}
\affiliation[2]{Kavli Institute for Theoretical Physics, Santa Barbara, CA 93106, USA}

\abstract{
We revisit ADE minimal string theories, focusing on the D- and E-series minimal models coupled to Liouville theory. Unlike the A-series, whose duals are solvable two-matrix models, these theories are conjectured to correspond to unsolvable four-matrix integrals.
We compute sphere four‑point and torus one‑point amplitudes in the AMS, DMS, and EMS via direct numerical integration over moduli space, confirming/disproving some known results and providing new data where matrix‑model predictions are unavailable.
For torus one-point amplitudes, we also present conjectural analytic formulas consistent with all our numerical checks.
From amplitudes with conformal boundaries, we find evidence for multi-matrix structure in the D-series, including scaled ramp behavior in cylinder diagrams and deviations from the ZZ-instanton sector of two-matrix models.
We also perform a preliminary positivity bootstrap to constrain critical points of the multi-matrix models relevant to the DMS string.
}

\begin{document}

\maketitle

\makeatletter
\g@addto@macro\bfseries{\boldmath}
\makeatother

\section{Introduction}

The moduli space of consistent string theories is not well understood \cite{Douglas:2003um,Denef:2008wq}. A class of theories under better control is low (target space) dimensional bosonic string theories, which take the form

\begin{equation}
\begin{array}{c}
\text{matter CFT} \\ \text{$c=c_{\mathrm{m}}$}
\end{array}
\ \otimes\  
\begin{array}{c} \text{Liouville CFT} \\ \text{$c = 26-c_{\mathrm{m}}$} \end{array} \ \otimes\  
\begin{array}{c} \text{$\mathfrak{b}\mathfrak{c}$-ghosts} \\ \text{$c= -26$} \end{array}\,, 
\label{eq:minimal string}
\end{equation}
where the worldsheet theory is defined by Liouville CFT coupled to a choice of matter CFT. Empirically, each choice of matter CFT gives rise to a string theory that is dual to a double-scaled random matrix integral. The most well-studied cases are: 
\begin{table}[H] 
\centering
\makebox[\textwidth][c]{
\begin{tabular}{|l|l|c|c|}
\hline
string theory & matter CFT & \#-matrices & \t{solvable?} \\  \hline\hline
Minimal String (MS/AMS) \cite{Kazakov:1986hu,Boulatov:1986sb,Gross:1989vs,Brezin:1990rb,Douglas:1989dd,Ginsparg:1993is,DiFrancesco:1993cyw,Seiberg:2003nm} & A-series minimal model & 2 & TR \\ \hline
VMS \cite{Collier:2023cyw} & timelike Liouville &  1 & TR \\ \hline
JT gravity \cite{Saad:2019lba} & AMS$_{2,p \to \infty}$ or VMS$_{b\to0}$ &  1 & TR \\ \hline
CLS \cite{Collier:2024kmo} & complex-Liouville &  2 & TR\\ \hline
$ c=1$ \cite{Klebanov:1991qa} & timelike boson &  $\infty$ & FF \\ \hline
DMS, EMS \cite{Kostov:1995xw,DiFrancesco:1990jd} & D,E-series minimal model & 4(+?)  & $\times$ \\ 
\hline
\end{tabular}
}
\end{table}
Almost all theories with known duals are solvable on the matrix side through topological recursion (TR) \cite{Eynard:2002kg,Chekhov:2006vd,Eynard:2007kz,Eynard:2015aea}\footnote{The $c=1$ string is dual to matrix quantum mechanics and solvable through a free-fermion (FF) description \cite{Moore:1991zv}.} which allows for efficient calculation of all amplitudes.
It is natural to expect that yet undiscovered low-dimensional string theories continue to admit matrix integral duals, but with analytic structures that go beyond the known solvable class of two-matrix models.

\textbf{In this work} we revisit a family of low-dimensional theories that lie on the edge of tractability: $D_{p,p'}$ and $E_{p,p'}$-series minimal models coupled to Liouville (with a focus on D-series). The matrix integral duals for these theories are expected to be unsolvable $4+$ matrix integrals, with the DMS conjecturally dual to \cite{DiFrancesco:1990jd}
\be
\mathcal{Z} = \int \prod_{i=1}^4 \d M_i \hspace{.03cm} e^{-N \Tr ( V_i(M_i)-h M_1(M_2+M_3+M_4))}\,,
\ee
for some tuned potential. We find evidence for this four-matrix structure.

The paper is organized as follows. In \textbf{section 2} we review basic CFT data and describe the string theories being studied. 

In \textbf{section 3} we analytically and numerically study sphere four-point and torus one-point amplitudes in the AMS/DMS/EMS. See tables \ref{table:torus1pt A(p,3)}, \ref{table:torus1pt A(p,4-5)}, \ref{table:torus1pt D(p,3-5-7)}, \ref{table:torus1pt E(12,pp)}, \ref{table:Sphere_table A}, \ref{table:Sphere_table D}, \ref{table:Sphere_table E}. We find the $(p,p')$ AMS sphere amplitudes match previous bulk computations in the literature. We also study higher genus amplitudes: we find the $(3,p')$ AMS torus one-point amplitude only partially agrees with results from matrix model computations available in the literature. 
Additionally, we compute many AMS/DMS/EMS amplitudes for which there are currently no MM predictions, and, based on numerical evidence, conjecture formulas for all torus one-point amplitudes in \eqref{eq:conjecture torus 1pt AMS}, \eqref{eq:conjecture torus 1pt DMS}, and \eqref{eq:conjecture torus 1pt EMS}.\footnote{We include \texttt{Mathematica} code in the source directory for the numerical evaluation of string amplitudes, as well as code for the positivity matrix bootstrap of the associated matrix models.}

In \textbf{section 4} we study amplitudes with conformal boundaries. Our first result is classifying inequivalent string theory boundaries in the DMS, following the classification in the AMS \cite{Seiberg:2003nm}. We explain how this leads to suggestive evidence of a four-matrix model dual. In section \ref{sec:4.3disk_diagrams}, assuming a similar dictionary to the AMS, we work out disk diagrams and obtain the density of states $\rho(E)$ of the DMS$_{p,p'}$. We find the density of states is equivalent to that of the AMS$_{p,p'}$. In section \ref{eqn:4.4_cylinders} we compute cylinder diagrams with fixed asymptotic boundary lengths, fixed energies, and ZZ-instantons. We find the cylinders deviate from the universal answers expected for a two-matrix model, which further indicates a multi-matrix structure. In particular, we find scaled linear ramp behavior in the double trumpet (see \eqref{eqn:DMS_doubletrumpet}) with a slope that deviates from the standard Hermitian two-matrix model expectation. Similarly, ZZ-instanton diagrams deviate from the known non-perturbative structure of two-matrix integrals, see \eqref{eqn:ZZ-ZZ_instantons}.

In \textbf{section 5} we first review the AMS duality to two-matrix integrals, and explain how to calculate cylinder diagrams from the matrix model. After reviewing the DMS conjecture, we perform a preliminary positivity matrix bootstrap in section \ref{sec:5.4_matrixbootstrap} to bound the space of possible couplings in which a critical point may reside, with the critical point indicating a continuum string theory. The final results of the bootstrap analysis is figures \ref{fig:PottsBootstrap} and \ref{fig:Mercedesbootstrap}.

\section{The worldsheet theory}
\label{sec:WS theory}

In this section, we briefly review the definition of the minimal string worldsheet CFT.
We begin by recalling the bootstrap solution of Liouville CFT and specify the conventions we will follow throughout the paper.
Next, we describe the A,D, and E series of the minimal model CFTs. 
We then combine these ingredients to present the critical background of minimal string theory.

\subsection{Liouville CFT}
\label{sec:Liouville CFT}

Liouville CFT is a family of non-perturbative solutions to the two-dimensional conformal bootstrap crossing equations, that exists for all complex values of the central charge $c$, except those in the interval $(-\infty,1]$.\footnote{Along this interval, a distinct solution exists, commonly referred to as ``timelike" Liouville CFT. We will not make use of this theory in this paper.}
Following standard convention in the literature, we parametrize the central charge of Liouville CFT as
\begin{align}\label{eq:c Liouv}
c=1+6Q^2 ~, \quad Q=b+b^{-1} ~,
\end{align}
where $b\in\mathbb{C}\backslash\{i\mathbb{R}\}$. 

Liouville CFT has a diagonal spectrum consisting of a continuum of scalar Virasoro primary operators, denoted by $V_P$, with holomorphic and anti-holomorphic conformal weights $h_P=\tilde{h}_P=\frac{c-1}{24}+P^2$. The parameter $P$ is commonly referred to as the Liouville momentum. 
Liouville CFT with central charge $c>1$ and with $P\in\mathbb{R}_{\geq0}$ is a unitary conformal field theory. However, as we will discuss shortly, it is possible and often convenient to analytically continue correlation functions of the theory to arbitrary values of the Liouville momenta. 

The three-point structure constants of Liouville CFT, denoted by $C_b(P_1,P_2,P_3)$, are defined by
\begin{align}
\langle V_{P_1}(z_1,\bar{z}_1) V_{P_2}(z_2,\bar{z}_2) V_{P_3}(z_3,\bar{z}_3) \rangle = \frac{C_b(P_1,P_2,P_3)}{|z_{12}|^{2(h_1+h_2-h_3)}|z_{13}|^{2(h_1+h_3-h_2)}|z_{23}|^{2(h_2+h_3-h_1)}} ~, 
\end{align}
where we introduced the short-hand notation $z_{ij}\equiv z_i-z_j$ and $h_j\equiv h_{P_j}$. These structure constants were bootstrapped by \cite{Dorn:1994xn,Zamolodchikov:1995aa,Teschner:1995yf,Teschner:2001rv} and are given by the celebrated DOZZ formula. 
In this paper, we will adopt a normalization convention for the Liouville vertex operators $V_P$ such that the structure constants take the following form (see for example \cite{Collier:2023cyw}, whose convention we follow)
\begin{align}\label{eq:DOZZ}
C_b(P_1,P_2,P_3) = \frac{\gb(2Q)\gb(\frac{Q}{2}\pm iP_1 \pm iP_2 \pm iP_3)}{\sqrt{2}\gb(Q)^3 \prod_{j=1}^3 \gb(Q\pm 2iP_j)} ~.
\end{align}
The notation $\pm$ in \eqref{eq:DOZZ} indicates a product over all possible sign choices. For instance, in \eqref{eq:DOZZ} both the numerator and the denominator contain nine factors of the function $\gb$.
Here, the double gamma function $\gb(z)$ is a meromorphic function that may be defined as the unique function that satisfies the periodicity functional relations
\begin{align}
\gb(z+b) = \frac{\sqrt{2\pi}b^{bz-\frac{1}{2}}}{\Gamma(bz)} \gb(z) ~, \quad\quad \gb(z+b^{-1}) = \frac{\sqrt{2\pi}b^{-b^{-1}z+\frac{1}{2}}}{\Gamma(b^{-1}z)} \gb(z) ~,
\end{align}
with a normalization $\gb(\frac{Q}{2})=1$. It has simple poles located at
\begin{align}\label{eq:gammab poles}
z = -mb - nb^{-1} ~, \quad m,n\in\mathbb{Z}_{\geq 0} ~.
\end{align}
Note that for rational $b^2$, these poles may overlap yielding higher-order poles. 
$\gb$ admits both an integral representation and --- more relevant for the purposes of this paper --- an infinite product formula that enables efficient numerical evaluation for complex values of $z$ and $b$. The explicit form of this product representation was derived in \cite{Alexanian_2023} and reviewed in \cite[Appendix B]{Collier:2024kwt}, to which we refer the reader for details.\footnote{The code accompanying this paper includes an implementation of this infinite product formula.} 

The two-point function on the plane may be obtained from \eqref{eq:DOZZ} by analytic continuation $P_3\to i\frac{Q}{2}$, upon which the Liouville vertex operator $V_{P_3}$ becomes the identity operator (see for example \cite{Eberhardt:2023mrq}). This results in
\begin{align}
\langle V_{P_1}(z_1,\bar{z}_1)V_{P_1}(z_2,\bar{z}_2) \rangle = \frac{1}{|z_{12}|^{2h_{1}}}\frac{1}{\rho_b(P)}\left( \delta(P_1-P_2) + \delta(P_1+P_2) \right) ~,
\end{align}
where
\begin{align}\label{eq:def rhob}
\rho_b(P) = 4\sqrt{2} \sinh(2\pi bP) \sinh(2\pi b^{-1}P) ~.
\end{align}

This data --- namely, the operator spectrum together with its structure constants --- is sufficient to calculate any $n$-point correlation function on any closed Riemann surface of genus $g$ through a Virasoro conformal block decomposition.
Let us describe the two simplest nontrivial correlation functions that we will study in detail in this paper.
The sphere four-point function may be decomposed as
\begin{align}\label{eq:Liouv sphere 4pt}
\langle V_{P_1}(0) V_{P_2}(z,\bar{z}) V_{P_3}(1) V'_{P_4}(\infty) \rangle = \int_{\mathbb{R}_{\geq 0}} \!\!\d P \rho_b(P) C_b(P_1,P_2,P) C_b(P_3,P_4,P) \nonumber\\ 
\times \mathcal{F}^{(b)}_{0,4}(P_1,P_2,P_3,P_4;P;z) \mathcal{F}^{(b)}_{0,4}(P_1,P_2,P_3,P_4;P;\bar{z}) ~,
\end{align}
where we have used conformal Killing group of the sphere to fix the positions of three vertex operators to $z_1=0$, $z_3=1$, $z_4=\infty$, and $V'_{P_4}(\infty)\equiv \lim_{z_4\to\infty} z^{2h_4} \bar{z}^{2 \tilde{h}_4} V_{P_4}(z_4,\bar{z}_4)$. 
Here, $\mathcal{F}^{(b)}_{0,4}(P_1,P_2,P_3,P_4;P;z)$ denoted the holomorphic sphere four-point non-degenerate Virasoro conformal block with external momenta $P_j$ with $j=1,\ldots,4$, internal momenta $P$, evaluated at the cross-ratio (or modulus) $z$. 
Similarly, the torus one-point function takes the form.
\begin{align}\label{eq:Liouv torus 1pt}
\langle V_{P_1}(0) \rangle_{T^2(\tau)} = \int_{\mathbb{R}_{\geq 0}} \!\!\d P \rho_b(P) C_b(P_1,P,P) \mathcal{F}^{(b)}_{1,1}(P_1;P;\tau) \mathcal{F}^{(b)}_{1,1}(P_1;P;\bar{\tau}) ~,
\end{align}
where we have used the conformal Killing group of the torus to fix the position of the vertex operator to $z_1=0$. Here, $\mathcal{F}^{(b)}_{1,1}(P_1;P;\tau)$ denotes holomorphic torus one-point non-degenerate Virasoro conformal block with external momenta $P_1$, internal momenta $P$, evaluated at the torus modulus $\tau$. 
These Virasoro conformal blocks $\mathcal{F}^{(b)}_{0,4}$, $\mathcal{F}^{(b)}_{1,1}$, as well as their higher-point and higher-genus counterparts, are not known in closed form in general. However, efficient recursion relations exist that allow their computation to very high precision \cite{Zamolodchikov:1984eqp,Zamolodchikov:1987avt,Hadasz:2009db,Cho:2017oxl}. In this paper, we employ these recursion relations using the conventions described in \cite[appendix C.2]{Collier:2023cyw}. For example, the torus block has the form
\be\label{eq:ellipticblocksdef}
\mF_{1,1}^{(b)}(P_1;P|\tau) = q^{h_{P}-\frac{c}{24}} \lr{\prod_{n=1}^\infty \frac{1}{1-q^n}} \mathcal{H}^{(b)}_{1,1}(P_1;P|\tau)\,,
\ee
with $q=e^{2\pi i \tau}$. Here, $\mathcal{H}^{(b)}_{1,1}$ is known as the elliptic conformal block, and it admits a power series expansion in $q$ of the form $\mathcal{H}^{(b)}_{1,1}=1+\frac{h_1(h_1-1)}{2h}q+\mathcal{O}(q^2)$, with $h_1=\frac{Q^2}{4}+P_1^2$ and $h=\frac{Q^2}{4}+P^2$. The coefficients in this elliptic block power series expansion can be computed efficiently using Zamolodchikov’s recursion relations.

Correlation functions in Liouville CFT, such as \eqref{eq:Liouv sphere 4pt} and \eqref{eq:Liouv torus 1pt}, are a priori defined for external momenta $P_j\in\mathbb{R}_{\geq 0}$. However, their analytic structure --- inherited from that of the DOZZ structure constants and the integral over intermediate momenta in the operator product expansion (OPE) --- allows for their continuation to more general complex values of the momenta $P_j$. 
As we will describe in section \ref{sec:stringbkg}, this continuation is necessary for the computation of observables in the full minimal string background.

\subsection{Minimal Model CFTs}
\label{sec:minimal model CFTs}

Let us now describe the three families of minimal model CFTs. These theories are labeled according to the ADE classification and are commonly referred to in the literature by the Dynkin diagram associated to each family. Each class corresponds to a distinct series of consistent two-dimensional rational CFTs, characterized by a finite spectrum of primary operators and specific fusion rules. In the following, we briefly review their defining CFT data. For a recent overview, see \cite{Ribault:2024rvk}.

Minimal models only exist for discrete values of $c_{\mathrm{m}}\leq 1$ and are characterized by two co-prime positive integers $(p,p')$ which have additional restrictions depending on if the model is of A, D, or E type. In all types, the $(p,p')$ model is equivalent to the $(p',p)$ model. 
Their spectra are compactly summarized through degenerate Virasoro representations $\mc{R}_{r,s}\otimes \widetilde{\mc{R}}_{r',s'}$ which are labeled by integers $r,s,r',s'$ and only exist for special values of $c_{\mathrm{m}}$. The central charge, the holomorphic and the anti-holomorphic conformal weights of a primary operator in this representation are given by
\begin{align}
c_{\mathrm{m}} &= 1-6 \frac{(p-p')^2}{p p'} = 1 - 6(\beta - \beta^{-1})^2 ~, \qq \beta^2 = \frac{p}{p'}\,, \label{eq:cmatter} \\
h_{r,s} &= \frac{c_{\mathrm{m}}-1}{24} + P_{r,s}^2 \,, \qq  P_{r,s}= \frac{1}{2}(r \beta^{-1} - s \beta)\,, \label{eq:hmatter} \\
\tilde h_{r',s'} &= \frac{c_{\mathrm{m}}-1}{24} + P_{r',s'}^2 \,. 
\end{align}
The degenerate representation has a holomorphic null-descendant at level $rs$, and an anti-holomorphic null-descendant at level $r' s'$.

\subsubsection{A-series}
\label{sec:Aseries CFT}

The first class of minimal model CFTs is the A-series parametrized by any co-prime positive integers $(p,p')$ with central charge \eqref{eq:cmatter} and Dynkin diagram $(A_{p-1},A_{p'-1})$. The spectrum is given by
\be \label{eq:A_spectrum}
\mathcal{S}_{p, p'}^{\mathrm{A} \text {-series }} = \frac{1}{2} \bigoplus_{\substack{r=1}}^{p-1} \bigoplus_{s=1}^{p'-1} \mathcal{R}_{r, s} \otimes \widetilde{\mathcal{R}}_{r, s} ~,
\ee
The representations are equivalent under $\mathcal{R}_{r, s} \otimes \widetilde{\mathcal{R}}_{r, s}\equiv \mathcal{R}_{p-r,p'-s}\otimes \widetilde{\mathcal{R}}_{p-r, p'-s}$ and are double-counted in the sum, which is corrected by the overall factor of $\frac{1}{2}$, resulting in $\frac{1}{2}(p-1)(p'-1)$ primary fields. We denote Virasoro primaries in these representations by $\widehat{V}_{r,s}$ subject to the same $\mathbb{Z}_2$ identification $\widehat{V}_{r,s}\equiv \widehat{V}_{p-r,p'-s}$. There are many choices of independent representatives. One possible choice makes use of the identity $p'(p - r) - p(p' - s) = - (p' r - p s)$, allowing us to restrict to $(r,s)$ satisfying, for example, $p' r - p s < 0$. Another, more explicit, possible choice is:
\begin{gather}
    (p \t{ even}, p' \t{ odd}) \qq r=1,2,\ldots,p-1\,, \qq  s=1,3,\ldots,p'-2 , ~\\
    (p \t{ odd}, p' \t{ even}) \qq r=1,3,\ldots,p-2\,, \qq  s=1,2,\ldots,p'-1 ~, \\
    (p \t{ odd}, p' \t{ odd}) \qq r=1,2,\ldots,p-1\,, \qq  s=1,3,\ldots,p'-2\,.
\end{gather} 
The conformal weights of these primary vertex operators are $h_{r,s}={\tilde{h}}_{r,s} = \frac{c_{\mathrm{m}}-1}{24} + P_{r,s}^2$ as in \eqref{eq:hmatter}.

\paragraph{Fusion rules and structure constants.}
The fusion rules are specified by
\be\label{eq:Afusion}
\widehat{V}_{r_1,s_1} \times \widehat{V}_{r_2,s_2} = \sum_{r_3 \overset{2}{=} |r_1-r_2|+1 }^{\t{min}(r_1+r_2,2 p-r_1-r_2)-1} ~~ \sum_{s_3 \overset{2}{=} |s_1-s_2|+1 }^{\t{min}(s_1+s_2,2 p'-s_1-s_2)-1} \widehat{V}_{r_3,s_3} ~.
\ee
Here, \eqref{eq:Afusion} is understood as a concise encoding of the fusion rules, rather than as an explicit operator product expansion (OPE), which would in particular include contributions from Virasoro descendants.\footnote{The operators are subject to the $\mathbb{Z}_2$ identification, but the fusion rule above allows us to input any two operators on the LHS without imposing this identification. The output on the RHS is not double-counted: if the fusion produces an operator $\widehat{V}_{r_3,s_3}$, its identified counterpart $\widehat{V}_{p - r_3,\,p' - s_3}$ does not appear separately.} Let us encode this fusion rule in an indicator function $f_{r_1,s_1;r_2,s_2;r_3,s_3} \in \{0,1\}$ which equals one if the representation labeled by $(r_3,s_3)$ appears on the RHS of the fusion, and zero otherwise. 
A more explicit factorized form is
\begin{align}
&f_{r_1,s_1;r_2,s_2;r_3,s_3}= f^{p}_{r_1,r_2,r_3} f^{p'}_{s_1,s_2,s_3}+f^{p}_{r_1,r_2,p-r_3} f^{p'}_{s_1,s_2,p'-s_3} ~, \quad \text{with} \vphantom{\frac{1}{1}}\\
&f_{r_1,r_2,r_3}^{p}=\begin{cases}
1&\text{if } r_3=|r_1-r_2|+1+2\mathbb{Z}_{\geq 0},~r_3\leq r_1+r_2-1,~r_3\leq 2 p-1-r_1-r_2 \,,\\
0&\text{otherwise}\,.
\end{cases}    
\end{align}
The reason for the two terms above is to take into account the $\mathbb{Z}_2$ identification of operators. 

The three-point function on the sphere takes the following form,
\begin{align} \label{eqn:Sphere_3pt_Aseries}
\langle \widehat{V}_{r_1,s_1}(z_1,\bar{z}_1) \widehat{V}_{r_2,s_2}(z_2,\bar{z}_2) \widehat{V}_{r_3,s_3}(z_3,\bar{z}_3) \rangle
= \frac{C^{\mathrm{A}}_{r_1,s_1;r_2,s_2;r_3,s_3}\times f_{r_1,s_1;r_2,s_2;r_3,s_3}}{|z_{12}|^{2(h_1+h_2-h_3)}|z_{13}|^{2(h_1+h_3-h_2)}|z_{23}|^{2(h_2+h_3-h_1)}} ~, 
\end{align}
where now $h_j=h_{r_j,s_j}$.
We have separated the coefficient into two pieces: $f_{r_1,s_1;r_2,s_2;r_3,s_3}$, an indicator function that enforces the fusion rules of the vertex operators, and $C^{\mathrm{A}}_{r_1,s_1;r_2,s_2;r_3,s_3}$, which we refer to as the structure constant.\footnote{Note that the structure constant $C^{\mathrm{A}}$ may be nonzero for vertex operators that do not satisfy the fusion rule; nevertheless, the corresponding three-point function vanishes \cite{Zamolodchikov:2005fy,Ribault:2024rvk}.} 

It is well-known that the A-series structure constant can be expressed in terms of a function analogous to the (time-like) DOZZ formula. In fact, in this paper we find it convenient to work in a convention such that this structure constant is precisely equal to the time-like DOZZ structure constant, in the form given in \cite{Collier:2023cyw}:
\begin{align}\label{eq:C_structureconstants_AMM}
C^{\mathrm{A}}_{r_1,s_1;r_2,s_2;r_3,s_3} 
&= 
\widehat{C}_\beta(P_{r_1,s_1},P_{r_2,s_2},P_{r_3,s_3}) \nonumber\\
&=
\frac{\sqrt{2}\gbeta(\beta+\beta^{-1})^3}{\gbeta(2(\beta +\beta^{-1}))}
\frac{\prod_{j=1}^{3}\gbeta(\beta+\beta^{-1} \pm 2P_{r_j,s_j})}{\gbeta(\frac{\beta+\beta^{-1}}{2}\pm P_{r_1,s_1}\pm P_{r_2,s_2}\pm P_{r_3,s_3}) }
 ~.
\end{align}
As a reminder, the notation $\pm$ means a product over all possible signs.
The function \eqref{eq:C_structureconstants_AMM} is real-valued for real $P_{r_j,s_j}$ and is manifestly symmetric under both permutations of the momenta and reflections $P_{r_j,s_j}\to -P_{r_j,s_j}$, and invariant under $\beta\to\beta^{-1}$.
Furthermore, this convention simplifies the explicit expressions for the various normalization constants associated with the on-shell string vertex operators, as well as other normalization constants in the string path integral. 
More importantly, we expect these conventions to help clarify the relation between the ADE minimal string theories and the more recent Virasoro minimal string and complex Liouville string. 
Additional details on the relation between the conventions in \eqref{eq:C_structureconstants_AMM} and the more standard minimal model CFT normalization are provided in Appendix~\ref{sec:Minimal model CFT conventions}. In these conventions the two-point function and norm of the state is
\begin{align} \label{eqn:two_pt_A_series}
\lb \widehat{V}'_{r,s}(\infty)  \widehat{V}_{r,s}(0)\rb = d(\beta)^2 \frac{\rho_b(i P_{r,s})}{(i P_{r,s})^2} ~,
\end{align}
where $d(\beta)$ is a momentum-independent constant, whose explicit expression is given in \eqref{eq:dbeta6simplify}. However, as will be explained in detail at the end of section \ref{sec:stringbkg}, this constant can be consistently absorbed into certain normalization constants of the full string theory. Accordingly, for the purposes of minimal string theory, the explicit form of $d(\beta)$ is not physically meaningful.

\paragraph{Correlation functions.} With this data, any correlation function on a Riemann surface of genus 
$g$ with $n$ punctures can be computed via a conformal block decomposition. For example, the torus one-point function takes the form\footnote{We have written the formula in a way that generalizes for all minimal models, but for the A-series we have that $h_{r,s}=\tilde h_{r,s}$ for all operators in the spectrum and so the anti-holomorphic block takes as input the same scaling dimensions as the holomorphic block.} 
\begin{align} \label{eqn:torus_1pt_MM}
\lb \widehat{V}_{r_1,s_1}(0) \rb_{T^2(\tau)} = &d(\beta)^{-2} \sum_{(r,s) \in \mathcal{S}^{\t{A}}} \frac{(iP_{r,s})^2}{\rho_b(iP_{r,s})}\, C^{\t{A}}_{r_1,s_1;r,s;r,s} \, f_{r_1,s_1;r,s;r,s} \nn \\
&\times \mF^{(i\beta),\text{deg}}_{1,1}(P_{r_1,s_1};P_{r,s}|\tau) \mF^{(i\beta),\text{deg}}_{1,1}(P_{r_1,s_1};P_{r,s}|\ol\tau) ~.
\end{align}
Here, $ \mF^{(i\beta),\text{deg}}_{1,1}(P_{r_1,s_1};P_{r,s}|\tau)$ denotes holomorphic torus one-point \emph{degenerate} Virasoro conformal block with degenerate external dimension $h_{P_{r_1,s_1}}$, and degenerate internal dimension $h_{P_{r,s}}$, evaluated at the torus modulus $\tau$.\footnote{The superscript $(i\beta)$ in the conformal block includes an $i$ because we parametrize the central charge of the block $\mF^{(b)}_{g,n}$ with a Liouville-like parameter $b$ as in \eqref{eq:c Liouv}, following the notation of \cite{Collier:2023cyw}.} 
We sum over primaries in the A-series spectrum \eqref{eq:A_spectrum}, taking care not to double count identical primaries. For each internal state $|h_{r,s}\rb$ in the loop, we include the appropriate structure constant $C^{\t{A}}_{r_1,s_1;r,s;r,s}$ and the measure factor $(iP_{r,s})^2\rho_b(iP_{r,s})^{-1}$ which simply reflects our normalization of the two-point function (see Appendix \ref{sec:Minimal model CFT conventions}).\footnote{The torus one-point is given by $\Tr \lr{ e^{-\beta H} \mathcal{O}} = \sum_{i} \frac{\lb i | e^{-\beta H} \mathcal{O} | i \rb}{\lb i | i \rb}$ where we sum over all states in the spectrum and we have not assumed $|i\rb$ are normalized. This gives the extra measure factor given by the two-point function described above.}
As a reminder, the structure constant multiplies the indicator function $f$ that encodes the fusion rule \eqref{eq:Afusion} and is therefore only non-zero when the fusion is allowed.

Similarly, the sphere four-point function takes the form
\begin{align}\label{eq:sphere_4pt_AMM}
\langle \widehat{V}_{r_1,s_1}&(0) \widehat{V}_{r_2,s_2}(z,\bar{z}) \widehat{V}_{r_3,s_3}(1) \widehat{V}'_{r_4,s_4}(\infty) \rangle \nn\\
&= d(\beta)^{-2} \sum_{(r,s) \in \mathcal{S}^{\t{A}}} \frac{(iP_{r,s})^2}{\rho_b(iP_{r,s})}\, C^{\t{A}}_{r_1,s_1;r_2,s_2;r,s} \, f_{r_1,s_1;r_2,s_2;r,s} \, C^{\t{A}}_{r_3,s_3;r_4,s_4;r,s} \, f_{r_3,s_3;r_4,s_4;r,s} \nn \\
&\quad \times \mF^{(i\beta),\text{deg}}_{0,4}(P_{r_1,s_1},\ldots,P_{r_4,s_4};P_{r,s}|\tau) \mF^{(i\beta),\text{deg}}_{0,4}(P_{r_1,s_1},\ldots,P_{r_4,s_4};P_{r,s}|\ol\tau) ~,
\end{align}
in the s-channel decomposition, where the first vertex operator is fused with the second, and the third with the fourth. Again, the allowed fusion \eqref{eq:Afusion} is imposed implicitly through the indicator functions $f$.

While recursion relations are known for non-degenerate Virasoro conformal blocks, we are not aware of analogous results for degenerate Virasoro blocks, apart from the sphere four-point blocks\footnote{There is a conjecture \cite{Ribault:2018jdv} that one may simply take the limit of the non-degenerate block to degenerate values for the external and internal weights to get degenerate blocks on the sphere. We have found this procedure consistent with crossing in all of our examples.} and the Virasoro torus characters, the latter of which are known in closed form.
In general, degenerate blocks contain additional towers of null descendants that must be removed from the spectrum. 
However, for the purposes of this paper --- namely the numerical evaluation of such correlation functions further integrated over the relevant moduli space of the string diagram --- we find it sufficient to approximate correlation functions in the minimal model CFTs by directly using the non-degenerate blocks. 
Indeed, as we will see in section \ref{sec:String Perturbation Theory} the error in this approximation occurs at level $rs$ for a state (external or internal) with weight $h_{r,s}$. In the elliptic frame for the conformal blocks, introduced in section \ref{sec:String Perturbation Theory}, this corresponds to an error in the conformal block expansion at order $q^{rs}$, where $q=\e^{-c\pi t}$ (with $c$ a constant) is a plumbing parameter for the Riemann surface in question. Thus, the approximation introduces only an exponentially small correction in the numerical calculation of string amplitudes performed in this paper.

\subsubsection{D-series}
\label{sec:Dseries CFT}

The D-series minimal CFTs exist for co-prime $(p,p')$ with even $p \geq 6$, and odd $p'\geq 3$, and with central \eqref{eq:cmatter} and Dynkin diagram $(A_{p'-1},D_{\frac{p}{2}+1})$.\footnote{For each $(p,p')$ model, the $(p',p)$ model is identical, and so it does not matter whether $p$ or $p'$ is taken to be the even integer.} There are two types of D-series minimal CFTs, depending on whether $p=0$ or $2$ mod $4$, with spectra given by
\be
\begin{aligned}\label{eq:D_spectrum}
& \mathcal{S}_{p, p'}^{\mathrm{D} \text {-series }} \underset{p=0 \,\t{mod}\,4}{=} \frac{1}{2} \bigoplus_{ r\overset{2}{=}1}^{p-1} \bigoplus_{s=1}^{p'-1} ~~\,\mathcal{R}_{r, s} \otimes \widetilde{\mathcal{R}}_{r, s} 
~~\,\oplus~\frac{1}{2} \bigoplus_{r\overset{2}{=}2}^{p-2} \bigoplus_{s=1}^{p'-1}~~\, \mathcal{R}_{r, s} \otimes \widetilde{\mathcal{R}}_{p-r, s} ~, \\
& \mathcal{S}_{p, p'}^{\mathrm{D} \text {-series }} \underset{p=2\,\t{mod}\,4}{=} \frac{1}{2} \bigoplus_{r\overset{2}{=}1}^{p-1} \bigoplus_{s=1}^{p'-1} \underbrace{\vphantom{\int}\mathcal{R}_{r, s} \otimes \widetilde{\mathcal{R}}_{r, s}}_{\t{``diagonal" primaries}} 
\oplus~\frac{1}{2} \bigoplus_{r\overset{2}{=}1}^{p-1} \bigoplus_{s=1}^{p'-1} \underbrace{\vphantom{\int}\mathcal{R}_{r, s} \otimes \widetilde{\mathcal{R}}_{p-r, s}}_{\t{``non-diagonal" primaries}}  ~.
\end{aligned}
\ee
The representations again satisfy $\mathcal{R}_{r, s} \otimes \widetilde{\mathcal{R}}_{r, s}\equiv \mathcal{R}_{p-r,p'-s}\otimes \widetilde{\mathcal{R}}_{p-r, p'-s}$ and $\mathcal R_{r,s} \otimes \widetilde{\mathcal R}_{p-r,s} \equiv \mathcal R_{p-r,p'-s} \otimes \widetilde{\mathcal R}_{r,p'-s}$ and are double-counted in the sum, which is corrected by the factor of $\frac{1}{2}$.

The new feature of D-series compared to the A-series is that there are additional ``non-diagonal" primary fields which, with the exception of the case $\mathcal{R}_{\frac{p}{2},s} \otimes \widetilde{\mathcal{R}}_{\frac{p}{2},s}$, have $h \neq \tl h$.\footnote{The ``diagonal" primaries of the $(p,p')$ D-series model consist of \emph{half} of the diagonal fields of the corresponding $(p,p')$ A-series model \eqref{eq:A_spectrum}.} The primary vertex operators will be labeled
\begin{align}\label{eq:Dseries vertexops}
\text{Diagonal: }& \mc{R}_{r,s} \otimes \widetilde{\mc{R}}_{r,s}
&:&~ \widehat{V}^D_{r,s}~~~r=1,3,\ldots,p-1,~~~s=1,2,\ldots,p'-1 
\nn \\
\text{Non-diagonal: }& \mc{R}_{r,s} \otimes \widetilde{\mc{R}}_{p-r,s} 
&:&~ \widehat{V}^N_{r,s}~~~r=
\begin{cases}
2,4,\ldots,p-2 &\text{ when }p=0\text{ mod }4\\
1,3,\ldots,p-1 &\text{ when }p=2\text{ mod }4
\end{cases} ~,
\nn \\
&&& \quad\quad~~ s=1,2,\ldots,p'-1 
\end{align}
The list double counts $\widehat{V}_{r,s}\equiv \widehat{V}_{p-r,p'-s}$ and a single set of independent operators must be chosen. A consistent choice is restricting to $s=1,3,\ldots,p'-2$ for both diagonal and non-diagonal.

For $p = 0~\text{mod}\,4$ the spectrum contains $\frac{1}{4}p (p'-1) + \frac{1}{4}(p-2)(p'-1) = \frac{1}{2}(p-1)(p'-1)$ primaries. 
For $p = 2~\text{mod}\,4$ the spectrum contains $\frac{1}{4}p (p'-1) + \frac{1}{4}p (p'-1) = \frac{1}{2}p(p'-1)$ primaries. 
The representation $R_{p/2,s}\otimes \widetilde{R}_{p/2,s}$ is present in both the diagonal and non-diagonal sectors. Thus, for each value of $s$ we have two distinct primaries, $\widehat{V}_{p/2,s}^D$ and $\widehat{V}_{p/2,s}^N$, with the same conformal weight and no spin ($h=\tilde{h}$). There are $(p'-1)/2$ such pairs in total.

\paragraph{Fusion rules and structure constants.}
The D-series fusion rule is given by
\begin{align} \label{eq:D_series_fusion}
\widehat{V}_{r_1,s_1}^{X_1} \times \widehat{V}_{r_2,s_2}^{X_2} &= \sum_{r\overset{2}{=}|r_1-r_2|+1}^{\min(r_1+r_2-1,2p-r_1-r_2-1)} \sum_{s\overset{2}{=}|s_1-s_2|+1}^{\min(s_1+s_2-1,2p'-s_1-s_2-1)} \widehat{V}_{r,s}^{X_3} ~,
\end{align}
where the superscript takes the values $X\in\{D,N\}$, which refer to the diagonal and non-diagonal primaries in the spectrum as in \eqref{eq:Dseries vertexops}. Additionally, the fusion rules \eqref{eq:D_series_fusion} satisfy conservation of diagonality
\be\label{eq:Dseries consevation of diagonality}
D\times D= D~, \quad D\times N= N~,\quad N\times N= D ~.
\ee
The structure constants are defined by
\begin{align}
\langle \widehat{V}^{X_1}_{r_1,s_1}&(z_1,\bar{z}_1) \widehat{V}^{X_2}_{r_2,s_2}(z_2,\bar{z}_2) \widehat{V}^{X_3}_{r_3,s_3}(z_3,\bar{z}_3) \rangle 
\nn\\
&= \frac{C^{\t{D},\, X_1 X_2 X_3}_{r_1,s_1;r_2,s_2;r_3,s_3}\times f_{r_1,s_1;r_2,s_2;r_3,s_3}}{z_{12}^{h_1+h_2-h_3}\bar{z}_{12}^{\tilde{h}_1+\tilde{h}_2-\tilde{h}_3}z_{13}^{h_1+h_3-h_2}\bar{z}_{13}^{\tilde{h}_1+\tilde{h}_3-\tilde{h}_2}z_{23}^{h_2+h_3-h_1}\bar{z}_{23}^{\tilde{h}_2+\tilde{h}_3-\tilde{h}_1}} ~, 
\end{align}
where again $f_{r_1,s_1;r_2,s_2;r_3,s_3}$ is an indicator function for the fusion rules \eqref{eq:D_series_fusion}. 
In view of \eqref{eq:Dseries consevation of diagonality}, there are two nontrivial structure constants, $C^{\t{D},\,DDD}$ and $C^{\t{D},\,DNN}$. 
The first takes precisely the same form as in the A-series. In our conventions, further explained in Appendix \ref{sec:Minimal model CFT conventions}, it is given by the timelike DOZZ formula
\begin{align} \label{eq:CDseries DDD}
C^{\t{D},\,DDD}_{r_1,s_1;r_2,s_2;r_3,s_3}
&= 
\widehat{C}_\beta(P_{r_1,s_1},P_{r_2,s_2},P_{r_3,s_3}) \nonumber\\
&=
\frac{\sqrt{2}\gbeta(\beta+\beta^{-1})^3}{\gbeta(2(\beta +\beta^{-1}))}
\frac{\prod_{j=1}^{3}\gbeta(\beta+\beta^{-1} \pm 2P_{r_j,s_j})}{\gbeta(\frac{\beta+\beta^{-1}}{2}\pm P_{r_1,s_1}\pm P_{r_2,s_2}\pm P_{r_3,s_3}) }
 ~.
\end{align} 
Note that, in particular, the D-series contains a subset of vertex operators that close under the OPE among themselves. We will make use of this fact to relate certain perturbative string amplitudes in the D-series to corresponding amplitudes in the A-series minimal string. 
The second nontrivial structure constant can also be brought to a form similar to the timelike DOZZ structure constant,
\begin{align}\label{eq:CDseries DNN}
C&^{\t{D},\,DNN}_{r_1,s_1;r_2,s_2;r_3,s_3} \nn\\
&= (-1)^{\frac{r_3}{2}-\frac{p}{4}}(-1)^{\frac{r_1-1}{2}}\frac{\sqrt{2}\gbeta(\beta+\beta^{-1})^3}{\gbeta(2(\beta +\beta^{-1}))} \nn\\
&\quad\times\frac{\gbeta(\beta+\beta^{-1} \pm 2 P_{r_1,s_1})\prod_{j=2}^{3}\big[\gbeta(\beta+\beta^{-1} \pm 2 P_{r_j,s_j})\gbeta(\beta+\beta^{-1} \pm 2 \widetilde{P}_{r_j,s_j})\big]^{\frac{1}{2}}}{ \gbeta(\frac{\beta+\beta^{-1}}{2} + P_{r_1,s_1}\pm P_{r_2,s_2}\pm P_{r_3,s_3}) \gbeta(\frac{\beta+\beta^{-1}}{2} - P_{r_1,s_1}\pm \widetilde{P}_{r_2,s_2}\pm \widetilde{P}_{r_3,s_3}) } ~.
\end{align}
Here, the first vertex operator with momenta $P_{r_1,s_1}$ is diagonal, while the second and third are non-diagonal. For the latter, in \eqref{eq:CDseries DNN} we have also introduced the ``left-moving" Liouville momenta,
\begin{align}
\widetilde{P}_{r,s} \equiv P_{p-r,s} ~,
\end{align}
with the RHS defined in \eqref{eq:hmatter}. 
The structure constant \eqref{eq:CDseries DNN} is manifestly invariant under $\beta\to\beta^{-1}$ and is reflection symmetric under $P_{r_j,s_j}\to-P_{r_j,s_j}$, $\widetilde{P}_{r_j,s_j}\to-\widetilde{P}_{r_j,s_j}$ for $j=2,3$. Although not immediately obvious, it is also invariant under the exchange $P_{r_j,s_j}\leftrightarrow \widetilde{P}_{r_j,s_j}$, and independently $P_{r_1,s_1}\to -P_{r_1,s_1}$. The normalization of the states is 
\be
\lb \widehat{V}^{'D}_{r,s}(\infty)  \widehat{V}^D_{r,s}(0)\rb = d(\beta)^{2} \frac{\rho_b(i P_{r,s})}{(i P_{r,s})^2}\,, \q \lb \widehat{V}^{'N}_{r,s}(\infty)  \widehat{V}^N_{r,s}(0)\rb = d(\beta)^{2} \frac{4\sqrt{2}\sin(2\pi bP_{r,s})\sin(2\pi b^{-1}\widetilde{P}_{r,s})}{(-1)^{\frac{r}{2}-\frac{p}{4}}P_{r,s}\widetilde{P}_{r,s}}\,.
\ee

\paragraph{Torus one-point and sphere four-point functions.} 
The torus one-point function takes a similar form as in \eqref{eqn:torus_1pt_MM}, except in the present case we also have a contribution from non-diagonal intermediate states,
\begin{align} \label{eqn:DseriesMM torus1pt}
\lb \widehat{V}^{D}_{r_1,s_1}(0) \rb_{T^2(\tau)} 
&= d(\beta)^{-2} \sum_{(r,s) \in \mathcal{S}^{\t{D},\,D}} \frac{(iP_{r,s})^2}{\rho_b(iP_{r,s})}\, C^{\t{D},\,DDD}_{r_1,s_1;r,s;r,s} \, f_{r_1,s_1;r,s;r,s} \nn \\
&\quad\quad\quad\quad\quad \times \mF^{(i\beta),\text{deg}}_{1,1}(P_{r_1,s_1};P_{r,s}|\tau) \mF^{(i\beta),\text{deg}}_{1,1}(P_{r_1,s_1};P_{r,s}|\ol\tau) \nn\\
&\quad +d(\beta)^{-2} \sum_{(r,s) \in \mathcal{S}^{\t{D},\,N}} \frac{(-1)^{\frac{r}{2}-\frac{p}{4}}P_{r,s}\widetilde{P}_{r,s}}{4\sqrt{2}\sin(2\pi bP_{r,s})\sin(2\pi b^{-1}\widetilde{P}_{r,s})} \, C^{\t{D},\,DNN}_{r_1,s_1;r,s;r,s} \, f_{r_1,s_1;r,s;r,s} \nn \\
&\quad\quad\quad\quad\quad \times \mF^{(i\beta),\text{deg}}_{1,1}(P_{r_1,s_1};P_{r,s}|\tau) \mF^{(i\beta),\text{deg}}_{1,1}(P_{r_1,s_1};\widetilde{P}_{r,s}|\ol\tau) ~.
\end{align}
Here, $\mathcal{S}^{\t{D},\,D}$ and $\mathcal{S}^{\t{D},\,N}$ denote the diagonal and non-diagonal part of the spectrum \eqref{eq:D_spectrum}, respectively.
Note that although the term inside the square root in the numerator of \eqref{eqn:DseriesMM torus1pt} is not sign-definite in general, this factor appears squared in \eqref{eqn:DseriesMM torus1pt}. 
In fact, this remains true for higher-genus and higher-point functions whenever the external vertex operator carries no spin, which is precisely the situation relevant for string theory, as discussed in section \ref{sec:stringbkg}.
Also note that the measure factor for the integration over non-diagonal fields reduces to that for diagonal fields, up to the sign $(-1)^{\frac{r}{2}-\frac{p}{4}}$, when taking $\widetilde{P}_{r,s}= P_{r,s}$. In particular, this is true for non-diagonal vertex operators with $r=\frac{p}{2}$, and the sign factor is precisely 1.

On the other hand, due to the conservation of diagonality \eqref{eq:Dseries consevation of diagonality}, the torus one-point function of a non-diagonal field vanishes
\be
\lb \widehat{V}^N_{r,s} \rb_{T^2(\tau)}=0 .
\ee
Similarly, the sphere four-point function takes the form
\begin{align}\label{eq:sphere_4pt_DMM}
\langle \widehat{V}^{X_1}_{r_1,s_1}&(0) \widehat{V}^{X_2}_{r_2,s_2}(z,\bar{z}) \widehat{V}^{X_3}_{r_3,s_3}(1) \widehat{V}^{'X_4}_{r_4,s_4}(\infty) \rangle \nn\\
&= d(\beta)^{-2} \sum_{(r,s) \in \mathcal{S}^{\t{D},\,D}} \frac{(iP_{r,s})^2}{\rho_b(iP_{r,s})}\, C^{\t{D},\,X_1 X_2 D}_{r_1,s_1;r_2,s_2;r,s} \, f_{r_1,s_1;r_2,s_2;r,s} \, C^{\t{D},\,X_3 X_4 D}_{r_3,s_3;r_4,s_4;r,s} \, f_{r_3,s_3;r_4,s_4;r,s} \nn \\
&\quad\quad \times \mF^{(i\beta),\text{deg}}_{0,4}(P_{r_1,s_1},\ldots,P_{r_4,s_4};P_{r,s}|\tau) \mF^{(i\beta),\text{deg}}_{0,4}(P^{X_1}_{r_1,s_1},\ldots,P^{X_4}_{r_4,s_4};P_{r,s}|\ol\tau) \nn \\
&\quad + d(\beta)^{-2} \sum_{(r,s) \in \mathcal{S}^{\t{D},\,N}} \frac{(-1)^{\frac{r}{2}-\frac{p}{4}}P_{r,s}\widetilde{P}_{r,s}}{4\sqrt{2}\sin(2\pi bP_{r,s})\sin(2\pi b^{-1}\widetilde{P}_{r,s})} \nn\\
&\quad\quad \times C^{\t{D},\,X_1 X_2 N}_{r_1,s_1;r_2,s_2;r,s} \, f_{r_1,s_1;r_2,s_2;r,s} \, C^{\t{D},\,X_3 X_4 N}_{r_3,s_3;r_4,s_4;r,s} \, f_{r_3,s_3;r_4,s_4;r,s} \nn \\
&\quad\quad \times \mF^{(i\beta),\text{deg}}_{0,4}(P_{r_1,s_1},\ldots,P_{r_4,s_4};P_{r,s}|\tau) \mF^{(i\beta),\text{deg}}_{0,4}(P^{X_1}_{r_1,s_1},\ldots,P^{X_4}_{r_4,s_4};\widetilde{P}_{r,s}|\ol\tau) ~,
\end{align}
where we have again separated the contributions from diagonal and non-diagonal intermediate states. Here, $P^{X=D}_{r,s}=P_{r,s}$ when the external state is diagonal, and $P^{X=N}_{r,s}=\widetilde{P}_{r,s}$ when it is non-diagonal.
Note that for the purposes of studying the minimal string theories, we will only be interested in the case where the external vertex operators carry no spin, $h=\tilde{h}$.

\subsubsection{E-series}
\label{sec:Eseries CFT}

The E-series minimal model CFTs exist for co-prime $(p,p')$ with $p=12,18,30$. Since the structure constants are only explicitly known for $p=12$ \cite{Nivesvivat:2025odb}, we restrict our attention to this case in this paper. The spectrum is 
\bal
\mc{S}^{\t{E-series}}_{12,p'}&=\frac{1}{2}\bigoplus_{s=1}^{p'-1}|\mc{R}_{1,s}\oplus \mc{R}_{7,s}|^2\oplus |\mc{R}_{4,s}\oplus \mc{R}_{8,s}|^2 \oplus|\mc{R}_{5,s}\oplus \mc{R}_{11,s}|^2\\
&=\bigoplus_{s=1}^{p'-1}|\mc{R}_{1,s}\oplus \mc{R}_{7,s}|^2 \oplus\frac{1}{2}\bigoplus_{s=1}^{p'-1} |\mc{R}_{4,s}\oplus \mc{R}_{8,s}|^2
\eal
A single copy of the spectrum is given by the following primary fields 
\bal \label{eqn:E_series_fields}
\text{Diagonal } \mc{R}_{r,s} \otimes \widetilde{\mc{R}}_{r,s}&:~ \widehat{V}^{D_r}_{r,s}~~~r=1,4,7,~~~s=1,2,\ldots,p'-1\\
\text{Non-diagonal } \mc{R}_{4,s} \otimes \widetilde{\mc{R}}_{8,s}&:~ \widehat{V}^{N_2}_{4,s}~~~s=1,2,\ldots,p'-1 \\
\text{Non-diagonal } \mc{R}_{1,s} \otimes \widetilde{\mc{R}}_{7,s}&:~ \widehat{V}^{N_3}_{1,s}~~~s=1,2,\ldots,p'-1 \\
\text{Non-diagonal } \mc{R}_{7,s} \otimes \widetilde{\mc{R}}_{1,s}&:~ \widehat{V}^{\widetilde{N}_{3}}_{7,s}~~~s=1,2,\ldots,p'-1\,,
\eal
where now there are many superscript labels: $D_r$ with $r=1,4,7$, $N_2, N_3,$ and $\widetilde{N}_3$. 
There are $3(p'-1)$ primaries with $h=\tilde h$ all of which are the diagonal fields. 

As the fusion rules and the structure constants are rather intricate, we defer their explicit form, along with a discussion of correlation functions, to Appendix \ref{app:E-series}.

\subsection{The string backgrounds}
\label{sec:stringbkg}

Let us now combine the ingredients of the preceding sections in order to form the worldsheet string theories \eqref{eq:minimal string}. 
The first elementary condition is that the total central charge of the worldsheet CFT \eqref{eq:minimal string} is equal to zero. This ensures the absence of a quantum Weyl anomaly, yielding a consistent critical string theory background. 
From parametrizations of the central charges in \eqref{eq:c Liouv} and \eqref{eq:cmatter}, we see that the condition $c+c_{\mathrm{m}}=26$ fixes $\beta = \pm b^{\pm1}$, allowing for four possible sign combinations. However, the underlying worldsheet CFT data discussed in the previous section are invariant under these sign transformations, which reflect a redundancy in the parametrization of the central charge. We will therefore simply choose
\begin{align}
\beta = b = \sqrt{\frac{p}{p'}} ~.
\end{align}
Henceforth, we will mostly use the parameter $b$. When possible we will take $p'>p$. Next, we discuss the physical, or on-shell, states of the string.

\subsubsection{Vertex operators}
\label{sec:onshell ops}

The on-shell vertex operators of the string theories are constructed by combining primary fields of the minimal model CFT with Liouville CFT primaries to obtain vertex operators of total dimension $(h,\tilde h)=(1,1)$. Including the conformal ghosts $\mathfrak{c\tilde{c}}$ then yields vertex operators of dimension $(0,0)$, inserted at fixed positions on the worldsheet, as usual in bosonic string theory. We now describe the on-shell vertex operators for each minimal string.

\paragraph{A-series.}
The A-series \eqref{eq:A_spectrum} has $\frac{1}{2}(p-1)(p'-1)$ primaries with $h=\tilde h$ and all of them can be dressed to give ``tachyon" vertex operators
\begin{gather} \label{eqn:Tachyons_Aseries}
\cT_{r,s} = \mathfrak{c\tilde{c}}\,\mathcal{N}^{(b)}_{r,s}\, \widehat V_{r,s} V_{i P_{r,s}}, \qq i P_{r,s}=\frac{i}{2} (r b-s b^{-1})\,,\\
(p \t{ even}, p' \t{ odd}): \qq r=1,2,\ldots,p-1, \qq  s=1,3,\ldots,p'-2\,. \nn \\ 
(p \t{ odd}, p' \t{ even}): \qq r=1,3,\ldots, p-2, \qq s=1,2,\ldots,p'-1\,. \nn \\ 
\,\,(p \t{ odd}, p' \t{ odd}): \qq r=1,2,\ldots, p-1, \qq s=1,3,\ldots,p'-2\,. \nn
\end{gather}
As noted at the end of section \ref{sec:Liouville CFT}, while the defining (external) spectrum of unitary Liouville CFT is given by $P\in\mathbb{R}_{\geq0}$, the CFT data admits analytic continuation in both the central charge and the Liouville momenta. In this sense, we define vertex operators with complex Liouville momenta, as in \eqref{eqn:Tachyons_Aseries}, and their associated correlation functions. The integration over intermediate states is still taken along the real line, as in \eqref{eq:Liouv sphere 4pt}, although we may deform this contour when necessary.

In \eqref{eqn:Tachyons_Aseries}, $\cN^{(b)}_{r,s}$ denotes an arbitrary normalization constant for the string on-shell vertex operators. As we will see later in sections \ref{sec:3pt amp} and \ref{sec:torus 1pt amp}, this normalization is in fact independent of both the external momenta $P_{r,s}$ (i.e. of the labels $(r,s)$).

\paragraph{D-series.} 
In the D-series minimal string, on-shell vertex operators are constructed by combining minimal model CFT primaries with $h=\tilde h$, from either the diagonal or non-diagonal sector, with Liouville CFT primaries. 
This leads to a total of $\frac{1}{4}(p'-1)(p+2)$ tachyon vertex operators:
\begin{align}
\cT_{r,s}^D &= \mathfrak{c\tilde{c}}\,\mc{N}^{(b),D}_{r,s} \, \widehat{V}^D_{r,s} V_{i P_{r,s}} ~, \qq r=1,3,\ldots, p-1 ~, \qq s=1,3,\ldots,p'-1 ~,\label{eqn:Tachyons_Dseries D} \\
\cT_{r,s}^N &= \mathfrak{c\tilde{c}}\,\mc{N}^{(b),N}_{r,s} \, \widehat{V}^N_{r,s} V_{i P_{r,s}} ~, \qq r= \frac{p}{2} ~, \qq\qq\qq\quad\, 
s=1,3,\ldots,p'-1 ~, \label{eqn:Tachyons_Dseries N}
\end{align}
where we have included arbitrary normalization constants $\mc{N}^{(b),D}_{r,s}$ and $\mc{N}^{(b),N}_{r,s}$ as above.

\paragraph{E-series.} For the $(12,p)$ E-series minimal string, the tachyon vertex operators are constructed from the E-series primaries in the diagonal sector \eqref{eqn:E_series_fields},
\begin{gather}\label{eqn:Tachyons_Eseries}
\cT_{r,s} = \mathfrak{c\tilde{c}}\,\mc{N}^{(b),E}_{r,s} \, \widehat{V}^{D_r}_{r,s} V_{i P_{r,s}} ~, \quad
r=1,4,7 ~, \qq s=1,2,\ldots,p-1 ~, \nn
\end{gather}
for a total of $3(p-1)$ on-shell string states.

\subsubsection{String amplitudes}

The main observables of the string theories \eqref{eq:minimal string} are computed as worldsheet CFT correlators of the on-shell string vertex operators, integrated over the moduli space $\mathcal{M}_{g,n}$ of an $n$-punctured genus-$g$ Riemann surface $\Sigma_{g,n}$ representing the corresponding string diagram.
These quantities are sometimes referred to as ``correlation numbers," and are defined as:
\begin{align}\label{eq:string amplitudes def}
\mathsf{N}^{(b)}_{g,n}(\mathbf{P_{r,s}}) \equiv C^{(b)}_{\Sigma_g} \int_{\mathcal{M}_{g,n}} \Big\langle \prod_{k=1}^{3g-3+n} \mathcal{B}_k \widetilde{\mathcal{B}}_k \prod_{j=1}^{n} \mathcal{T}_{r_j,s_j} \Big\rangle_{\Sigma_{g,n}} ~,
\end{align}
where $\mathbf{P_{r,s}}\equiv(P_{r_1,s_1},\ldots,P_{r_n,s_n})$ is a shorthand notation for the discrete momenta of the string on-shell tachyon vertex operators $\mathcal{T}_{r_j,s_j}$ for the appropriate minimal string background.\footnote{Note that the label $P$ on the LHS is somewhat superfluous, as we only need the labels $(r_j,s_j)$.} For the D-series minimal string, we further decorate the label $P^X_{r_j,s_j}$ with $X=D$ or $N$ to indicate whether the tachyon vertex operator is built from a primary in the diagonal or non-diagonal sector of the minimal model CFT. 
In \eqref{eq:string amplitudes def}, $\mathcal{B}_k,\widetilde{\mathcal{B}}_k$ denote the holomorphic and anti-holomorphic $\mathfrak{b}$-ghost insertions associated with the moduli of the Riemann surface $\Sigma_{g,n}$, while $C^{(b)}_{\Sigma_g}$ denotes the normalization constants of the string path integral on a genus-$g$ Riemann surface $\Sigma_{g}$  \cite{Polchinski:1998rq}.

Lastly, we note that the constant ($\beta$-dependent) prefactors $d(\beta)^{-2}$, appearing for example in \eqref{eqn:torus_1pt_MM}, can be consistently absorbed into the normalization of the on-shell string vertex operators $\mathcal{N}^{(b)}_{r,s}$ and the string path integral normalization constants $C^{(b)}_{\Sigma_g}$. 
To see this, note that an $n$-point string amplitude at genus $g$, when expressed in terms of a chosen conformal block decomposition, depends on these normalization factors as $C_{\Sigma_g}^{(b)}\big(\prod_{j=1}^n\mathcal{N}^{(b)}_{r_j,s_j}\big)\left(d^{-2}\right)^{\# \text{edges}}$. Here, $\#$ edges denotes the number of internal lines in the conformal block decomposition, i.e. the number of internal Liouville momenta to be integrated. This number is equal to $3g-3+n$. Thus, the total prefactor can be rewritten as as $\big(C_{\Sigma_g}^{(b)}(d^{-2})^{3g-3}\big)\big(\prod_{j=1}^n\mathcal{N}^{(b)}_{r_j,s_j}d^{-2}\big)$. 
Therefore, we see that the dependence on $d(\beta)$ can be fully reabsorbed into a redefinition of $\mathcal{N}^{(b)}_{r,s}$ and $C_{\Sigma_g}^{(b)}$.

\section{Perturbative string amplitudes}
\label{sec:String Perturbation Theory}

\subsection{Sphere three-point diagram}
\label{sec:3pt amp}

The simplest nontrivial amplitude to consider is the three-point sphere diagram, which requires no moduli integration. The amplitude is then proportional to the product of the three-point structure constants of the constituent CFTs. 

\paragraph{A-series.} We obtain \cite{Seiberg:2003nm}
\begin{align}
\mathsf{N}^{(b),\,\t{A}}_{0,3}&(P_{r_1,s_1},P_{r_2,s_2},P_{r_3,s_3}) \nn\\
&= C^{(b)}_{\t{S}^2} \Big(\prod_{j=1}^3 \mathcal{N}^{(b)}_{r_j,s_j}\Big) C^{\mathrm{A}}_{r_1,s_1;r_2,s_2;r_3,s_3} C_b(iP_{r_1,s_1},iP_{r_2,s_2},iP_{r_3,s_3}) f_{r_1,s_1;r_2,s_2;r_3,s_3} \nn\\
&= C^{(b)}_{\t{S}^2} \Big(\prod_{j=1}^3 \mathcal{N}^{(b)}_{r_j,s_j}\Big) \widehat{C}_b(P_{r_1,s_1},P_{r_2,s_2},P_{r_3,s_3}) C_b(iP_{r_1,s_1},iP_{r_2,s_2},iP_{r_3,s_3}) f_{r_1,s_1;r_2,s_2;r_3,s_3} \nn\\
&= C^{(b)}_{\t{S}^2} \Big(\prod_{j=1}^3 \mathcal{N}^{(b)}_{r_j,s_j}\Big)f_{r_1,s_1;r_2,s_2;r_3,s_3} ~,
\end{align}
where we used the fact that the timelike DOZZ structure constant $\widehat{C}_b$ is, up to a factor of $i$ in its arguments, the reciprocal of the spacelike DOZZ structure constant $C_b$. Note that this is precisely the same cancellation that occurs in the Virasoro minimal string \cite{Collier:2023cyw}. 
As is standard in the minimal string theory literature, we next fix the arbitrary on-shell string normalization constants by requiring that the three-point amplitude is equal to the indicator function enforcing the fusion rule,
\begin{align}
\mathsf{N}^{(b),\,\t{A}}_{0,3}(P_{r_1,s_1},P_{r_2,s_2},P_{r_3,s_3}) \overset{!}{=} f_{r_1,s_1;r_2,s_2;r_3,s_3} ~.
\end{align}
This in particular implies that $\mathcal{N}^{(b)}_{r,s} = \mathcal{N}^{(b)}$ is independent of the external momenta $P_{r,s}$ (or the labels $(r,s)$) and that
\begin{align}\label{eq:fixCS2}
C^{(b)}_{\t{S}^2} = \big(\mathcal{N}^{(b)}\big)^{-3} ~.
\end{align}

\paragraph{D-series.} The D-series proceeds in the same way. As discussed in \eqref{eqn:Tachyons_Dseries N}, there are two kinds of tachyon operators, each originating from the diagonal and non-diagonal sectors, respectively. 
Tachyons in the diagonal sector have precisely the same three-point amplitude as in the A-series. Thus, we arrive at the same conclusion as in the preceding paragraph: We normalize the three-point amplitude to be exactly the indicator function that enforces the fusion rule, $\mathsf{N}^{(b),\,\t{D}}_{0,3}(P^D_{r_1,s_1},P^D_{r_2,s_2},P^D_{r_3,s_3}) \overset{!}{=} f_{r_1,s_1;r_2,s_2;r_3,s_3}$, and we arrive at the same condition \eqref{eq:fixCS2} relating the sphere string path integral normalization and the D-series vertex operator normalization constants.
The other nonzero three-point amplitudes involve one diagonal and two non-diagonal tachyon on-shell states. As shown in \eqref{eqn:Tachyons_Dseries N}, the latter have $r=\frac{p}{2}$ and thus $\widetilde{P}_{r,s} = P_{p-r,s} = P_{r,s}$. 
Thus, the relevant minimal model structure constant $C^{\t{D},\,DNN}_{r_1,s_1;r_2,s_2;r_3,s_3}$ reduces to the timelike Liouville structure constant, up to a sign factor $(-1)^{\frac{r_1-1}{2}}$ as seen from \eqref{eq:CDseries DNN}. We therefore obtain
\begin{align}
\mathsf{N}^{(b),\,\t{D}}_{0,3}(P^D_{r_1,s_1},P^N_{r_2,s_2},P^N_{r_3,s_3}) 
= C^{(b)}_{\t{S}^2} \mathcal{N}^{(b)}\Big(\prod_{j=2}^3 \mathcal{N}^{(b), N}_{r_j,s_j}\Big) (-1)^{\frac{r_1-1}{2}} \times f_{r_1,s_1;r_2,s_2;r_3,s_3} ~,
\end{align}
where we have used the fact that the normalization for diagonal vertex operators $\mathcal{N}^{(b)}$ is independent of the momenta. In the present case, we will find it convenient then to choose a normalization convention such that the amplitude is
\begin{align}
\mathsf{N}^{(b),\,\t{D}}_{0,3}(P^D_{r_1,s_1},P^N_{r_2,s_2},P^N_{r_3,s_3}) 
\overset{!}{=} 
(-1)^{\frac{r_1-1}{2}} \times f_{r_1,s_1;r_2,s_2;r_3,s_3} ~.
\end{align}
With this choice, we arrive at the condition that the arbitrary normalization constants for on-shell string vertex operators in the non-diagonal sector $\mathcal{N}^{(b), N}_{r,s} = \mathcal{N}^{(b)}$ are also independent of the momenta. Additionally, we arrive at the same condition \eqref{eq:fixCS2} relating these vertex operator normalizations to the sphere string path integral normalization.

\paragraph{E-series.}
For the $(12,p')$ E-series note that the only scalar vertex operators come from the Diagonal sector \eqref{eqn:Tachyons_Eseries}, unlike in the D-series.
Furthermore, the three-point structure constant of these diagonal operators is equal to the timelike DOZZ formula multiplied by the fusion rule indicator function, up to a multiplicative constant factor $c$ arising from the constant prefactor in \eqref{eqn:E_series_C_appendix} and \eqref{eqn:E_series_C_diagonal_appendix}.
We demand the three-point function be proportional to the indicator function
\begin{align}
\mathsf{N}^{(b),\,\t{E}}_{0,3}(P^D_{r_1,s_1},P^D_{r_2,s_2},P^D_{r_3,s_3}) 
\overset{!}{=} 
c_{r_1,s_1;r_2,s_2;r_3,s_3} \times f_{r_1,s_1;r_2,s_2;r_3,s_3} ~,
\end{align}
which gives \eqref{eq:fixCS2}.

\subsection{Torus one-point diagram}
\label{sec:torus 1pt amp}

Consider a torus with modular parameter $\tau$ and $q=e^{2\pi i \tau}$. The torus one-point diagram for any tachyon operator described in section \ref{sec:stringbkg} in any ADE minimal string theory is given by \cite{Polchinski:1998rq}
\begin{align} \label{eqn:String_torus_1pt}
\mathsf{N}_{1,1}^{(b)} (P_{r_1,s_1}) &= C_{T^2}\frac{(2\pi)^2}{2} \int_{F_0} \d^2 \tau \left\lb b \tilde b c \tilde c \cT_{r_1,s_1}(0) \right\rb_{T^2(\tau)} \nn \\
&= C_{T^2}\cN^{(b)} \frac{(2\pi)^2 }{2} \int_{F_0} \d^2 \tau |\eta(\tau)|^4 \Big\lb V_{i P_{r_1,s_1}} (0) \Big\rb_{T^2(\tau)}^{\t{L}}  ~ \Big\lb \widehat{V}^X_{r_1,s_1} (0) \Big\rb_{T^2(\tau)}^{\t{MM}} ~.
\end{align}
Here, $F_0=\{\tau\in\mathbb{C}\mid |\Re\tau|\leq\frac{1}{2}\,, |\tau|\geq1\}$ is the fundamental domain of the torus moduli space.
In the first line we have used that on the torus the ghost contribution is $\lb b \tilde b c \tilde c (0) \rb_{T^2(\tau)} = |\eta(q)|^4$. We have left the minimal model one-point function in a generic form so that it is applicable for all of the ADE minimal string theories. The one-point functions in the integrand can be decomposed into Virasoro conformal blocks as shown in \eqref{eq:Liouv torus 1pt}, \eqref{eqn:torus_1pt_MM}, and \eqref{eqn:DseriesMM torus1pt}. 
In particular, following \cite{Collier:2023cyw} we find it convenient to work with the elliptic conformal blocks defined in \eqref{eq:ellipticblocksdef}.
For instance, in the case of the A-series the torus one-point diagram decomposed into elliptic conformal blocks takes the following form,
\begin{align}\label{eq:torus1ptdiagramexplicit}
\mathsf{N}&_{1,1}^{(b)} (P_{r_1,s_1}) 
=\frac{C_{T^2}\cN^{(b)}}{d(b)^2} \frac{(2\pi)^2 }{2} \int_{F_0} \d^2 \tau \nn\\
&\times
\int_{\mathcal{C}} \d P \rho_b(P) C_b(iP_{r_1,s_1},P,P) |q|^{2P^2} \mathcal{H}^{(b)}_{1,1}(iP_{r_1,s_1};P|q) \mathcal{H}^{(b)}_{1,1}(iP_{r_1,s_1};P|\bar{q}) \nonumber\\
&\times \!\!\sum_{(r,s) \in \mathcal{S}^{\t{A}}} \frac{(iP_{r,s})^2}{\rho_b(iP_{r,s})}\, C^{\t{A}}_{r_1,s_1;r,s;r,s} \, f_{r_1,s_1;r,s;r,s} |q|^{2P_{r,s}}
\mathcal{H}^{(ib)}_{1,1}(P_{r_1,s_1};P_{r,s}|q) \mathcal{H}^{(ib)}_{1,1}(P_{r_1,s_1};P_{r,s}|\ol q) ~.
\end{align}
Similarly, in the D-series the torus one-point diagram is given by
\begin{align}\label{eq:torus1ptdiagramexplicit D}
\mathsf{N}_{1,1}^{(b)} (P^D_{r_1,s_1}) 
&=\frac{C_{T^2}\cN^{(b)}}{d(b)^2}  \frac{(2\pi)^2 }{2} \int_{F_0} \d^2 \tau \nn\\
&\quad\times
\int_{\mathcal{C}} \d P \rho_b(P) C_b(iP_{r_1,s_1},P,P) |q|^{2P^2} \mathcal{H}^{(b)}_{1,1}(iP_{r_1,s_1};P|q) \mathcal{H}^{(b)}_{1,1}(iP_{r_1,s_1};P|\bar{q}) \nonumber\\
&\quad\times \left[ \sum_{(r,s) \in \mathcal{S}^{\t{D},\,D}} \frac{(iP_{r,s})^2}{\rho_b(iP_{r,s})}\, C^{\t{D},\,DDD}_{r_1,s_1;r,s;r,s} \, f_{r_1,s_1;r,s;r,s}
\right. 
\nn \\
&\quad\times
|q|^{2P_{r,s}} \mathcal{H}^{(ib)}_{1,1}(P_{r_1,s_1};P_{r,s}|q) \mathcal{H}^{(ib)}_{1,1}(P_{r_1,s_1};P_{r,s}|\ol q) 
\vphantom{\sum_{(r,s) \in \mathcal{S}^{\t{D},\,D}} \frac{(iP_{r,s})^2}{\rho_b(iP_{r,s})}}
\nn \\
&\quad +\sum_{(r,s) \in \mathcal{S}^{\t{D},\,N}} \frac{(-1)^{\frac{r}{2}-\frac{p}{4}}P_{r,s}\widetilde{P}_{r,s}}{4\sqrt{2}\sin(2\pi bP_{r,s})\sin(2\pi b^{-1}\widetilde{P}_{r,s})} \, C^{\t{D},\,DNN}_{r_1,s_1;r,s;r,s} \, f_{r_1,s_1;r,s;r,s}
\nn \\
&\quad\times 
\left.
q^{P_{r,s}} \overline{q}^{\widetilde{P}_{r,s}} \mathcal{H}^{(ib)}_{1,1}(P_{r_1,s_1};P_{r,s}|q) \mathcal{H}^{(ib)}_{1,1}(P_{r_1,s_1};\widetilde{P}_{r,s}|\ol q) 
\vphantom{\sum_{(r,s) \in \mathcal{S}^{\t{D},\,D}} \frac{(iP_{r,s})^2}{\rho_b(iP_{r,s})}}
\right] ~.
\end{align}
Lastly, for conciseness, we have provided the explicit expression for the torus one-point string amplitude in appendix \ref{sec:Minimal model CFT conventions}; see \eqref{eq:torus1ptdiagramexplicit E}.

\paragraph{Liouville integration contour.} We must specify the contour of integration $\mathcal{C}$ for the Liouville one-point function in \eqref{eq:torus1ptdiagramexplicit}. 
When the external momenta is real $P_{\t{ext}}\in \mathbb{R}$, the corresponding operator lies within the spectrum of Liouville theory, and the contour over intermediate states is simply $P\in\mathbb{R_{\geq0}}$. If desired, we may extend this contour to the full real line due to the reflection symmetry of the DOZZ structure constant. 
However, in the context of minimal string theory, as discussed in section \ref{sec:onshell ops}, dressing the minimal model fields requires taking $P_{\t{ext}} \in i \mathbb{R}$, which lies outside the original spectrum of Liouville theory. 
The correlation functions for such imaginary external momenta are defined via analytic continuation from real values of $P_{\t{ext}}$. 
This can be achieved by starting with $P_{\t{ext}}\in \mathbb{R}$ with the contour along the real axis and smoothly continuing  $P_{\t{ext}}\to iP_{r,s}$ into the complex plane. 

In this continuation, for certain values of $i P_{r,s}$, poles of the integrand arising from the DOZZ structure constants may cross the contour over intermediate states.
These residues must be included, resulting in additional discrete contributions to the Liouville correlation function. 
These poles in the integrand of the Liouville correlator arise from the double gamma functions $\Gi_b(z)$, which have simple poles at $z=-m b^{-1}-nb$, $m,n\in\mb{Z}_{\geq 0}$. This implies $C_{b}(P_{\t{ext}},P,P)$ in the integrand has an infinite set of poles in the $P$ plane located at
\begin{figure}
    \centering
    \includegraphics[width=0.5\linewidth]{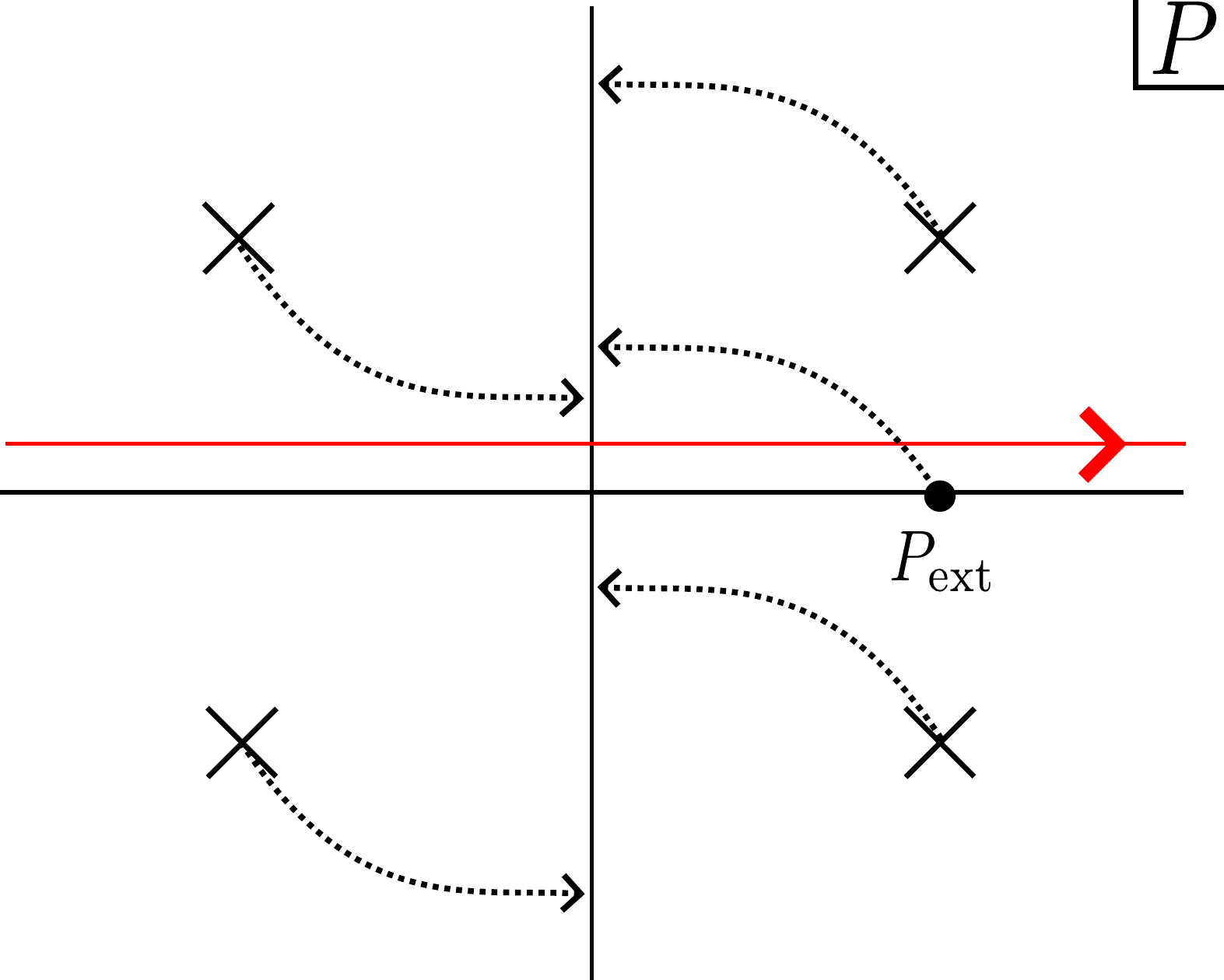}
    \caption{Integration contour (red) for the Liouville torus one-point function. As we analytically continue the external momenta to be purely imaginary poles in the integrand can cross the contour and their residues must be included. In the figure we have drawn a case where no poles cross.}
    \label{fig:contour}
\end{figure}
\be
(\pm) P= (\pm) \frac{P_{\rm ext}}{2} -\frac{i}{2}(m+\frac{1}{2}) b^{-1}-\frac{i}{2}(n+\frac{1}{2})b\,,
\ee
where this equation should be read as four equations with all possible sign choices. For fixed $m,n$, we have four poles, two above and two below the real axis. As we analytically continue $P_{\t{ext}}\to \pm i P_{r,s}=\pm\frac{i}{2}(r b^{-1}-s b)$ two of the poles move towards the real axis, while two move away. This is true regardless the choice of sign of $P_{r,s}$, which only determines which of the four poles move towards or away from the real axis. When a pole crosses the real axis, we pick up its residue, see figure \ref{fig:contour}. Due to the $P\to-P$ symmetry of the integrand \eqref{eq:Liouv torus 1pt} the residue of the pole crossing from above the real axis is identical to the pole crossing from below. We thus focus on the poles that potentially cross from below, the location of these poles at the endpoint of the analytic continuation is
\be
P_{m,n,\pm}=\pm \frac{i}{4}(r b^{-1}-s b) -i\frac{2m+1}{4} b^{-1}-i\frac{2n+1}{4}b \in i \mb{R}\,,
\ee
if either of $P_{m,n,\pm} \in i \mathbb{R}_{>0}$ the associated pole has crossed the integration contour and we arrive at the correct contour for the one-point function
\be
\int_{\mc{C}} \d P = \frac{1}{2} \lr{\int_{-\infty}^\infty \d P + 2 \sum_{P_{m,n,\pm} \in \text{pole crossing}} (-2\pi i)\mathop{\mathrm{Res}}_{P = P_{m,n,\pm}}}\,.
\ee
The factor of two is from the upper pole crossing, and the sign is due to the contour orientation. 

\paragraph{Regularization of divergent amplitudes.} When a pole crosses the real axis, the resulting discrete residue contribution can render the string amplitude divergent. For example, the torus one-point moduli integral \eqref{eqn:String_torus_1pt} behaves as
\be\label{eq:torus1ptdivergence}
\int_{F_0} \d^2 \tau \,  (q \ol q)^{ 2 P_{m,n,\pm}^2 } q^{P_{\t{int}}^2} \bar{q}^{\widetilde{P}_{\t{int}}^2} (1+\mathcal{O}(q \ol q)) ~,
\ee
where $P_{\t{int}}$ and $\widetilde{P}_{\t{int}}$ denote the left and right moving (discrete) Liouville momenta of an intermediate state in the minimal model CFT sector, and $q=\e^{2\pi i \tau}$. 
We see that \eqref{eq:torus1ptdivergence} diverges at large imaginary $\tau$ whenever the exponent has a negative real part. In practice, these divergences arise with increasing frequency once we move beyond the $(2, p)$ models, where the dressing momenta tend to lie deeper in the complex plane.

Divergences of this kind are familiar in string theory and have been consistently addressed in various contexts (see for example \cite{Berera:1992tm,Witten:2013pra,Sen:2019jpm,Sen:2020cef,Eberhardt:2022zay,Balthazar:2017mxh}). In our case, we will regularize the divergence by implementing the following replacement rule,
\be\label{eq:regularization}
\int_{0}^\infty \d t \,\e^{a t} \rightarrow 
-\frac{1}{a}
\ee
even in cases where $a>0$.
For the moduli space integral over the fundamental domain $F_0$, this prescription can be conveniently implemented by integrating over an ``inverse" fundamental domain 
\be \label{eqn:torus_reg}
\int_{F_0} \d^2 \tau ~(\ldots) \to
-\int_{-\frac{1}{2}}^{\frac{1}{2}} \d \tau_1 \int_{-\infty}^{\sqrt{1-\tau_1^2}} \d \tau_2 ~(\ldots)\,.
\ee
We use this regularization in the direct numerical evaluation of the torus one-point diagram presented below.

\paragraph{Analytical results.}
The torus one-point function \eqref{eqn:String_torus_1pt} is difficult to evaluate analytically and few results are known. In the $(p,2)$ A-series Minimal string the integral has been performed using the higher equations of motion of Liouville \cite{Belavin:2010sr,Artemev:2022sfi} (see also \cite{Artemev:2025pvk}). In our conventions the answer is
\be
(p,2) ~~\t{AMS:} \qq \mathsf{N}_{1,1}^{(b)} (P_{r,1}) = 2\pi^2 \frac{C_{T^2}\cN^{(b)}}{d(b)^2} \frac{1}{192\sqrt{p p'}} \times (p-r)r ~.
\ee
The other case that can be evaluated is the tachyon $\cT_{1,1}$ for the minimal model identity operator in any ADE minimal string
\begin{align} \label{eqn:torus_1pt_analytics}
(p,p') \q &\t{ DMS:} \qq \mathsf{N}_{1,1}^{(b)}(P_{1,1}) = (\t{const})\times(p'-1)\frac{(p+2)}{2} ~, \nn  \\ 
(12,p') \q &\t{ EMS:} \qq \mathsf{N}_{1,1}^{(b)}(P_{1,1}) = (\t{const})\times6(p'-1) ~, \nn \\
(18,p') \q &\t{ EMS:} \qq \mathsf{N}_{1,1}^{(b)}(P_{1,1}) = (\t{const})\times7(p'-1) ~,\nn \\
(30,p') \q &\t{ EMS:} \qq \mathsf{N}_{1,1}^{(b)}(P_{1,1}) = (\t{const})\times8(p'-1) ~.
\end{align}
We review this calculation in appendix \ref{app:torus_1pt}. The answer is proportional to the product of the number of nodes in the Dynkin diagrams associated with each model. The constant in our conventions is 
\be
(\t{const})=2\pi^2 \frac{C_{T^2}\cN^{(b)}}{d(b)^2} \frac{1}{192\sqrt{p p'}} ~.
\ee

Remarkably, we find that the result of the numerical calculation presented below takes the following form 
\begin{align}\label{eq:Trs def torus1pt}
\mathsf{N}_{1,1}^{(b)}(P_{r,s})=2\pi^2 \frac{C_{T^2}\cN^{(b)}}{d(b)^2} \frac{1}{192\sqrt{p p'}}\,T_{r,s} ~, \qq T_{r,s} \in\mathbb{Z} \t{ for A and D} ~,
\end{align}
and $T_{r,s} \in\mathbb{Z}$, $\mathbb{Z}\sqrt{2}$ or $\mathbb{Z}\sqrt{6}$ for the E-series minimal string.

\paragraph{Numerical results: AMS.}
We now turn to the direct numerical evaluation of the torus one-point diagram \eqref{eqn:String_torus_1pt}. For the A-series minimal string, the explicit expression for the moduli space integral of the worldsheet CFT correlator, decomposed into Virasoro conformal blocks, is given in \eqref{eq:torus1ptdiagramexplicit}.
Our numerical results for the integer $T_{r,s}$ in \eqref{eq:Trs def torus1pt} for several A${}_{p,p'}$ minimal string theories are shown in tables \ref{table:torus1pt A(p,3)} and \ref{table:torus1pt A(p,4-5)}. 
Note that in these tables we do not report the numerical error from our integration, but simply present the integer closest to the numerical result. 
The discrepancy between the computed values and the reported integers is below $10^{-6}$, and in most cases significantly smaller.\footnote{For example, in the $A_{8,3}$ minimal string $T_{7,1}=-10.0000000003$.
} 
Results shown in red ${\color{BrickRed} T_{r,s}}$ indicate that the amplitude has a divergent moduli space integral, which we regulate according to the procedure described in the preceding paragraph, see \eqref{eqn:torus_reg}. 

For $(p,3)$ AMS, a basis of independent on-shell vertex operators is given by $\cT_{r,1}$ with $r=1,\ldots,p-1$. This set of operators is complete under $\cT_{r,s}\leftrightarrow \cT_{p-r,3-s}$. 
Furthermore, $\mathsf{N}_{1,1}^{(b)}(P_{2n,1})=0$ follows simply by minimal model fusion rules. Table \ref{table:torus1pt A(p,3)} lists the non-trivial torus one-point string amplitudes. 
For $(p,4)$ AMS, an independent set of on-shell vertex operators is $r=1,3,5,\ldots,p-2$ and $s=1,2,3$. The  minimal model fusion rules imply that $\mathsf{N}_{1,1}^{(b)}(P_{r,2})=0$. Table \ref{table:torus1pt A(p,4-5)} lists the remaining non-trivial torus one-point string amplitudes. 
For $(p \t{ odd},5)$ AMS, an independent set of on-shell vertex operators is $r=1,3,\ldots,p-2$ and $s=1,2,3,4$, while for $(p \t{ even},5)$ we have $r=1,2,3,\ldots,p-1$ and $s=1,3$. The minimal model fusion rules imply that only $\mathsf{N}_{1,1}^{(b)}(P_{\t{odd},\t{odd}})$ are non-vanishing. 
Table \ref{table:torus1pt A(p,4-5)} lists the non-trivial torus one-point string amplitudes. 

We comment that the integers $T_{r,s}$ appear to satisfy a positivity condition: whenever the Liouville integration contour does not receive additional discrete contributions from crossed poles, $T_{r,s}$ is positive. This appears to be true in all cases we computed in the A-series minimal string. However, this does not seem to hold for the D- or E-series minimal string results discussed in the next paragraph.

\begin{table}
\centering
\begin{tabular}{|c|l|}
\hline
$(p,p')$ & AMS tachyon one-point string amplitude \\
\hline
4,3 & $T_{1,1}=6$, $T_{3,1}=-2$ \\
5,3 & $T_{1,1}=8$, $T_{3,1}=4$ \\
7,3 & $T_{1,1}=12$, $T_{3,1}=16$, $T_{5,1}=-4$ \\
8,3 & $T_{1,1}=14$, $T_{3,1}=22$, $T_{5,1}=6$, $T_{7,1}=-10$ \\
10,3 & $T_{1,1}=18$, $T_{3,1}=34$, $T_{5,1}=26$, $T_{7,1}=-6$, $T_{9,1}=-14$ \\
11,3 & $T_{1,1}=20$, $T_{3,1}=40$, $T_{5,1}=36$, $T_{7,1}=8$, $T_{9,1}=-20$ \\
13,3 & $T_{1,1}=24$, $T_{3,1}=52$, $T_{5,1}=56$, $T_{7,1}=36$, $T_{9,1}=-8$, $T_{11,1}=-28$ \\
14,3 & $T_{1,1}=26$, $T_{3,1}=58$, $T_{5,1}=66$, $T_{7,1}=50$, $T_{9,1}=10$, $T_{11,1}=-30$ \\
 & $T_{13,1}=-22$ \\
16,3 & $T_{1,1}=30$, $T_{3,1}=70$, $T_{5,1}=86$, $T_{7,1}=78$, $T_{9,1}=46$, $T_{11,1}=-10$ \\
 & $T_{13,1}=-42$, $T_{15,1}=-26$ \\
17,3 & $T_{1,1}=32$, $T_{3,1}=76$, $T_{5,1}=96$, $T_{7,1}=92$, $T_{9,1}=64$, $T_{11,1}=12$ \\
 & $T_{13,1}=-40$, $T_{15,1}=-44$\\ 
19,3 & $T_{1,1}=36$, $T_{3,1}=88$, $T_{5,1}=116$, $T_{7,1}=120$, $T_{9,1}=100$, $T_{11,1}=56$ \\
 & $T_{13,1}=-12$, $T_{15,1}=-56$, $T_{17,1}=-52$\\
20,3 & $T_{1,1}=38$, $T_{3,1}=94$, $T_{5,1}=126$, $T_{7,1}=134$, $T_{9,1}=118$, $T_{11,1}=78$ \\
 & $T_{13,1}=14$, $T_{15,1}=-50$, $T_{17,1}=-66$, $T_{19,1}=-34$\\ 
\hline 
\end{tabular}
\caption{Results of the numerical calculation of the integers $T_{r,s}$, defined in \eqref{eq:Trs def torus1pt} and proportional to the torus one-point string amplitude, in the $(p,3)$ A-series minimal string, for a range of values of $p$.}
\label{table:torus1pt A(p,3)}
\end{table}

\begin{table}
\centering
\begin{tabular}{|c|l|}
\hline
$(p,p')$ & AMS tachyon one-point string amplitude \\
\hline
5,4 & $T_{1,1}=12$, $T_{3,1}=-6$, \\ 
& $T_{1,3}=-12$, $T_{3,3}=2$ \\
7,4 & $T_{1,1}=18$, $T_{3,1}=12$, $T_{5,1}=-18$, \\  & $T_{1,3}=\myred{-30}$, $T_{3,3}=-4$, $T_{5,3}=6$ \\
9,4 & $T_{1,1}=24$, $T_{3,1}=30$, $T_{5,1}=-12$, $T_{7,1}=\myred{-30}$  \\ & $T_{1,3}=\myred{-56}$, $T_{3,3}=-18$, $T_{5,3}=4$, $T_{7,3}=10$ \\
11,4 & $T_{1,1}=30$, $T_{3,1}=48$, $T_{5,1}=18$, $T_{7,1}=-36$, $T_{9,1}=\myred{-42}$ \\ &$T_{1,3}=\myred{-90}$, $T_{3,3}=\myred{-40}$, $T_{5,3}=-6$, $T_{7,3}=12$, $T_{9,3}=14$ \\
13,4 & $T_{1,1}=36$, $T_{3,1}=66$, $T_{5,1}=48$, $T_{7,1}=-18$, $T_{9,1}=\myred{-60}$, $T_{11,1}=\myred{-54}$ \\ & $T_{1,3}=\myred{-132}$, $T_{3,3}=\myred{-70}$, $T_{5,3}=-24$, $T_{7,3}=6$, $T_{9,3}=20$, $T_{11,3}=18$ \\
15,4 & $T_{1,1}=42$, $T_{3,1}=84$, $T_{5,1}=78$, $T_{7,1}=24$, $T_{9,1}=-54$, $T_{11,1}=\myred{-84}$ \\ & 
$T_{13,1}=\myred{-66}$ \\ & 
$T_{1,3}=\myred{-182}$, $T_{3,3}=\myred{-108}$, $T_{5,3}=\myred{-50}$, $T_{7,3}=-8$, $T_{9,3}=18$, $T_{11,3}=28$ \\
& $T_{13,3}=22$ \\
17,4 & $T_{1,1}=48$, $T_{3,1}=102$, $T_{5,1}=108$, $T_{7,1}=66$, $T_{9,1}=-24$, $T_{11,1}=\myred{-90}$,\\
& $T_{13,1}=\myred{-108}$, $T_{15,1}=\myred{-78}$ \\
& $T_{1,3}=\myred{-240}$, $T_{3,3}=\myred{-154}$, $T_{5,3}=\myred{-84}$, $T_{7,3}=-30$, $T_{9,3}=8$, $T_{11,3}=30$ \\
& $T_{13,3}=36$, $T_{15,3}=26$ \\
\hline
6,5 & $T_{1,1}=20$, $T_{3,1}=-12$, $T_{5,1}=\myred{-28}$ \\ 
& $T_{1,3}=-26$, $T_{3,3}=6$, $T_{5,3}=-2$ \\
7,5 & $T_{1,1}=24$, $T_{3,1}=0$, $T_{5,1}=\myred{-32}$ \\ 
& $T_{1,3}=\myred{-36}$, $T_{3,3}=8$, $T_{5,3}=12$ \\ 
8,5 & $T_{1,1}=28$, $T_{3,1}=12$, $T_{5,1}=\myred{-36}$, $T_{7,1}=\myred{-44}$\\ & $T_{1,3}=\myred{-54}$, $T_{3,3}=2$, $T_{5,3}=18, T_{7,3}=-6$ \\
9,5 & $T_{1,1}=32$, $T_{3,1}=24$, $T_{5,1}=-40$, $T_{7,1}=\myred{-64}$\\ & $T_{1,3}=\myred{-80}$, $T_{3,3}=-12$, $T_{5,3}=16$, $T_{7,3}=4$ \\
11,5 & $T_{1,1}=40$, $T_{3,1}=48$, $T_{5,1}=-24$, $T_{7,1}=\myred{-80}$, $T_{9,1}=\myred{-96}$\\ & $T_{1,3}=\myred{-132}$, $T_{3,3}=-40$, $T_{5,3}=12$, $T_{7,3}=24$, $T_{9,3}=-4$ \\
12,5 & $T_{1,1}=44$, $T_{3,1}=60$, $T_{5,1}=-4$, $T_{7,1}=\myred{-76}$, $T_{9,1}=\myred{-108}$, $T_{11,1}=\myred{-76}$\\ & $T_{1,3}=\myred{-158}$, $T_{3,3}=\myred{-54}$, $T_{5,3}=10$, $T_{7,3}=34$, $T_{9,3}=18$, $T_{11,3}=-14$ \\
13,5 & $T_{1,1}=48$, $T_{3,1}=72$, $T_{5,1}=16$, $T_{7,1}=\myred{-72}$, $T_{9,1}=\myred{-120}$, $T_{11,1}=\myred{-128}$\\
& $T_{1,3}=\myred{-192}$, $T_{3,3}=\myred{-76}$, $T_{5,3}=0$, $T_{7,3}=36$, $T_{9,3}=32$, $T_{11,3}=-12$ \\
\hline
\end{tabular}
\caption{Results of the numerical calculation of the integers $T_{r,s}$, defined in \eqref{eq:Trs def torus1pt} and proportional to the torus one-point string amplitude, in the $(p,4)$ and $(p,5)$ A-series minimal string, for a range of values of $p$.}
\label{table:torus1pt A(p,4-5)}
\end{table}

\paragraph{Numerical results: DMS.}
In the D-series minimal string, non-diagonal (but scalar, c.f. \eqref{eqn:Tachyons_Dseries N}) on-shell vertex operators have vanishing torus one-point amplitudes due to the minimal model fusion rules and their conservation of diagonality.
Therefore, we restrict attention to diagonal tachyon vertex operators for $(p \t{ even}, p' \t{ odd})$ labeled by $r=1,3\ldots,p-1$ and $s=1,3,\ldots,p'-2$. This is a set of independent tachyon vertex operators, accounting for the usual $\mathbb{Z}_2$ identification of operators described in section \ref{sec:minimal model CFTs}. 
Our numerical results for the integer $T_{r,s}$ for several D${}_{p,p'}$ minimal string theories are shown in table \ref{table:torus1pt D(p,3-5-7)}.

\begin{table}
\centering
\begin{tabular}{|c|l|}
\hline
$(p,p')$ & DMS tachyon one-point string amplitude \\
\hline
8,3 & $T_{1,1}=10$, $T_{3,1}=14$, $T_{5,1}=6$, $T_{7,1}=10$\\
10,3 &  $T_{1,1}=12$, $T_{3,1}=20$, $T_{5,1}=16$, $T_{7,1}=0$, $T_{9,1}=-28$ \\
14,3 & $T_{1,1}=16$, $T_{3,1}=32$, $T_{5,1}=36$, $T_{7,1}=28$, $T_{9,1}=8$, $T_{11,1}=0$, $T_{13,1}=-44$ \\
16,3 & $T_{1,1}=18$, $T_{3,1}=38$, $T_{5,1}=46$, $T_{7,1}=42$, $T_{9,1}=26$, $T_{11,1}=-2$ \\ 
&$T_{13,1}=-42$, $T_{15,1}=26$ \\
20,3 & $T_{1,1}=22$, $T_{3,1}=50$, $T_{5,1}=66$, $T_{7,1}=70$, $T_{9,1}=62$, $T_{11,1}=42$, $T_{13,1}=10$ \\ 
& $T_{15,1}=-10$, $T_{17,1}=-66$, $T_{19,1}=34$ \\
22,3 & $T_{1,1}=24$, $T_{3,1}=56$, $T_{5,1}=76$, $T_{7,1}=84$, $T_{9,1}=80$, $T_{11,1}=64$, $T_{13,1}=36$ \\ 
& $T_{15,1}=-4$, $T_{17,1}=-56$, $T_{19,1}=0$, $T_{21,1}=-76$\\
\hline
6,5 & $T_{1,1}=16$, $T_{3,1}=0$, $T_{5,1}=\myred{-56}$ \\ 
&$T_{1,3}=-16$, $T_{3,3}=0$, $T_{5,3}=-4$\\
8,5 & $T_{1,1}=20$, $T_{3,1}=12$, $T_{5,1}=\myred{-36}$, $T_{7,1}=\myred{44}$ \\ & $T_{1,3}=\myred{-18}$, $T_{3,3}=10$, $T_{5,3}=18$, $T_{7,3}=6$\\
12,5 & $T_{1,1}=28$,  $T_{3,1}=36$, $T_{5,1}=4$, $T_{7,1}=\myred{4}$, $T_{9,1}=\myred{-108}$, $T_{11,1}=\myred{76}$ \\ 
&$T_{1,3}=\myred{-70}$, $T_{3,3}=\myred{-18}$, $T_{5,3}=14$, $T_{7,3}=26$, $T_{9,3}=18$, $T_{11,3}=14$\\
14,5 & $T_{1,1}=32$, $T_{3,1}=48$, $T_{5,1}=24$, $T_{7,1}=-16$, $T_{9,1}=\myred{-120}$, $T_{11,1}=0$ \\ &$T_{13,1}=\myred{-184}$, $T_{1,3}=\myred{-120}$, $T_{3,3}=\myred{-56}$, $T_{5,3}=-12$, $T_{7,3}=12$, $T_{9,3}=16$\\
&$T_{11,3}=0$, $T_{13,3}=-36$\\
16,5 & $T_{1,1}=36$, $T_{3,1}=60$, $T_{5,1}=44$, $T_{7,1}=-12$, $T_{9,1}=\myred{-108}$, $T_{11,1}=\myred{-28}$ \\
&$T_{13,1}=\myred{-204}$, $T_{15,1}=\myred{108}$\\
&$T_{1,3}=\myred{-162}$, $T_{3,3}=\myred{-86}$, $T_{5,3}=-30$, $T_{7,3}=6$, $T_{9,3}=22$, $T_{11,3}=18$\\
&$T_{13,3}=-6$, $T_{15,3}=22$ \\
\hline
6,7 & $T_{1,1}=24$, $T_{3,1}=0$, $T_{5,1}=\myred{-132}$\\
& $T_{1,3}=-16$, $T_{3,3}=0$, $T_{5,3}=-40$\\
& $T_{1,5}=\myred{-32}$, $T_{3,5}=0$, $T_{5,5}=4$\\
8,7 & $T_{1,1}=30$, $T_{3,1}=-6$, $T_{5,1}=\myred{-126}$, $T_{7,1}=\myred{102}$\\
& $T_{1,3}=-36$, $T_{3,3}=4$, $T_{5,3}=-12$, $T_{7,3}=\myred{36}$\\
& $T_{1,5}=\myred{-70}$, $T_{3,5}=-18$, $T_{5,5}=6$, $T_{7,5}=2$\\
10,7 & $T_{1,1}=36$, $T_{3,1}=12$, $T_{5,1}=\myred{-96}$, $T_{7,1}=0$, $T_{9,1}=\myred{-276}$\\
& $T_{1,3}=\myred{-48}$, $T_{3,3}=16$, $T_{5,3}=24$, $T_{7,3}=0$, $T_{9,3}=\myred{-104}$\\
& $T_{1,5}=\myred{-132}$, $T_{3,5}=-60$, $T_{5,5}=-16$, $T_{7,5}=0$, $T_{9,5}=-12$\\
12,7 & $T_{1,1}=42$, $T_{3,1}=30$, $T_{5,1}=\myred{-66}$, $T_{7,1}=\myred{-30}$, $T_{9,1}=\myred{-294}$, $T_{11,1}=\myred{174}$\\
& $T_{1,3}=\myred{-76}$, $T_{3,3}=12$, $T_{5,3}=44$, $T_{7,3}=20$, $T_{9,3}=\myred{-60}$, $T_{11,3}=\myred{68}$\\
& $T_{1,5}=\myred{-170}$, $T_{3,5}=\myred{-78}$, $T_{5,5}=\myred{-14}$, $T_{7,5}=22$, $T_{9,5}=30$, $T_{11,5}=10$\\
\hline
\end{tabular}
\caption{Results of the numerical calculation of the integers $T_{r,s}$, defined in \eqref{eq:Trs def torus1pt} and proportional to the torus one-point string amplitude, in the $(p,p')$ D-series minimal string, for a range of values of $p$ and $p'$.}
\label{table:torus1pt D(p,3-5-7)}
\end{table}

\paragraph{Numerical results: EMS.} 
In the $(12,p)$ E-series minimal string, the fusion rules imply that $\mathsf{N}_{1,1}^{(b)}(P_{1,\text{even}})=0$ and $\mathsf{N}_{1,1}^{(b)}(P_{4,s})=0$. Our numerical results for the remaining non-vanishing torus one-point amplitudes are shown in table \ref{table:torus1pt E(12,pp)}.

\begin{table}
\centering
\begin{tabular}{|c|l|}
\hline
$(p,p')$ & EMS tachyon one-point string amplitude \\
\hline
12,5 & $T_{1,1}=24$, $T_{1,3}=\myred{-36}$\\
& $T_{7,1}=\myred{-36\sqrt{2}}$, $T_{7,2}=2\sqrt{6}$, $T_{7,3}=30 \sqrt{2}$, $T_{7,4}=4\sqrt{6}$\\
12,7 &  $T_{1,1}=36$, $T_{1,3}=\myred{-48}$, $T_{1,5}=\myred{-132}$\\
&  $T_{7,1}=\myred{138\sqrt{2}}$, $T_{7,2}=-22\sqrt{6}$, $T_{7,3}=-12 \sqrt{2}$, $T_{7,4}=4\sqrt{6}$ \\
& $T_{7,5}=-18 \sqrt{2}$, $T_{7,6}=-18\sqrt{6}$ \\
12,11 &  $T_{1,1}=60$, $T_{1,3}=-72$, $T_{1,5}=\myred{-204}$, $T_{1,7}=\myred{-336}$, $T_{1,9}=\myred{-372}$\\
&  $T_{7,1}=\myred{510\sqrt{2}}$, $T_{7,2}=62\sqrt{6}$, $T_{7,3}=\myred{192 \sqrt{2}}$, $T_{7,4}=28\sqrt{6}$ \\
& $T_{7,5}=18 \sqrt{2}$, $T_{7,6}=-6\sqrt{6}$, $T_{7,7}=-12\sqrt{2}$, $T_{7,8}=8\sqrt{6}$ \\
& $T_{7,9}=54\sqrt{2}$, $T_{7,10}=\myred{70\sqrt{6}}$ \\
12,13&  $T_{1,1}=72$, $T_{1,3}=-60$, $T_{1,5}=\myred{-216}$, $T_{1,7}=\myred{-372}$, $T_{1,9}=\myred{-528}$ \\ 
&  $T_{1,11}=\myred{-492}$\\
&  $T_{7,1}=\myred{756\sqrt{2}}$, $T_{7,2}=\myred{82\sqrt{6}}$, $T_{7,3}=\myred{354 \sqrt{2}}$, $T_{7,4}=\myred{68\sqrt{6}}$ \\
&$T_{7,5}=96 \sqrt{2}$, $T_{7,6}=6\sqrt{6}$, $T_{7,7}=-18\sqrt{2}$, $T_{7,8}=-8\sqrt{6}$ \\
&$T_{7,9}=12\sqrt{2}$, $T_{7,10}=26\sqrt{6}$, $T_{7,11}=\myred{90\sqrt{2}}$, $T_{7,12}=\myred{108\sqrt{6}}$ \\
\hline
\end{tabular}
\caption{Results of the numerical calculation of the integers $T_{r,s}$, defined in \eqref{eq:Trs def torus1pt} and proportional to the torus one-point string amplitude, in the $(12,p')$ E-series minimal string, for a range of values of $p'$.}
\label{table:torus1pt E(12,pp)}
\end{table}

\paragraph{Comparison to the Matrix Model}
The only result for matrix model torus one-point string amplitudes we are aware of beyond the A-series $(p,2)$ minimal string is \cite{Spodyneiko:2014lla}, where the A-series $(p,3)$ case was solved and a conjecture for general $(p,p')$ was proposed. 
For the $(p,3)$ string, the matrix model computation yields \cite{Spodyneiko:2014lla} 
\be \label{eqn:spod}
\text{$(p,3)$ MM:}~~~T_{r,1} =1+2p r-3r^2\,.
\ee
This formula does not match our numerical results. 
Specifically, we find that it holds for small values $r$, but fails for large values. 
The cases in which our amplitude disagrees with the result of \cite{Spodyneiko:2014lla} turn out to correspond precisely to situations where, from the string theory point of view, a pole in the intermediate Liouville momentum crosses the integration contour, and the corresponding residue must be included as discussed earlier in this section. 
Including this pole is crucial for modular invariance. 

We have found the following formula from inspecting the numerical data for the $(p,3)$ AMS:
\bal
(p,3) \t{ conjecture:} \qq T_{r,1} =
\begin{cases}
1+2pr-3r^2&\text{when }3r < 2p\\
4p^2-10pr+6r^2&\text{when }3r > 2p
\end{cases}
\eal
which agrees with all the numerical results. It would be interesting to derive this formula from the matrix model.  

\newpage~\newpage

\paragraph{Conjectural formulas.}

Motivated by the numerical results presented above, we conjecture the following formulas. They reproduce all entries in the displayed tables and have also been verified against a substantially larger numerical dataset for AMS, DMS, and EMS models with higher values of $p$ and $p'$.

It is useful to introduce the common kernel
\bal
K_{r,s}^{p,p'}(m,n) = 2pp'-rp'-sp-6|mp'-np| \,.
\eal

For the $(p,p')$ AMS, choosing the odd-odd representative $r,s\in 2\mathbb Z+1$, the torus one-point amplitude is
\bal \label{eq:conjecture torus 1pt AMS}
T_{r,s}^{\rm A}=\sum_{m \in \mathcal I_p(r)} \sum_{n \in \mathcal I_{p'}(s)}K_{r,s}^{p,p'}(m,n)\,.
\eal
where
\bal
\mathcal I_p(r) =  \Big\{\frac{r+1}{2},\frac{r+3}{2},\ldots,p-\frac{r+1}{2}\Big\}
\eal
which consists precisely of those $m$ satisfying the fusion condition $f^p_{r,m,m}=1$. Thus, the sum is indexed by the scalar primaries whose self-fusion contains the external field $\widehat V_{r,s}$.

This formula $T_{r,s}^{\rm A}$ can be derived using the higher equations of motion; see, for example, \cite{Collier:2024kwt}.\footnote{We thank Lorenz Eberhardt for pointing out this fact.}

For the $(p,p')$ DMS with $p$ even, we conjecture from our numerical results that, for $r,s\in 2\mathbb Z+1$, again the amplitude can similarly be organized as a weighted sum over fusion-allowed scalar fields:
\bal \label{eq:conjecture torus 1pt DMS}
T_{r,s}^{\rm D}=\sum_{\overset{m\in\mathcal I_p(r)}{m \text{ odd}}} \sum_{n\in\mathcal I_{p'}(s)} K_{r,s}^{p,p'}(m,n) + (-1)^{\frac{r-1}{2}} \sum_{n\in\mathcal I_{p'}(s)} K_{r,s}^{p,p'}(\frac{p}{2},n)\,.
\eal
The first sum is indexed by the diagonal fields $\widehat V^D_{m,n}$, whose first Kac label is restricted to odd $m$. The second sum is indexed by the scalar fields $\widehat V^N_{p/2,n}$ in the non-diagonal sector.

For the $(12,p')$ EMS, the conjectures admit a similar organization using the same kernel:
\bal \label{eq:conjecture torus 1pt EMS}
T_{1,s}^{\rm E}&=
2\sum_{m\in\{1,4,7\}}\sum_{n\in\mathcal I_{p'}(s)}K_{1,s}^{12,p'}(m,n) \,,&& s\in2\mathbb Z+1 \,,\\
T_{7,s}^{\rm E}&=
\left\{
\begin{aligned}
&\sqrt{2}\,\nu\sum_{n\in\mathcal I_{p'}(s)}
\left[K_{7,s}^{12,p'}(4,n)+2K_{7,s}^{12,p'}(7,n)\right] \,,\\
&\sqrt{6}\,\mu\sum_{n\in\mathcal I_{p'}(p'-s)}K_{5,p'-s}^{12,p'}(4,n) \,,
\end{aligned}\right.
&&
\begin{aligned}&
\vphantom{\sqrt{2}\,\nu
\sum_{n\in\mathcal I_{p'}(s)}\left[
K_{7,s}^{12,p'}(4,n)+2K_{7,s}^{12,p'}(7,n)
\right] \,,}s\in2\mathbb Z+1 \,,\\
&\vphantom{\sqrt{6}\,\mu\sum_{n\in\mathcal I_{p'}(p'-s)}K_{5,p'-s}^{12,p'}(4,n) \,,}s\in2\mathbb Z \,.
\end{aligned}
\eal
The signs $\mu$ and $\nu$ are
\bal
\mu&=\begin{cases}
+1 \text { for } p'=1,7, 17, 23 ~\,\text{ mod }24\\
-1 \text { for } p'=5,11,13,19 \text{ mod }24\\
\end{cases} \,, \\
\nu&=\begin{cases}
+1 \text { for } p'=1,5, 19, 23 ~\,\text{ mod }24\\
-1 \text { for } p'=7,11,13,17 \text{ mod }24\\
\end{cases} \,.
\eal
The structure of these formulas are related to the E-series fusion rules summarized in Appendix \ref{app:E-series}.

\subsection{Sphere $n$-point amplitudes}
The general sphere $n$-point amplitude is not analytically computable in any of the models under consideration.
We begin by outlining several simplifications and structural relations among the amplitudes.  
We then present direct numerical results for several sphere four-point amplitudes in ADE minimal string theories.

\paragraph{Relation between AMS and DMS sphere amplitudes.} 
A subset of sphere amplitudes involving tachyons in the $(p,p')$ DMS are equivalent to similar sphere amplitude in the $(p,p')$ AMS. A general amplitude for the AMS and DMS on any Riemann surface will have identical Liouville and ghost contributions, and therefore the problem reduces to comparing minimal model correlation functions on the surface. 

On the sphere (i.e. at tree-level), there are no intermediate states running in loops, and so the problem reduces to examining the OPE of the external tachyons to determine if the amplitudes differ. The fusion rules for diagonal fields are identical in the A- and D-series.\footnote{The fusion rules of the D-series spectrum form a closed subset of those of the A-series spectrum.} The structure constants are also identical for these fields, as discussed in section \ref{sec:minimal model CFTs}. Thus, we have that
\be\label{eq:DMS allD same as AMS}
\mathsf{N}_{0,n}^{(b), \,\mathrm{DMS}}(P^D_{r_1,s_1},\ldots,P^D_{r_n,s_n}) =\mathsf{N}_{0,n}^{(b), \,\mathrm{AMS}}(P_{r_1,s_1},\ldots,P_{r_n,s_n}) ~.
\ee

\paragraph{Conservation of diagonality at any genus.} Now let us consider the special case with non-diagonal tachyons $\mc{T}^N_{\frac{p}{2},s}$ in the DMS. First, recall that the D-series fusion rules obey a $\mb{Z}_2$ structure:
\be
D\times D=D,~~~D\times N=N,~~~N\times N=D.
\ee
Consequently, there must be an even number of non-diagonal insertions otherwise the correlator on an arbitrary surface $\Sigma_g$ vanishes
\begin{align}\label{eq:DMS with odd Ns}
\mathsf{N}_{g,n}^{(b), \,\mathrm{DMS}} \big( \underbrace{P^N_{\frac{p}{2},s_1}, \ldots, P^N_{\frac{p}{2},s_{k}}}_{\t{odd number}}, P^D_{r_{k+1},s_{k+1}}, \ldots, P^D_{r_{n},s_{n}} \big) = 0 ~.
\end{align}
This can be seen through the conformal block decomposition of the amplitude. An amplitude on a general surface can be decomposed into three-holed spheres, where on each sphere one hole is an external state while the other two are intermediate states. This decomposition makes it clear that inserting intermediate states always introduces an even number of non-diagonal states. With an odd number of external tachyons $\mc{T}^N$ the conformal block decomposition must consequently have an odd$+$even number of non-diagonal indices in the sum over structure constants $C^{N D D}\ldots C^{\ldots}$. Since the only non-vanishing structure constant involving $\mc{T}^N$ fields is $C^{D N N}$ the product of structure constants with an odd number of labels $N$ will vanish.

\paragraph{Sphere four-point amplitudes.} 
We now turn to the direct worldsheet calculation of sphere four-point amplitudes in the ADE minimal string theories. 
Our procedure closely parallels the torus one-point amplitude calculation presented in the previous section.
On the sphere, three of the four on-shell vertex operator insertions can be fixed using the conformal Killing group, leaving the position of the fourth operator as the nontrivial modulus to be integrated over.
The amplitude in any ADE minimal string then takes the general form
\begin{align} \label{eqn:String_sphere_4pt}
\mathsf{N}_{0,4}^{(b)}&(P^{X_1}_{r_1,s_1},P^{X_2}_{r_2,s_2},P^{X_3}_{r_3,s_3},P^{X_4}_{r_4,s_4}) \nn\\
&= C^{(b)}_{S^2} \Big( \prod_{j=1}^{4} \mathcal{N}^{(b)} \Big) \int_{\mathbb{C}} \d^2 z \, \Big\lb V_{i P_{r,s}}(0) V_{i P_{r_2,s_2}}(z,\bar{z}) V_{i P_{r_3,s_3}}(1) V'_{i P_{r_4,s_4}}(\infty) \Big\rb_{S^2}^{\t{L}} \nn\\
&\quad\times \Big\lb \widehat{V}^{X_1}_{r_1,s_1}(0) \widehat{V}^{X_2}_{r_2,s_2}(z,\bar{z}) \widehat{V}^{X_3}_{r_3,s_3}(1) \widehat{V}^{'X_4}_{r_4,s_4}(\infty) \Big\rb_{S^2}^{\t{MM}} ~.
\end{align}
The explicit conformal block decomposition in the so-called s-channel, where the OPE is taken between the first and second operators and between the third and fourth, is given in \eqref{eq:Liouv sphere 4pt} for the Liouville CFT correlator.
For the minimal model CFTs, the corresponding expressions are shown in \eqref{eq:sphere_4pt_AMM} and \eqref{eq:sphere_4pt_DMM} for the A- and D-series, respectively. Although not presented explicitly, the E-series conformal block decomposition takes a similar form as that of the D-series.

For the purpose of performing the moduli space integral efficiently, we follow the same procedure used in the numerical evaluation of four-point sphere amplitudes in the VMS \cite{Collier:2023cyw} and CLS \cite{Collier:2024kwt}. 
The first step is to divide the moduli space (i.e. the complex plane) into six regions, each of which can be mapped to a compact, “lemon-shaped” domain near the origin using the crossing symmetries of the constituent worldsheet CFTs (see also \cite{Chang:2014jta,Balthazar:2017mxh,Balthazar:2022atu,Rodriguez:2023kkl}). 
The second step is to map the lemon-shaped region of the complex plane to the familiar keyhole domain  $F_0=\{t\in\mathbb{C}\mid |\Re t|\leq\frac{1}{2}\,, |t|\geq1\}$ in the upper half-plane, via the same change of variables that brings the conformal block decomposition into the form of \emph{elliptic} conformal blocks \cite{Collier:2023cyw}. 
The so-called elliptic nome variable $q$ is defined as $q=\e^{i \pi t}$, where $t\in F_0$ is related to the cross-ratio $z\in\mathbb{C}$ via $t=iK(1-z)/K(z)$, where $K(z)={}_2F_1(\frac{1}{2},\frac{1}{2},1|z)$ \cite{Zamolodchikov:1987avt}. For a more detailed description of this procedure, we refer the reader to \cite[section 7]{Collier:2023cyw}.
For instance, after applying this procedure, the sphere four-point amplitude in the A-series minimal string takes the following explicit form:
\begin{align}\label{eq:sphere4pt amp explicit A}
\mathsf{N}_{0,4}^{(b)}&(P_{r_1,s_1},P_{r_2,s_2},P_{r_3,s_3},P_{r_4,s_4}) \nn\\
&= \frac{\pi^2 C^{(b)}_{S^2} (\mathcal{N}^{(b)})^4}{d(b)^2}  
\int_{F_0} \!\d t
\int_{\mathcal{C}} \d P \rho_b(P) C_b(iP_{r_1,s_1},iP_{r_2,s_2},P) C_b(iP_{r_3,s_3},iP_{r_4,s_4},P) 
\nn\\
&\quad
\times|16q|^{2P^2} \mathcal{H}^{(b)}_{0,4}(iP_{r_j,s_j};P|q) \mathcal{H}^{(b)}_{0,4}(iP_{r_j,s_j};P|\bar{q}) \nonumber\\
&\quad\times \!\!\sum_{(r,s) \in \mathcal{S}^{\t{A}}} \frac{(iP_{r,s})^2}{\rho_b(iP_{r,s})}\, C^{\t{A}}_{r_1,s_1;r_2,s_2;r,s} \, f_{r_1,s_1;r_2,s_2;r,s} \,C^{\t{A}}_{r_3,s_3;r_4,s_4;r,s} \, f_{r_3,s_3;r_4,s_4;r,s} 
\nn\\
&\quad
\times |16q|^{2P_{r,s}} \mathcal{H}^{(ib)}_{0,4}(P_{r_j,s_j};P_{r,s}|q) \mathcal{H}^{(ib)}_{0,4}(P_{r_j,s_j};P_{r,s}|\ol q) 
\nn\\
&\quad + \big( \text{5 perms of } \{1,2,3\} \big) ~.
\end{align}
Here, $\mathcal{H}^{(b)}_{0,4}(P_j;P|\bar{q})$ is the sphere four-point elliptic Virasoro conformal block. See for example \cite[appendix C.2]{Collier:2023cyw} for the explicit recursions, whose conventions we follow. 
The sphere four-point amplitudes in the D- and E-series minimal strings take a similar form. For conciseness, we present their explicit expressions in \ref{sec:Minimal model CFT conventions}, see \eqref{eq:sphere4pt amp explicit DE}.

In this case, we find that the result of the direct numerical integration over moduli space for the tree-level four-point string amplitude takes the following strikingly simple form:
\begin{align}\label{eq:Trs def sphere4pt}
\mathsf{N}_{0,4}^{(b)}(P_{r_1,s_1},P_{r_2,s_2},P_{r_3,s_3},P_{r_4,s_4}) =  \frac{C^{(b)}_{S^2} (\mathcal{N}^{(b)})^4}{d(b)^{2}} \frac{\pi^2}{4 \sqrt{p p'}} \times T_{r_1,s_1;r_2,s_2;r_3,s_3;r_4,s_4} ~,
\end{align}
where, in the case of the D-series minimal string, we further decorate $T^{XXXX}_{r_1,s_1;r_2,s_2;r_3,s_3;r_4,s_4}$ with superscripts to indicate the sector of each tachyon vertex operator. 
As in the torus one-point amplitudes, in the A- and D-series minimal strings we find that $T_{r_1,s_1;r_2,s_2;r_3,s_3;r_4,s_4}\in\mathbb{Z}$. 
For the E-series, many amplitudes also yield integer $T_{r_1,s_1;r_2,s_2;r_3,s_3;r_4,s_4}$, but not always.

\paragraph{Numerical results: AMS.}
The results of the direct numerical integration of \eqref{eq:sphere4pt amp explicit A} in the A-series minimal string are presented in table~\ref{table:Sphere_table A}.
The integration strategy, including the regularization procedure applied when necessary (as in \eqref{eq:regularization}), follows precisely the same approach used for the torus one-point amplitude. 

Certain sphere four-point amplitudes diverge in a manner that cannot be regularized by the simple prescription used in the previous section for the torus one-point amplitude.
This divergence occurs when a pole from one of the DOZZ structure constants crosses the Liouville integration contour $\mathcal{C}$ and subsequently collides with a pole from the second DOZZ structure constant.
The resulting pinch of the contour leads to a genuine divergence of the Liouville correlator, and thus of the full string amplitude.

In more detail, consider the structure of poles in a given OPE channel, say the $12\to34$ channel. 
The integrand of the Liouville four-point function has poles at
\begin{align}\label{eq:polesL4pt}
\pm P=&\pm\tfrac{i}{2}(r_1 b^{-1} -s_1 b)\pm\tfrac{i}{2}(r_2 b^{-1} -s_2 b)-i (m+\tfrac{1}{2})b^{-1} - i (n+\tfrac{1}{2})b \nn \\
\pm P=&\pm\tfrac{i}{2}(r_3 b^{-1} -s_3 b)\pm\tfrac{i}{2}(r_4 b^{-1} -s_4 b)-i (m+\tfrac{1}{2})b^{-1} - i (n+\tfrac{1}{2})b
\end{align}
where $m,n\in \mb{Z}_{\geq 0}$. All signs are independent, resulting in 16 distinct families of poles. 
A divergence arises when a pole from one DOZZ factor (say, first line of \eqref{eq:polesL4pt}) crosses the real axis and collides with a pole from the other DOZZ factor (second line of \eqref{eq:polesL4pt}). Thus, this pinching divergence occurs when the condition
\begin{align}
\pm\frac{r_1 b^{-1} -s_1 b}{2}\pm\frac{r_2 b^{-1} -s_2 b}{2}\pm\frac{r_3 b^{-1} -s_3 b}{2}\pm\frac{r_4 b^{-1} -s_4 b}{2}= k b^{-1}+lb,~~~k,l\in \mb{Z}_{\geq 0}
\end{align}
is satisfied. This well-known condition corresponds to the saturation of the Liouville background charge and generalizes to higher-point and higher-genus cases, see e.g. \cite{Dotsenko:1984ad,Goulian:1990qr,Collier:2024kwt}.
As an example, consider the four-point amplitude $\mathsf{N}_{0,4}^{(b)}(P_{3,1},P_{3,1},P_{3,1},P_{3,1})$ in the $(4,3)$ AMS, the pole
\bal
P=\tfrac{i}{2}(r_1 b^{-1} -s_1 b)+\tfrac{i}{2}(r_3 b^{-1} -s_3 b)-i (0+\tfrac{1}{2})b^{-1} - i (0+\tfrac{1}{2})b
\eal
crosses the real axis and collides with 
\bal
P=-\tfrac{i}{2}(r_3 b^{-1} -s_3 b)-\tfrac{i}{2}(r_4 b^{-1} -s_4 b)-i (-2+\tfrac{1}{2})b^{-1} - i (-1+\tfrac{1}{2})b ~,
\eal
leading to a divergence.

We denoted such divergent cases in table~\ref{table:Sphere_table A} by $\myinf$.
In \cite{Belavin:2006ex}, a regularization procedure was proposed that involves an analytic continuation from the generalized minimal models, which are conformal field theories that can be used as matter sectors to define an a priori distinct minimal string theory from the ones considered here.
We will briefly discuss this approach later in this section.
For $(5,2)$, $(7,2)$, $(4,3)$, and $(5,3)$ models, the table includes all independent four-point amplitudes. 
For the remaining models, only a subset of the independent amplitudes are shown in the table, and in particular we do not display these divergent cases.

\begin{table}
\centering
\begin{tabular}{|c|l|}
\hline
$(p,p')$ & AMS tachyon four-point string amplitude \\
\hline
5,2 & $T_{1,1;1,1;1,1;1,1}=-2$, $T_{1,1;1,1;2,1;2,1}=2$, $T_{1,1;2,1;2,1;2,1}=4$, $T_{2,1;2,1;2,1;2,1}=6$ \\
7,2 & $T_{1,1;1,1;1,1;1,1}=-6$, $T_{1,1;1,1;2,1;2,1}=-2$, $T_{1,1;1,1;3,1;3,1}=2$, $T_{1,1;2,1;2,1;3,1}=2$ \\
& $T_{1,1;2,1;3,1;3,1}=4$, $T_{1,1;3,1;3,1;3,1}=6$, $T_{2,1;2,1;2,1;2,1}=4$,  $T_{2,1;2,1;2,1;3,1}=6$\\
& $T_{2,1;2,1;3,1;3,1}=8$, $T_{2,1;3,1;3,1;3,1}=10$, $T_{3,1;3,1;3,1;3,1}=12$\\
4,3 & $T_{1,1;1,1;1,1;1,1}=4$, $T_{1,1;1,1;2,1;2,1}=2$, $T_{2,1;2,1;2,1;2,1}=2$, $T_{1,1;2,1;2,1;3,1}=-2$ \\
& $T_{1,1;1,1;3,1;3,1}=\myred{-4}$, $T_{2,1;2,1;3,1;3,1}=\myinf$, $T_{3,1;3,1;3,1;3,1}=\myinf$ \\
 5,3 & $T_{1,1;1,1;1,1;1,1}=2$, $T_{1,1;1,1;2,1;2,1}=4$, $T_{2,1;2,1;2,1;2,1}=4$, $T_{1,1;1,1;3,1;3,1}=-2$ \\
& $T_{2,1;2,1;3,1;3,1}=0$, $T_{2,1;2,1;2,1;4,1}=-4$, $T_{1,1;3,1;3,1;3,1}=-4$, $T_{1,1;2,1;2,1;3,1}=2$ \\
& $T_{3,1;3,1;3,1;3,1}=\myinf$, $T_{1,1;2,1;3,1;4,1}=\myinf$, $T_{2,1;3,1;3,1;4,1}=\myinf$, $T_{1,1;1,1;4,1;4,1}=\myinf$ \\
& $T_{2,1;2,1;4,1;4,1}=\myinf$, $T_{3,1;3,1;4,1;4,1}=\myinf$, $T_{4,1;4,1;4,1;4,1}=\myinf$\\
8,3 & $T_{1,1;1,1;1,1;1,1}=-4$, $T_{1,1;1,1;2,1;2,1}=2$, $T_{1,1;1,1;3,1;3,1}=4$, $T_{1,1;1,1;4,1;4,1}=-2$\\
& $T_{1,1;2,1;2,1;3,1}=6$, $T_{1,1;2,1;3,1;4,1}=4$, $T_{1,1;3,1;3,1;3,1}=8$, $T_{1,1;3,1;4,1;4,1}=2$\\
& $T_{2,1;2,1;2,1;2,1}=10$, $T_{2,1;2,1;2,1;4,1}=8$, $T_{2,1;2,1;3,1;3,1}=10$, $T_{2,1;2,1;4,1;4,1}=6$\\
& $T_{2,1;3,1;3,1;4,1}=8$, $T_{2,1;4,1;4,1;4,1}=4$, $T_{3,1;3,1;3,1;3,1}=12$, $T_{3,1;3,1;4,1;4,1}=6$\\
&$T_{4,1;4,1;4,1;4,1}=2$, $T_{1,1;3,1;3,1;5,1}=2$, , $T_{1,1;4,1;4,1;5,1}=-4$, $T_{3,1;4,1;4,1;5,1}=-2$\\
10,3 & $T_{1,1;1,1;1,1;1,1}=-8$, $T_{1,1;1,1;2,1;2,1}=-2$, $T_{1,1;1,1;3,1;3,1}=4$, $T_{1,1;1,1;4,1;4,1}=2$\\
& $T_{1,1;2,1;3,1;4,1}=6$, $T_{1,1;3,1;3,1;5,1}=6$, $T_{1,1;2,1;4,1;5,1}=2$, $T_{2,1;3,1;4,1;5,1}=10$\\
 & $T_{1,1;1,1;5,1;5,1}=-4$, $T_{1,1;3,1;5,1;5,1}=2$, $T_{3,1;3,1;5,1;5,1}=10$, $T_{3,1;5,1;5,1;5,1}=6$\\
 & $T_{1,1;5,1;5,1;5,1}=-2$, $T_{5,1;5,1;5,1;5,1}=2$, $T_{5,1;5,1;1,1;7,1}=-8$\\
12,5 & $T_{1,1;1,1;1,1;1,1}=-4$, $T_{1,1;1,1;4,1;4,1}=-6$, $T_{4,1;4,1;4,1;4,1}=-10$, $T_{7,3;7,3;7,3;7,3}=34$\\
& $T_{4,1;4,1;7,3;7,3}=2$, $T_{4,2;4,2;4,2;4,2}=22$, $T_{4,1;4,1;4,2;4,2}=2$, $T_{4,1;4,2;4,2;4,3}=\myred{-16}$\\
12,7 & $T_{1,1;1,1;1,1;1,1}=4$, $T_{1,1;1,1;4,1;4,1}=\myred{-18}$, $T_{1,1;1,1;4,2;4,2}=6$, $T_{1,1;1,1;4,3;4,3}=-2$\\
 & $T_{4,1;4,1;4,1;4,1}=\myred{-62}$, $T_{4,2;4,2;4,2;4,2}=20$, $T_{4,3;4,3;4,3;4,3}=-12$, $T_{4,2;4,2;7,4;7,4}=20$\\
  & $T_{4,3;4,3;7,4;7,4}=12$, $T_{7,4;7,4;7,4;7,4}=42$, $T_{7,4;7,4;7,5;7,5}=\myred{-4}$, $T_{7,5;7,5;7,5;7,5}=\myred{-40}$\\
\hline 
\end{tabular}
\caption{Results of the numerical calculation of the $T_{r_1,s_1;r_2,s_2;r_3,s_3;r_4,s_4}$, defined in \eqref{eq:Trs def sphere4pt} and proportional to the sphere four-point string amplitude, in the $(p,p')$ A-series minimal string, for a range of values of $p$ and $p'$.}
\label{table:Sphere_table A}
\end{table}

\paragraph{Numerical results: DMS.}
Next, we present our results for the direct numerical integration of the sphere four-point amplitude in the D-series minimal string.
As discussed above, amplitudes involving four tachyons in the diagonal sector of the DMS are manifestly equal to the corresponding amplitudes in the AMS; see \eqref{eq:DMS allD same as AMS}.
Likewise, amplitudes with four tachyons in the non-diagonal sector of the DMS are equal to those in the AMS with tachyons in the diagonal sector and the same Liouville momenta $P_{r_j,s_j}$. 
To see this, observe that in the minimal model CFT correlator, the intermediate states in the conformal block decomposition are diagonal, due to the fusion rule $N\times N = D$. Moreover, the fusion rule \eqref{eq:D_series_fusion} implies that the resulting set of diagonal intermediate states is the same as if the external operators had been diagonal themselves (with $D\times D = D$). 
The only distinction in the structure constants arises from the sign factor $(-1)^{\frac{r_i-1}{2}}$ in \eqref{eq:CDseries DNN}, where $r_i$ denotes the label of the intermediate state with momenta $P_{r_i,s_i}$ and takes values $r_i=1,3,\dots, p-1$ according to \eqref{eq:D_series_fusion}. 
This sign alternates between intermediate states, but crucially, it appears twice in the full four-point amplitude: once from the fusion of the first pair of external operators, and once from the second.
As a result, the sign contribution gets squared and thus cancels out.
Therefore, the resulting minimal model correlator (and hence the full string amplitude) is identical to that of the A-series.

Therefore, in view of the conservation of diagonality \eqref{eq:DMS with odd Ns}, the nontrivial and potentially distinct sphere four-point amplitudes from those of the AMS are those involving exactly two tachyons from the diagonal sector and two from the non-diagonal sector.
Numerical results for these amplitudes are shown in table~\ref{table:Sphere_table D}. 
Indeed, we find that some of these amplitudes differ from their A-series counterparts with the same momenta but all tachyons in the diagonal sector.
For instance, we find $T_{4,1;4,1;3,1;3,1}^{\rm NNDD}=-2$, whereas $T_{4,1;4,1;3,1;3,1}^{\rm DDDD}=6$ as shown in table~\ref{table:Sphere_table A}. 
We note that table~\ref{table:Sphere_table D} shows results for only a subset of the possible four-point amplitudes of type NNDD.

\begin{table}
\centering
\begin{tabular}{|c|l|}
\hline
$(p,p')$ & DMS tachyon four-point string amplitude \\
\hline
8,3 & $T_{4,1;4,1;1,1;1,1}^{\rm NNDD}=-2$, $T_{4,1;4,1;1,1;3,1}^{\rm NNDD}=-2$, $T_{4,1;4,1;3,1;3,1}^{\rm NNDD}=-2$, $T_{4,1;4,1;1,1;5,1}^{\rm NNDD}=-4$\\
& $T_{4,1;4,1;3,1;5,1}^{\rm NNDD}=-2$\\
10,3 & $T_{5,1;5,1;1,1;1,1}^{\rm NNDD}=-4$, $T_{5,1;5,1;1,1;3,1}^{\rm NNDD}=-2$, $T_{5,1;5,1;3,1;3,1}^{\rm NNDD}=-2$, $T_{5,1;5,1;1,1;5,1}^{\rm NNDD}=-2$\\
& $T_{5,1;5,1;1,1;1,1}^{\rm NNDD}=-4$, $T_{5,1;5,1;5,1;5,1}^{\rm NNDD}=-2$, $T_{5,1;5,1;1,1;7,1}^{\rm NNDD}=8$\\
\hline 
\end{tabular}
\caption{Results of the numerical calculation of the $T^{\mathrm{NNDD}}_{r_1,s_1;r_2,s_2;r_3,s_3;r_4,s_4}$, defined in \eqref{eq:Trs def sphere4pt} and proportional to the sphere four-point string amplitude, in the $(p,p')$ D-series minimal string, for a range of values of $p$ and $p'$. We only calculate correlators that differ from the AMS.}
\label{table:Sphere_table D}
\end{table}

\paragraph{Numerical results: EMS.}
Lastly, in table~\ref{table:Sphere_table E} we present numerical results for the sphere four-point amplitude in the $(12,p)$ E-series minimal string. 
In contrast to the torus one-point amplitudes shown in table~\ref{table:torus1pt E(12,pp)}, where deviations from integrality could always be accounted for by simple irrational prefactors such as $\sqrt{2}$ or $\sqrt{6}$, we have not yet identified a similarly simple irrational factor multiplying an integer in the E-series four-point cases.

\begin{table}
\centering
\begin{tabular}{|c|l|}
\hline
$(p,p')$ & EMS tachyon four-point string amplitude \\
\hline
12,5 & $T_{1,1;1,1;1,1;1,1}=-4$, $T_{1,1;1,1;4,1;4,1}=-6$, $T_{4,1;4,1;4,1;4,1}=-18$, $T_{7,3;7,3;7,3;7,3}=36$\\
& $T_{4,1;4,1;7,3;7,3}=6$, $T_{4,2;4,2;4,2;4,2}=6$, $T_{4,1;4,1;4,2;4,2}=6.01091$\\
& $T_{4,1;4,3;4,2;4,2}=\myred{0.296287}$\\
12,7 & $T_{1,1;1,1;1,1;1,1}=4$, $T_{1,1;1,1;4,1;4,1}=\myred{-18}$, $T_{1,1;1,1;4,2;4,2}=6$, $T_{1,1;1,1;4,3;4,3}=-2$\\
& $T_{4,1;4,1;4,1;4,1}=\myred{-54}$, $T_{4,2;4,2;4,2;4,2}=27$, $T_{4,3;4,3;4,3;4,3}=3$, $T_{4,2;4,2;7,4;7,4}=12$\\
& $T_{4,3;4,3;7,4;7,4}=0$, $T_{7,4;7,4;7,4;7,4}=12$, $T_{7,4;7,4;7,5;7,5}=-0.0246077$\\
& $T_{7,5;7,5;7,5;7,5}=\myred{-48}$ \\
\hline 
\end{tabular}
\caption{Results of the numerical calculation of the $T_{r_1,s_1;r_2,s_2;r_3,s_3;r_4,s_4}$, defined in \eqref{eq:Trs def sphere4pt} and proportional to the sphere four-point string amplitude, in the $(12,p)$ E-series minimal string, for a range of values of $p$. We expect the non-integer values in the table are actually algebraic numbers.}
\label{table:Sphere_table E}
\end{table}

\paragraph{Minimal string amplitudes from GMS amplitudes.}
Certain four-point diagrams in the $(p,p')$ A-series minimal string were analytically computed in \cite{Belavin:2006ex} through a subtle limiting process which we now outline. 
In \cite{Belavin:2006ex}, the worldsheet theory studied is given by a generalized minimal model (GMM) coupled to Liouville. GMMs are 2d CFTs that in a certain sense generalize the minimal model CFTs summarized in this work. Whereas minimal models exist for $b=\sqrt{p/p'}$, GMMs are defined for a continuous central charge labelled by $b$ with a continuous set of non-degenerate primary operators $\widehat V_{\alpha}$, along with a discrete set of degenerate primary operators $\widehat V_{r,s}$. The four-point string amplitude for one degenerate and three non-degenerate tachyons $\lb \cT_{r,s} \cT_{\alpha_1} \cT_{\alpha_2} \cT_{\alpha_3} \rb$ was computed using the Liouville higher equations of motion \cite{Zamolodchikov:2003yb}. It was argued that under certain conditions the amplitudes can be analytically continued $\cT_{\alpha_i} \to \cT_{r_i,s_i}$ with $b \to \sqrt{p/p'}$ to recover the corresponding $(p,p')$ AMS amplitudes. The final result is\footnote{This formula has some assumptions. For instance, when $s_i=1$, the assumption is $r_1+r_4\leq r_2+r_3$, $r_1\leq \ldots\leq r_4$ (see for example \cite[eq. (2.26)]{Belavin:2013nba}).}
\be
\mathsf{N}&_{0,4}^{(b),\,\text{BZ}}(P_{r_1,s_1},P_{r_2,s_2},P_{r_3,s_3},P_{r_4,s_4}) \nn\\
&\quad= f(b) \lr{
r_1s_1 (r_1 p'+s_1 p) - \sum_{i=2}^4 \sum_{r \overset{2}{=}1-r_1}^{r_1-1} \sum_{s \overset{2}{=}1-s_1}^{s_1-1} |(r_i-r)p' - (s_i-s)p|  }
\ee
These results were matched from the matrix integral side for $(2,p)$ models in \cite{Belavin:2008kv} and for $(3,p)$ models in \cite{Belavin:2013nba}. In particular, this procedure and resulting formula yields finite answers even when the naive string amplitude appears infinite, which we denoted by $\myinf$ in the tables. We tested this formula against our numerics and found agreement for \emph{more general} $(p,p')$ A-series minimal string models.

Following these methods, five-point correlators were computed for $(2,p)$ models on the worldsheet in \cite{Artemev:2022nvl} and in the matrix model in \cite{Tarnopolsky:2009ec}, but the calculations do not match and the problem remains open.

\section{Towards more general observables: conformal boundaries on the worldsheet}
\label{sec:conformalboundaries}

In this section we discuss minimal string theory in the presence of boundaries. We first discuss boundary conditions in the A- and D-series minimal CFTs, as well as in Liouville CFT. The boundary conditions for rational diagonal CFTs were classified in \cite{Cardy:1984bb}, while rational non-diagonal theories were considered in \cite{Behrend:1998fd,Behrend:1998mu}. We first start with the A-series which is diagonal, and then discuss the D-series. The E-series boundary structure constants have not yet been worked out in detail.

\subsection{Conformal boundary conditions}

\subsubsection{A-series minimal model boundaries}
In the $(p,p')$ A-series CFT, there are $\frac{1}{2}(p-1)(p'-1)$ distinct conformal boundary conditions in one-to-one correspondence with the diagonal primary fields \cite{Cardy:1989ir,Lewellen:1991tb}. The Cardy boundary states take the form
\begin{align}\label{eq:Acardy}
| r,s \rb =  \sum_{\substack{r' = 1,2,\ldots, p-1  \\  s' = 1,3,\ldots, p'-2} } \frac{S_{r',s'}^{r,s}}{(S_{1,1}^{r',s'})^{1/2}} 
\times\frac{1}{N_{r',s'}}|\widehat{V}_{r',s'}\rrangle\,, \qq
\end{align}
where we assume $p'$ is odd and the step size of $s$ is two. The sum is over Ishibashi states $|\widehat{V}_{r',s'}\rrangle$ \cite{Ishibashi:1988kg} with the factor of $N_{r',s'}^{-1}$ an artifact of our normalization convention for the minimal model operators. The modular S-matrix takes the form
\be
S^{r',s'}_{r,s} = \sqrt{\frac{8}{p p'}} (-1)^{1+r s' + s r'} \sin \lr{ \frac{\pi p'}{p} r' r } \sin \lr{ \frac{\pi p}{p'} s' s }\,.
\ee
In our conventions, it is more natural to rewrite the Cardy state \eqref{eq:Acardy} as
\begin{gather}
| r,s \rb = d(\beta)^{-2} \!\!\!\sum_{\substack{r' = 1,2,\ldots, p-1  \\  s' = 1,3,\ldots, p'-2} }\frac{(iP_{r',s'})^2}{\rho_{\beta}(iP_{r',s'})} \widehat{\Psi}_{r,s}(P_{r',s'}) |\widehat{V}_{r',s'}\rrangle ~, \\ 
\widehat{\Psi}_{r,s}(P_{r',s'}) = \frac{N_{r',s'}S_{r',s'}^{r,s}}{(S_{1,1}^{r',s'})^{1/2}} = d(\beta)\frac{2\sqrt{2}(pp')^{\frac{1}{4}}}{\sqrt{(r'\beta^{-2} - s'\beta^{2})^2}}S^{r,s}_{r',s'} ~,
\end{gather}
where we multiplied both numerator and denominator by $N^2_{r',s'}$ to write the answer in terms of the measure used earlier, using \eqref{eq:measuresimplify 0}. The disk one-point function $\widehat{\Psi}_{r,s
}(P_{r',s'})$ in the presence of the $|r,s\rb$ boundary condition \cite{Cardy:1989ir,Lewellen:1991tb}:
\begin{align}
\lb \widehat{V}_{r',s'}(z,\overline{z}) \rb_{|r,s\rb} = \frac{\widehat{\Psi}_{r,s
}(P_{r',s'})}{|z-\overline z|^{2 h_{r,s}}} ~,
\end{align}
where we use complex coordinates for the upper-half plane (UHP), with the boundary along the real axis. As in the case of the closed string sector,  the various prefactors of $d(\beta)$ could be reabsorbed into suitable string path-integral normalization constants (with boundaries) and are not physically meaningful.

The cylinder amplitude sandwiched between boundary states $a=|r_1,s_1\rb$ and $b=|r_2, s_2\rb$ with interval length $\pi$ and thermal periodicity $2\pi t$ in the open string channel is \cite{Runkel:1998he,Runkel:1999dz,Martinec:2003ka,mtms}
\be \label{eqn:matter_cylinder}
Z_{a|b}(q) = \sum_{\substack{r = 1,2,\ldots, p-1  \\  s = 1,3,\ldots, p'-2}} n^{a,b}_{r,s} ~\chi^{(i b)}_{r,s}(q)\,, \qq  n^{a,b}_{r,s} = f_{r_1,s_1;r_2,s_2;r,s} = \mathcal{N}_{r_1,s_1;r_2,s_2}^{r,s}
\ee
where $n^{a,b}_{r,s}$ is the indicator function for whether the open string state is in the spectrum, and in this case it is given precisely by the fusion rule indicator function.
The characters in the open channel use $q = e^{-2\pi t}$. The degenerate Virasoro characters are
\begin{align}
\chi^{(i b)}_{r, s}(q) &=\frac{1}{\eta(q)} \sum_{k \in \mathbb{Z}}q^{a_{r, s}(k)}-q^{a_{r,-s}(k)}, \quad a_{r, s}(k)=\frac{\left(2 p^{\prime} p k+p' r-p s\right)^2}{4 p^{\prime} p}\,, \nn\\
\chi^{(i b)}_{r, s}(q')&=\frac{1}{\sqrt{t} \eta(q)} \sum_{k \in \mathbb{Z}}(q')^{a_{r, s}(k)}-(q')^{a_{r,-s}(k)}\,,
\end{align}
where we have given the characters in both channels $q=e^{-2\pi t}, q'=e^{-2\pi/t}$, and a superscript for later clarity.\footnote{The cylinder amplitude is easy to understand, the only possible states that propagate in the open channel are those that are in the fusion of the $a$ and $b$ boundary conditions. This is enforced by the indicator function $n^{a,b}_{r,s}$. The boundary condition we will be most interested in is the identity boundary $|1,1\rb$, which only supports the identity operator propagating in the open-channel on the cylinder with two identity boundaries, and so \eqref{eqn:matter_cylinder} only has a contribution from $r=s=1$.}

There are also boundary primary operators $\psi^{a b}_i(x)$, which are restricted to live on the boundary (with $x\in\mathbb{R}$ in the UHP) and can change the boundary condition from $a$ to $b$ \cite{Lewellen:1991tb}. In this work, we will not make use of such boundary-changing operators. The only case of interest is the \emph{boundary} identity operator, which in our conventions is normalized as follows,
\be
\lb \mathbb{1}(x) \rb_{|r,s\rb}= \frac{N_{1,1}S^{1,1}_{r,s}}{\lr{S^{1,1}_{1,1}}^{1/2}} = \widehat{\Psi}_{r,s}(P_{1,1})~,
\ee
which is the same as the bulk identity one-point function.

\subsubsection{D-series minimal model boundaries}
\begin{figure}
    \centering
    \includegraphics[width=0.8\linewidth]{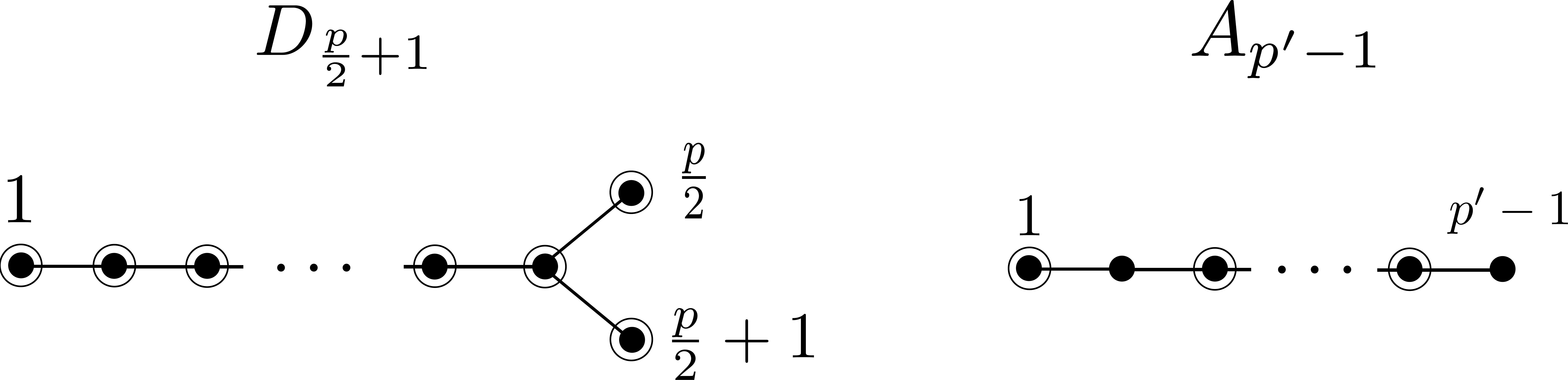}
    \caption{Dynkin diagrams that are used to label the Cardy boundary states of the $(p,p')$ D-series minimal model with even $p$. The Cardy states is labelled by $|\alpha,\beta\rb$ with $\alpha$ picked out of ant node of the $D_{\frac{p}{2}+1}$ diagram, and $\beta$ from any of the odd-nodes of the $A_{p'-1}$ diagram. The ``non-invariant" boundary states come from picking either of two off-shoot nodes from the first diagram. For the $(p,p')$ A-series we would have $(A_{p-1},A_{p'-1})$ and boundary states would be labelled by any node of the first and odd nodes of the second diagram. }
    \label{fig:Dynkin_diagram}
\end{figure}

The Cardy states in the D-series are worked out in \cite{Behrend:1998fd,Behrend:1998mu,Behrend:1999bn,Runkel:1999dz}. While the spectrum is no longer diagonal, boundary states are still in one-to-one correspondence with primaries that satisfy $h=\tilde h$. There are $\frac{1}{4}(p'-1)(p+2)$ distinct boundary conditions, where as a reminder $p\geq 6$ is even. The cylinder amplitude in the open channel is identical to earlier
\be
Z_{a|b}(q') =  \sum_{\substack{r = 1,2,\ldots, p-1  \\  s = 1,3,\ldots, p'-2}} n^{a,b}_{r,s} \chi_{r,s}^{(i b)}(q')\,,
\ee
 The Cardy states are most easily labeled by choices of nodes in the Dynkin diagrams \cite{Runkel:1999dz} $(A_{p'-1},D_{\frac{p}{2}+1})$ with two integers $|\alpha, \beta\rb$ and are split into two categories known as (invariant) i-type and (non-invariant) n-type boundaries, see figure \ref{fig:Dynkin_diagram}.\footnote{The terminology comes from the $\mathbb{Z}_2$ symmetry of the $D$ Dynkin diagram. An n-type boundary condition is given by choosing one of the two end nodes on the Dynkin diagram which swap under the obvious reflection of the diagram.} The classification is
\begin{align} \label{eqn:D-series_bdycnds}
&\t{i-type:} \qquad |\alpha,\beta\rb, \qquad \alpha =1,2,\ldots, \frac{p}{2}-1, \qquad \beta = 1,3,\ldots, p'-2 \,,\\
&\t{n-type:} \qquad |\alpha,\beta\rb, \qquad \alpha = \frac{p}{2}, \frac{p}{2}+1, \qquad \beta =1,3,\ldots, p'-2\,.
\end{align}
States in the spectrum on the cylinder take a complicated form
\begin{align}\label{eqn:cylinder_Dseries}
    \t{i-i boundary:} \qquad n_{(r,s)}^{a,b} &= \mc{N}^{(r_i,s_i)}_{(\alpha,\beta),(\alpha',\beta')} + \mc{N}^{(r_i,s_i)}_{(\alpha,\beta),(p-\alpha',\beta')} \\
    \t{n-i boundary:} \qquad n_{(r,s)}^{a,b} &= \mc{N}^{(r_i,s_i)}_{(\frac{p}{2},\beta),(\alpha',\beta')} \nn \\
    \t{n-n boundary:} \qquad n_{(r,s)}^{a,b} &=\begin{cases}
        \delta_{r_i \t{\,mod\,} 4 =1} \times \mc{N}^{(r_i,s_i)}_{(\frac{p}{2},\beta),(\frac{p}{2},\beta')} ,\qquad \alpha = \alpha' \nn\\
        \delta_{r_i \t{\,mod\,} 4 =3} \times \mc{N}^{(r_i,s_i)}_{(\frac{p}{2},\beta),(\frac{p}{2},\beta')} ,\qquad \alpha \neq \alpha'
    \end{cases}
\end{align}

\paragraph{Identity boundary state.}
The boundary state we primarily focus on in this work is the analogue of the D-series identity boundary state given by $|\alpha,\beta\rb=|1,1\rb$ which is an i-type boundary.\footnote{In the work \cite{Runkel:1999dz} it is denoted by $|\omega\rb$ to distinguish it from the A-series identity.} The cylinder diagram between two identity branes is
\begin{align} \label{eqn:cylinder_Dseries_identity}
 Z_{1|1}(q')=\chi^{(i b)}_{1,1}(q) +  \chi^{(i b)}_{p-1,1}(q)\,,
\end{align}
with the same conventions as in \eqref{eqn:matter_cylinder}. The identity boundary in the D-series supports two boundary primary operators: the identity operator $\mathbb{1}$, and a primary with dimension $h=\frac{1}{4}(p'-2)(p-2)$. 

%
\paragraph{Extra boundary state.}
There is another boundary state that will be important later given by the n-type $a=|\frac{p}{2},1\rb$.\footnote{We could have also used $|\frac{p}{2}+1,1\rb$ without much modification. In the full string theory we find evidence that these states are BRST equivalent up to terms involving the identity state $|1,1\rb$, tensored with an appropriate Liouville states of course.} The cylinder diagram in the open string channel is 
\begin{align} \label{eqn:cylinder_Dseries_extra}
 Z_{a|a}(q)=\sum_{\substack{r = 1,2,\ldots, p-1  \\  s = 1,3,\ldots, p'-2}}  \delta_{r \t{\,mod\,} 4 =1} \times \mc{N}^{(r,s)}_{(\frac{p}{2},1),(\frac{p}{2},1)}  \times \chi^{(i b)}_{r,s}(q) 
\end{align}

\paragraph{Disk one-point functions.}The only non-vanishing one-point functions on the disk involve the boundary identity operator (or a bulk operator that has the boundary identity operator in its boundary OPE). We use conventions where the one-point function of the identity operator in the D-series is \cite{Runkel:1998he,Runkel:1999dz}
\begin{gather} \label{eqn:disk_bdy_1pt_MM}
\lb \mathbb{1}(x) \rb_{|1,1\rb}= \lr{S_{1,1}^{1,1}}^{1/2} N_{1,1}\,, \\
\lb \mathbb{1}(x) \rb_{|\frac{p}{2},1\rb}= \lb \mathbb{1}(x) \rb_{|\frac{p}{2}+1,1\rb}=\frac{1}{2}\lr{S^{\frac{p}{2},1}_{1,1}}^{1/2} N_{1,1}\,,
\end{gather}
where the first line is identical to A-series result, but the second is slightly different.\footnote{The odd type boundary state has a different normalization for the one-point function, see \cite[eq. (70)]{Runkel:1999dz}.} The one-point function of diagonal and non-diagonal fields for the identity boundary condition is\footnote{We use different normalization from \cite{Runkel:1999dz}. Our conventions are that bulk fields have Kronecker-delta normalized two-point functions, and the boundary identity one-point function is scaled to match the similar A-series answer \cite{Lewellen:1991tb,Runkel:1998he}.} \cite{Runkel:1999dz}
\begin{align} \label{eqn:MM_1pt}
\lb \widehat{V}^D_{r,s}(z,\overline{z}) \rb_{|1,1\rb} &= \frac{1}{|z-\overline z|^{2 h_{r,s}}} \lr{S^{r,s}_{1,1}}^{1/2} N_{r,s}\,,\\
\lb \widehat{V}^N_{r,s}(z,\overline{z}) \rb_{|1,1\rb} &= 0\,. 
\end{align}
The non-diagonal field has vanishing one-point function since the boundary identity is not in its OPE. The bulk one-point function for the n-type boundaries is more complicated and given in \cite{Runkel:1999dz}.

\subsubsection{Liouville boundary conditions}
The boundary conditions of Liouville fall into two classes: FZZT and ZZ. They are given by
\be
|\t{FZZT}^{(b)}(s)\rb = \int_0^\infty \d P \rho_0^{(b)}(P) \Psi^{(b)}(s;P) | V_P\rrangle\,,
\ee
with $| V_P\rrangle$ Ishibashi states for Liouville primaries $V_P$, and $s\in \mathbb{R}$. The one-point function of a primary $V_P$ evaluated on a disk bounded by the FZZT brane after stripping off the position dependence is
\be \label{eqn:Liouville_1pt}
\Psi^{(b)}(s;P)=\frac{2 \sqrt{2} \cos \lr{4\pi s P}}{\rho_0^{(b)}(P)}\,.
\ee
The ZZ-branes come in a discrete set indexed by two integers $(m,n)$ and take the form
\be
|\t{ZZ}^{(b)}_{(m,n)}\rb = \int_0^\infty \d P \rho_0^{(b)}(P) \Psi^{(b)}_{(m,n)} | V_P\rrangle\,.
\ee
The disk one-point function of primary $V_P$ with ZZ boundary condition is 
\be
\Psi^{(b)}_{(m,n)}(P)=\frac{4 \sqrt{2} \sinh \lr{2\pi m b P} \sinh \lr{2\pi n b^{-1} P}}{\rho_0^{(b)}(P)}\,.
\ee
\paragraph{Cylinder Amplitudes.} 
There are three cylinder amplitudes we can evaluate given by all pairings of FZZT and ZZ boundary conditions. We quote cylinder amplitudes in both the open and closed channels since both representations will be useful in applications later. The FZZT-FZZT amplitude is best expressed in the closed channel with thermal periodicity $\pi$ and interval length $2\pi t$. We quote the necessary results following conventions of \cite{Collier:2023cyw}
\begin{align} \label{eqn:Liouville_FZZT_cylinder}
\lb \t{FZZT}^{(b)}(s_1)| e^{-\pi t (L_0+ \bar L_0 - \frac{c}{12})} &| \t{FZZT}^{(b)}(s_2) \rb \nn\\
& = \frac{1}{2} \int_\Gamma \d P  \rho_0^{(b)}(P) \Psi^{(b)}(s_1;P) \Psi^{(b)}(s_2;P) \nn \\ 
&= \frac{1}{\sqrt{2}} \int_\Gamma \d P \frac{\cos \lr{4\pi s_1 P} \cos \lr{4\pi s_2 P}}{\sinh \lr{2\pi b P} \sinh \lr{2\pi b^{-1} P}} \chi_P^{(b)}(q) ~. 
\end{align}
with $q=e^{-2\pi t}$. The contour $\Gamma$ runs along $\mathbb{R}\pm i \epsilon$ slightly above or slightly below the real axis, and avoids the singularity at $P=0$. The non-degenerate Virasoro characters $\chi_P^{(b)}(q)$ are listed below.

For the ZZ-ZZ boundaries with interval length $\pi$ and time periodicity $2\pi t$ the open string channel expression is more useful 
\be \label{eqn:ZZ_liouville}
\lb\t{ZZ}_{(m, n)}^{(b)}| \mathrm{e}^{-\pi t(L_0+ \bar L_0-\frac{c}{12})} |\t{ZZ}_{(m', n')}^{(b)}\rangle=\sum_{r\overset{2}{=} \left|m-m'\right|+1}^{m+m'-1} \sum_{s\overset{2}{=} \left|n-n'\right|+1}^{n+n'-1} \chi_{(r, s)}^{(b)}\lr{q'} ~,
\ee
where $q'=e^{-2\pi/t}$ is now in the open channel, and $\chi_{(r, s)}^{(b)}$ are degenerate Virasoro characters which have a single null vector and are given, along with non-degenerate characters, by
\begin{gather}
    \chi_P^{(b)}(q) = \frac{q^{P^2}}{\eta(q)} ~, \qq \chi_{(r, s)}^{(b)}\lr{q} = \frac{q^{-\frac{1}{4}(r b + s b^{-1})^2}-q^{-\frac{1}{4}(r b - s b^{-1})^2}}{\eta(q)}~.
\end{gather}

\subsection{String boundary conditions}

\paragraph{AMS boundary conditions.} We discuss the valid string theory boundary conditions. With FZZT boundaries for Liouville the set of boundary conditions is 
\be
| \t{FZZT}^{(b)}(s) \rb \otimes |r',s' \rb\,,
\ee
where $r',s'$ label the Cardy state in the matter sector. Seiberg-Shih \cite{Seiberg:2003nm} showed the above states are over complete as boundary conditions in the full string theory. A non-identity matter state can be re-written as superposition of FZZT states tensored with the matter identity state $|1,1\rb$ up to BRST exact terms
\be \label{eqn:Seiberg-Shih-equivalence}
| s \rb_{\t{FZZT}} \otimes |r,s\rb =\sum_{u\overset{2}{=}1-r}^{r-1}\sum_{v\overset{2}{=}1-s}^{s-1} \, |s + \tfrac{i}{2} (u b^{-1} +v b)\rb_{\t{FZZT}} \otimes |1,1\rb + (\t{BRST exact}) ~.
\ee 
We have modified the notation for an FZZT state in the equation for clarity.  It is thus sufficient to restrict to the matter identity state. For ZZ boundaries a basis of independent states is similarly
\begin{gather}
|\t{ZZ}^{(b)}_{(m,n)} \rb \otimes |1,1\rb, \nn \\ m=1,\ldots, p-1, \qq n = 1,\ldots, p'-1\,, \qq \t{requiring:}~~ m p' - n p >0\,.
\end{gather}
There are $(p-1)(p'-1)/2$ such states known as ZZ-instantons, and they match the number of physical vertex operators.

\paragraph{DMS boundary conditions.} In the DMS with FZZT boundaries in the Liouville sector the states are identical
\be
| \t{FZZT}^{(b)}(s) \rb \otimes |r',s' \rb\,,
\ee
where the matter states now take values from \eqref{eqn:D-series_bdycnds}. One of our results, explained in detail in appendix \ref{app:seiberg_shih_D_series}, is the Seiberg-Shih equivalence for D-series boundary states,\footnote{In the Seiberg-Shih equivalence all matter states $|r,s\rb$ can be reduced to $|1,1\rb$. In the D-series we find this is also the case for the second index $|a,b\rb \to |a,1\rb$, the first index has more structure as we explain below.} and so it is sufficient to only consider the following sets of states\footnote{For the n-type boundaries we can consider matter states of the form $|\t{FZZT}\rb\otimes(|\frac{p}{2},1\rb \pm |\frac{p}{2}+1,1\rb)$. It turns out that the $+$ combination is BRST equivalent to an $|\t{FZZT}\rb \otimes |1,1\rb$ state. Interestingly the $-$ state is orthogonal to all $|\t{FZZT}\rb \otimes |1,1\rb$ states, i.e. the cylinder diagram vanishes. This is slightly unusual from the dual matrix integral perspective, since in general we expect the cylinder diagram not to vanish between resolvents of different matrices that are coupled, see for example \cite{Daul:1993bg}. We thus think a more natural state to consider is $|\t{FZZT}\rb\otimes|\frac{p}{2},1\rb$. We can also take the matter state to be $|\frac{p}{2}+1,1\rb$ but this is BRST equivalent to the state we choose up to $|\t{FZZT}\rb\otimes|1,1\rb$ states.}
\be
| \t{FZZT}^{(b)}(s) \rb \otimes |1,1 \rb\,,~~~ | \t{FZZT}^{(b)}(s) \rb \otimes  |\frac{p}{2},1 \rb  \, .
\ee
For ZZ states in the DMS we find a similar structure. There are many possible representations of the $\frac{1}{4}(p+2)(p'-1)$ basis states. One of them is
\begin{gather} \label{eqn:ZZ_bdy_states}
|\t{ZZ}^{(b)}_{(r,s)} \rb \otimes |1,1\rb \q  \bigcup \q |\t{ZZ}^{(b)}_{(1,s)} \rb \otimes |\frac{p}{2},1\rb \,,\nn\\
r = 1,2,\ldots, \frac{p}{2}, \qq s=1,3,\ldots, p'-2 \,.
\end{gather}

\subsection{Disk diagrams} \label{sec:4.3disk_diagrams}
We first explain what different string theory boundary conditions map to on the matrix model side.\footnote{The relation between boundary conditions and matrix integral insertions is well established for the AMS, and we will make the assumption that the same relations continue to hold for the DMS.} To explain how bulk boundary conditions relate to matrix integral observables it is useful to introduce the classical Lagrangian with the FZZT boundary term on a disk with metric $\tilde g$ \cite{Fateev:2000ik, Teschner:2000md}
\be
I_{L}[\varphi] = \frac{1}{4\pi} \int_M \sqrt{\tilde{g}} \lr{\tilde{g}^{i j} \partial_i \varphi \partial_j \varphi + Q \tilde{R} \varphi + \mu e^{2 b \varphi}} + \frac{1}{2\pi} \int_{\partial M} \sqrt{\tilde{h}} \lr{ Q \tilde{K} \varphi + 2\pi \mu_B e^{b \varphi} }\,.
\ee
The dimensionful bulk cosmological constant is $\mu$ while the adjustable FZZT boundary cosmological constant is related to the parameter $s$ used earlier by \cite{Fateev:2000ik}
\be\label{eqn:muB_s_relation}
\mu_B=\lr{\frac{\mu}{\sin(\pi b^2)}}^{1/2}\cosh(2\pi b s)\in [E_0,\infty)\,,
\ee
with $s=\frac{1}{2\pi b} \arccosh\lr{\frac{\mu_B}{E_0}}\in \mathbb{R}$ and we have implicitly defined a parameter $E_0$ for later. 

\paragraph{Fixed length boundary.}
The natural observable in a matrix integral is to evaluate the expectation value of $\Tr\lr{e^{-\beta H_i}}$ where $H_i$ is one of the matrices. This amplitude is interpreted gravitationally as disk with a boundary of length $\beta$. The natural measure for the length of the boundary in Liouville is through the integral $\int_{\partial M} \sqrt{\tilde{h}} e^{b \varphi}$ which can be achieved by an inverse Laplace transform motivated by the following transform of the classical Lagrangian
\be
\int_{i\mathbb{R}} \d \mu_B \exp\lr{\mu_B \beta-\int_{\partial M} \sqrt{\tilde{h}} \hspace{.04cm}\mu_B e^{b \varphi} } = \delta\lr{\beta - \int_{\partial M} \sqrt{\tilde{h}} e^{b \varphi} }\,,
\ee
This is not entirely correct, see for example \cite{Kostov:2003uh}. The calculation on the disk will show that the transform actually implements
\be
-\int_{i\mathbb{R}}\frac{\d \mu_B}{2\pi i}\, \e^{\mu_B \beta} | \t{FZZT}^{(b)}\lr{s\lr{\mu_B}} \rb \otimes |1,1\rb_{\t{MM}} \qq \longleftrightarrow \qq \frac{1}{\beta} \Tr\lr{e^{-\beta H}}\,,
\ee
with an extra factor of $\frac{1}{\beta}$. To get rid of the prefactor, we have to mark the boundary with an operator insertion, which can be done by taking a derivative\footnote{The factor of $\frac{1}{\beta}$ can be understood as follows: the boundary has no marked points (operator insertions), and so field configurations in the path integral that are related by a rotation are gauge equivalent and are not double counted. This makes the disk without a marked point a factor of $\frac{1}{\beta}$ smaller than the disk with the point. To add a marked point we must insert an operator on the boundary, the most natural non-vanishing string theory boundary operator is given by dressing the minimal model identity by a Liouville operator $c e^{b \varphi} \mathbb{1}(x)$, which is just the Liouville boundary cosmological constant. This operator can be brought down by taking a derivative. This does not dress the operator with the ghost, but the correct amplitude with the ghost dressing does not modify any equations.}
\be \label{eqn:fixed_length_bc}
-\int_{i\mathbb{R}}\frac{\d \mu_B}{2\pi i}\, \e^{\mu_B \beta} \partial_{\mu_B} | \t{FZZT}^{(b)}\lr{s\lr{\mu_B}} \rb \otimes |1,1\rb_{\t{MM}} \qq \longleftrightarrow \qq \Tr\lr{e^{-\beta H}}\,,
\ee
This is also known as a marked disk boundary condition. The matter doesn't play any non-trivial role.

\paragraph{Fixed energy boundaries.} 
Applying the identity $-\frac{1}{2\pi i}\int_{i\mathbb{R}} \d E e^{\beta E} \Tr \frac{1}{-E-H} = \Tr \lr{e^{-\beta H}}$ to \eqref{eqn:fixed_length_bc}, it follows that marked and unmarked FZZT boundaries map to matrix integral insertions
\begin{align}
\partial_{\mu_B} &| \t{FZZT}^{(b)}\lr{s\lr{\mu_B}} \rb \otimes |1,1\rb_{\t{MM}} \qq \longleftrightarrow \qq -\Tr\lr{\frac{1}{\mu_B+H}}\,,\\
&| \t{FZZT}^{(b)}\lr{s\lr{\mu_B}} \rb \otimes |1,1\rb_{\t{MM}} \qq \longleftrightarrow \qq - \Tr \log \lr{\mu_B+H}\,.
\end{align}
Resolvents are typically defined as $R(\tilde{E})=\Tr (\tilde{E}-H)^{-1}$ and so the FZZT parameter maps to $\mu_B=-\tilde{E}$.

\paragraph{Punctured disk.} The disk one-point string diagrams of bulk tachyons $\mc{T}^X_{r,s} = \mathcal{N}^{(b)} V_{i P_{r,s}} \widehat{V}^X_{r,s}$ for both AMS and DMS with an FZZT boundary parameterized by $s$ is \cite{Polchinski:1998rq} 
\begin{align} \label{eqn:Disk_1_pt}
Z_{\t{disk}}(s;P_{r,s}) &= \int \frac{\d^2 z}{\t{vol}(\t{PSL}(2;\mathbb{R}))} \lb c \ol c \mathcal{T}^X_{r,s}(z, \ol z) \rb_s\\ 
&= \mathcal{N}^{(b)}\frac{C_{D^2}^{(b)}}{\t{vol}(\t{U}(1))} \lb V_{i P_{r,s}} \rb^{\t{L}}_s \times \lb \widehat{V}^X_{r,s} \rb^{\t{MM}}_{|1,1\rb} \times \lb c \ol c \rb \nn \\
&=\mathcal{N}^{(b)}\frac{C_{D^2}^{(b)}}{2\pi} \Psi^{(b)}(s;i P_{r,s}) \widehat{\Psi}_{1,1}(P_{r,s})  \delta_{X,D} \nn \\
&= \mathcal{N}^{(b)}\frac{C_{D^2}^{(b)}N_{1,1}}{2\pi} \times \frac{2 \sqrt{2} \cos \lr{4\pi i s P_{r,s}}}{\rho_0^{(b)}(i P_{r,s})} \times \lr{S^{r,s}_{1,1}}^{1/2} \,\delta_{X,D}
\,, \nn 
\end{align}
where we used the minimal model one-point function \eqref{eqn:MM_1pt} and Liouville one-point function \eqref{eqn:Liouville_1pt}. The D-series tachyons have an extra index $X=D,N$ compared to the A-series where we can set $X=D$ and hence the Kronecker delta is absent. It is interesting that the non-diagonal fields in the D-series have vanishing one-point functions. The ghost partition function amounts to a normalization constant that will always be absorbed into the disk normalization $C_{D^2}^{(b)}$.

\paragraph{Punctured disk: Cosmological constant operator.}
There is a special tachyon operator which corresponds to inserting the bulk cosmological constant operator. In the Lagrangian formulation this is achieved by differentiating the empty disk partition function w.r.t. the bulk cosmological constants. This is the identity tachyon considered above
\begin{align} \label{eqn:cc_1pt}
Z_{\t{disk}}(s;P_{1,1}) =\mathcal{N}^{(b)}\frac{C_{D^2}^{(b)}N_{1,1} \lr{S^{1,1}_{1,1}}^{1/2}}{4\pi} \times \frac{\cosh \lr{2\pi s (b-b^{-1})}}{\sin(-\frac{\pi}{b^2} ) \sin(\pi b^{2})} \times \mu^{\frac{1-b^2}{2 b^2}}\,,
\end{align}
where we have set $P_{1,1}=\frac{1}{2}(b-b^{-1})$. We have also temporarily restored the bulk cosmological constant $\mu$ which we have fixed to unity throughout the work, and we have included the normalization from the minimal model for consistency with other formulas. See \cite{Seiberg:2003nm,mtms} for these amplitudes in conventions for Liouville with the classic form of the DOZZ formula.

\paragraph{Empty disk.} The simplest string amplitude with boundary is the empty disk diagram. The amplitude is
\begin{align}
&Z_{\t{disk}} = \frac{Z_{\t{CFT}}}{\t{vol}(\t{PSL}(2;\mathbb{R}))} 
\q = \q\raisebox{-.6cm}{\includegraphics[width=1.5cm]{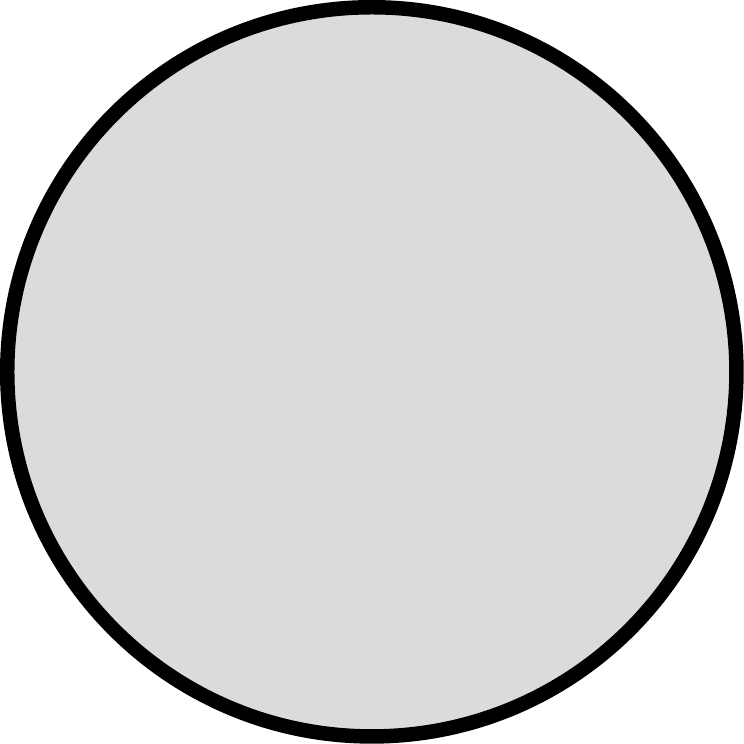}}\\ 
&Z_{\t{CFT}} = Z_{\t{Liouville}} Z_{\t{MM}} Z_{\t{ghost}}\,,\nn
\end{align}
where we must specify a choice of disk boundary conditions for the Liouville and minimal model partition functions. The volume of the conformal killing group is divided out, and is infinite and famously difficult to regularize \cite{Liu:1987nz,Tseytlin:1987ww,Eberhardt:2021ynh}.\footnote{At a technical level, the infinite volume comes from integrating over zero modes in the ghost path integral, which have been separated out of the ghost partition function in the above formula \cite{Polchinski:1998rq}.}  In minimal string theory the above amplitude is finite because the Liouville disk with FZZT/ZZ boundaries has a divergence that cancels the volume \cite{Zamolodchikov:1982vx,Alexandrov:2003nn,Mahajan:2021nsd}.\footnote{The divergence of the Liouville partition function can be seen in the classical limit $b\to 0$ at the one-loop level due to the appearance of zero modes around classical saddlepoints \cite{Zamolodchikov:1982vx,Mahajan:2021nsd}.} 
An alternative way to compute the empty disk diagram is through the so-called ``dilaton equation", which in particular relates the one-point disk diagram (free of ambiguities) to the empty disk diagram \cite{Collier:2023cyw,Collier:2024kmo}. In other minimal string theories, this dilaton equation can be derived either from the higher equations of motion in Liouville theory or from an analogous relation satisfied by the dual matrix integral. However, since the precise form of the dilaton equation is not currently known for the ADE minimal string theories, we will instead follow the standard prescription of integrating a finite one-point diagram to define the empty disk amplitude \cite{Seiberg:2003nm,mtms}
\be
\partial_\mu Z_{\t{disk}}(s) = Z_{\t{disk}}(s;P_{1,1}) ~,
\ee
where the one-point function is given by \eqref{eqn:cc_1pt}, and the above equation can be integrated with respect to $\mu$ to define the empty disk. On the right hand-side we should also note that the parameter $s$ depends on both the bulk and boundary cosmological constants through \eqref{eqn:muB_s_relation}. The equation can be integrated to give 
\begin{gather} \label{eqn:empty_disk_result}
Z_{\t{disk}}(s) =C_{D^2}^{\t{empty}} \mu^{\frac{1+b^2}{2b^2}}\lr{Q \cosh \big(2\pi s \hat{Q}\big)-\hat{Q} \cosh \big(2\pi s Q\big)}\,,
\end{gather}
where $Q=b^{-1}+b$ and $\hat{Q}=b^{-1}-b$, and we have grouped the various normalization constants into
\be
C_{D^2}^{\t{empty}} = \mathcal{N}^{(b)}\frac{C_{D^2}^{(b)} N_{1,1} \lr{S^{1,1}_{1,1}}^{1/2}}{4\pi (1+b^2)}\frac{b^3}{\sin(-\frac{\pi}{b^2} ) \sin(\pi b^{2})} ~.
\ee
This answer is identical for both A and D series minimal strings with identity boundary conditions for the matter. In principle, there is an integration constant in this formula. We note that the final result \eqref{eqn:empty_disk_result} indeed looks very similar to what one would get from a dilaton equation \cite{Collier:2023cyw,Collier:2024kwt} evaluation of the empty disk.

\paragraph{Marked disks.}
 We now compute the amplitude with the insertion of the boundary cosmological constant dressed with minimal model boundary identity operators. Naively, fixing only one boundary operator using the CKG leaves a residual infinite volume factor to divide by, but it is expected that the resulting amplitudes are finite. They are defined through
\begin{align}
Z_{\t{disk}}^{\t{M}}(s) &\equiv \partial_{\mu_B} Z_{\t{disk}}(s) \q = \q\raisebox{-.7cm}{\includegraphics[width=1.5cm]{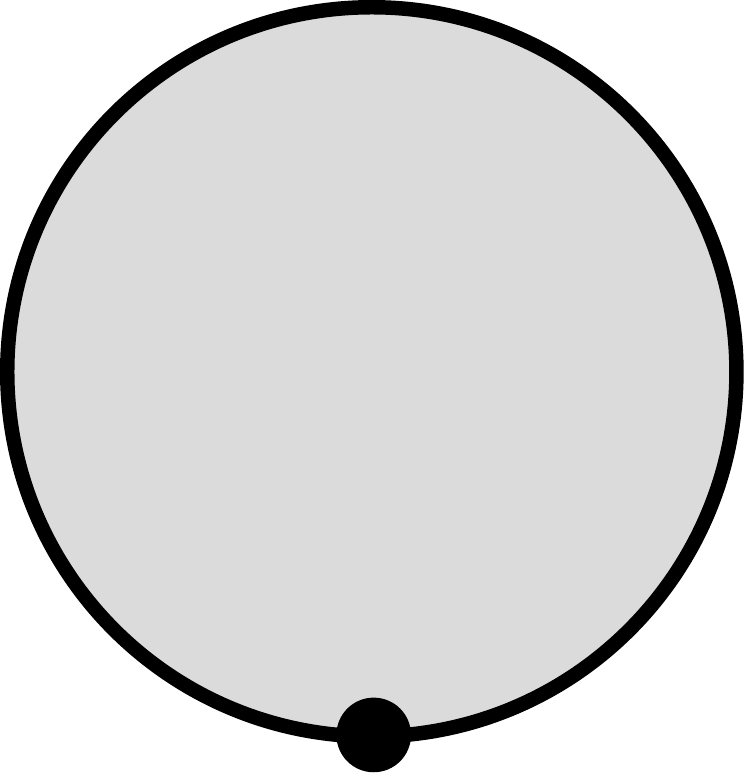}}  \nn\\
&=C_{D^2}^{\t{M}} \cosh \lr{\frac{2\pi}{b} s}
\end{align}
These amplitudes are also known as marked disks \cite{mtms} since a point has been marked on the boundary by the operator insertion. We have defined the normalization $C_{D^2}^{\t{M}}=-2\hat{Q} \mu^{\frac{1}{2 b^2}} C_{D^2}^{\t{empty}} $.

\paragraph{Thermal partition function.}
We now compute the fixed length amplitude defined in \eqref{eqn:fixed_length_bc}. This calculation was recently explained in \cite{mtms}.
We define the dimensionful scale 
\be
E_0=\sqrt{\frac{\mu}{\sin(\pi b^2)}}\,.
\ee
The partition function takes the form for both AMS and DMS
\begin{align} \label{eqn:Z_marked_beta}
Z^M_{\t{disk}}(\beta) &= -\frac{1}{2\pi i}\int_{-i \infty}^{i \infty} \d\mu_B e^{\beta \mu_B} Z^{\t{M}}_{\t{disk}}(s) \nn\\
&= -\frac{C_{D^2}^{\t{M}}}{2\pi i} \int_{-E_0}^{-\infty} \d\mu_B e^{\beta \mu_B} \t{Disc}\Big[\cosh \lr{\frac{1}{b^2}\arccosh\lr{\frac{\mu_B}{E_0}}} \Big] \nn \\
&= \frac{C_{D^2}^{\t{M}}}{\pi} \sin\lr{\frac{\pi}{b^2}} \int_{E_0}^{\infty} \d E e^{-\beta E} \sinh \lr{\frac{1}{b^2}\arccosh\lr{\frac{E}{E_0}}}\,.
\end{align}
In the second line we deformed the contour to wrap the branch cut of $Z^M_{\t{disk}}(s)$ that runs along the negative real axis from $(-\infty,-E_0]$, giving us a discontinuity across the cut. In the last line we evaluated the discontinuity and used the more natural variable $E=-\mu_B$. From the above it is manifest that the disk level density of states is 
\be \label{eqn:bulk_dos}
\rho_0(E)= \mc{N} \sinh \lr{\frac{1}{b^2}\arccosh\lr{\frac{E}{E_0}}}\,,
\ee
with the edge of the spectrum at $E_0$ and the prefactor $\mc{N}$ normalization dependent. Thus, following this procedure we obtain that the density of states for the AMS and DMS is identical. This is slightly surprising since they are different string theories, and the DMS is expected to be dual to a multi-matrix integral so it isn't clear which matrix this is supposed to be the density of states of. Regardless, this agrees with the proposal of \cite{mtms} that the matter CFT is not relevant for determining the density of states, which is determined purely by the Liouville sector.

\paragraph{DMS: Extra boundary condition.} In the DMS we have an extra boundary condition $\left| \t{FZZT}^{(b)}(s) \right\rb \otimes  |\frac{p}{2},1 \rb $ we must consider. The only difference is the matter sector, which has a different normalization for the boundary identity one-point function \eqref{eqn:disk_bdy_1pt_MM}. Carrying out the computation of the marked FZZT boundary we would arrive at \eqref{eqn:Z_marked_beta} with the normalization slightly modified, so we get the same density of states \eqref{eqn:bulk_dos} for both boundary conditions. We have not computed one-point functions of bulk tachyons with this boundary condition, the bulk one-point functions in the matter CFT are contained in \cite{Runkel:1999dz}.

\subsection{Cylinder diagrams} \label{eqn:4.4_cylinders}
In this section, we compute cylinder amplitudes with both FZZT and ZZ boundary conditions. For clarity, we leave details of the calculations to appendix \ref{app:cylinder_computations}.

\paragraph{$|\text{FZZT}(s)\rangle\times|1,1\rangle$ string cylinder diagram.} We first consider the case where both boundaries are FZZT labeled by $s,s' \in \mathbb{R}$. In the open channel with periodic direction $2\pi t$ and interval length $\pi$ we have
\be
Z(s_1,s_2)=\int_0^\infty \frac{\d t}{t} Z^{MM}(q) Z^L(q) Z_{\mathfrak{b,c}}(q)= ~\raisebox{-.7cm}{\includegraphics[width=4.5cm]{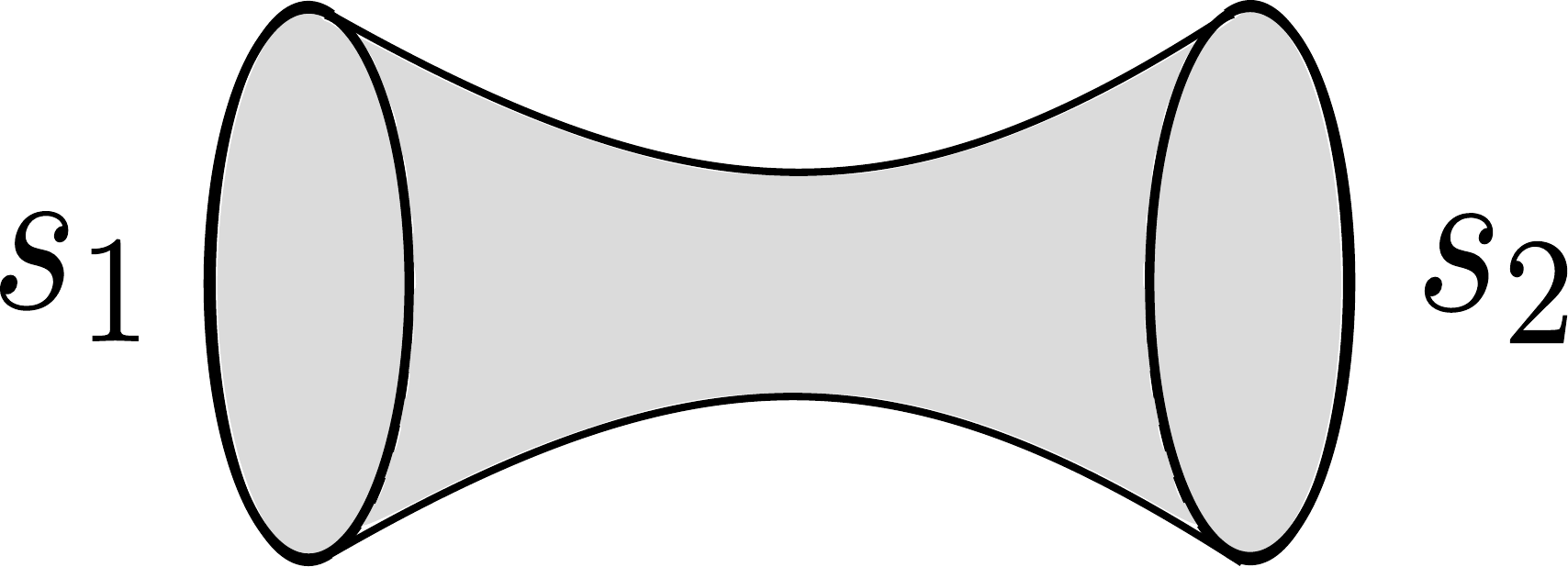}} 
\ee
with $q=e^{-2\pi t}$. In the open channel the ghosts give $Z_{\mathfrak{b,c}}(q)=\eta(q)^2$ and the matter is \eqref{eqn:matter_cylinder} for the A-series and \eqref{eqn:cylinder_Dseries_identity} for the D-series. The Liouville contribution \eqref{eqn:Liouville_FZZT_cylinder} is written in the closed channel, and one can either transform it into the open channel, or transform both the ghost and matter into the closed channel. The latter is easier, and we simply send $q\to q'=e^{-2\pi/t}$ for the matter and ghosts in the above formula. Using modular properties of the Dedekind eta the measure factor $1/t$ cancels giving the closed channel integral
\begin{align} \label{eqn:unmarked_cyl_general}
Z(s_1, s_2) &= \int_0^\infty \d t \int_\Gamma \d P \frac{\cos \lr{4\pi s_1 P} \cos \lr{4\pi s_2 P}}{\sqrt{2}\sinh \lr{2\pi b P} \sinh \lr{2\pi \frac{P}{b}}} \chi_P^{(b)}(q) \eta(q)^2 \nn \\
&\times \begin{cases}
\chi^{(i b)}_{1,1}(q') \qq \qq \qq ~~~~ \t{A-series identity cylinder}\,, \\
\chi^{(i b)}_{1,1}(q')+\chi^{(i b)}_{p-1,1}(q') \qq \t{D-series identity cylinder}\,.
\end{cases}     
\end{align}
We see the only difference between the A and D series is the D-series has an extra piece in the integrand. The DMS amplitudes with matter boundary state $|1,1\rb$ satisfy an interesting identity at the cylinder level that relates them to AMS amplitudes
\be \label{eqn:DMS=AMScyls}
Z^{\t{DMS}}_{(p,p')} &= Z_{(p,p')}^{\t{AMS}} + Z_{(2,p')}^{\t{AMS}} ~,
\ee
which holds for fixed length, marked, and unmarked boundaries. This will be explained later in this section. DMS amplitudes involving the other matter boundary state $|\frac{p}{2},1\rb$ have a more complicated relation to AMS cylinders than the above. 

\paragraph{FZZT cylinder: AMS.}
The integrals for the $(p,p')$ A-series were carried out in \cite{Martinec:2003ka,Kutasov:2004fg,mtms}, we explain it in more detail in the appendix \ref{app:cylinder_computations}. We will first do the modular $t$ integral, followed by marking the cylinder boundaries, and then finally integrating over the Liouville momenta $P$. A useful integral is
\be
\frac{1}{\sqrt{2}}\int_0^\infty \d t  \hspace{.04cm}\eta(q)^2 \chi_P^{(b)}(q) \chi^{(i b)}_{r,s}(q') = \begin{cases}
    \frac{\sinh (2\pi b^{-1} P(p-r)) \sinh (2\pi b P s) }{P \sinh (2\pi p' b P)} \qq \t{if} \qq r p' > s p \,,\\
    \frac{\sinh (2\pi b^{-1} P r) \sinh (2\pi b P (p'-s) )}{P \sinh (2\pi p' b P)} \qq \t{if} \qq r p' < s p\,,
\end{cases}
\ee
Using this the integral over $t$ can be performed where we take $p'>p$
\be
Z(s_1,s_2) =2\int_0^\infty \d P
\frac{\cos(4\pi s_1 P)\cos(4\pi s_2 P)\sinh(2\pi \frac{P}{b} (p-1)) }{P \sinh(2\pi p' b P)\sinh(2\pi \frac{P}{b})}\,.
\ee
The factor of two is from restricting the integral over the positive real axis. Taking derivatives with respect to $\mu_B=E_0 \cosh (2\pi b s)$ we get the marked cylinder
\begin{align} \label{eqn:Z^M_Cylinder_AMS}
&Z^{\t{M}}(s_1,s_2) = ~\raisebox{-.7cm}{\includegraphics[width=4cm]{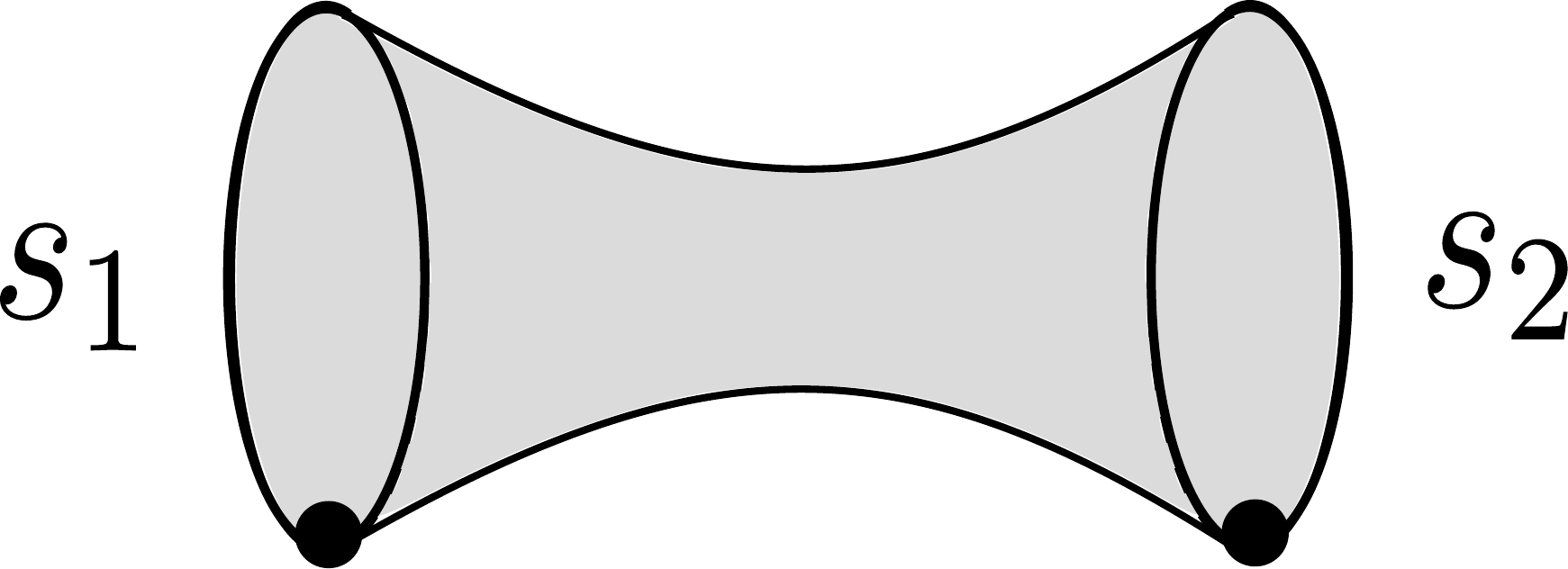}} \\
&=\frac{8}{ (b E_0)^2}\int_0^\infty \d P
\frac{P \sin(4\pi s_1 P)\sin(4\pi s_2 P)}{\sinh(2\pi p' b P)  \sinh(2\pi \frac{P}{b})} \frac{\sinh(2\pi \frac{P}{b} (p-1))}{\sinh (2\pi b s_1) \sinh (2\pi b s_2)}\nn\\
&= \frac{1}{(2\pi b E_0)^2 }\csch\lr{2\pi b s_1} \csch\lr{2\pi b s_2} \partial_{s_1}\partial_{s_2}\log \lr{ \frac{\cosh\lr{2\pi \frac{b}{p} s_1}-\cosh\lr{2\pi \frac{b}{p} s_2}}{\cosh\lr{2\pi b s_1}-\cosh\lr{2\pi b s_2}}}\,, \nn
\end{align}
which we represent in a compact form. We assumed $p'>p$, but the case $p'<p$ is given by the above with the swap $p\leftrightarrow p'$ inside the logarithm but not outside it. This is the universal answer for the marked cylinder in double scaled two-matrix integrals. The unmarked cylinder can be obtained by integrating twice and is given in the appendix.

In the $(2,p')$ model the answer simplifies
\begin{align} \label{eqn:1MM_R_2}
Z^{\t{M}}(s_1,s_2) &= \frac{1}{16 E_0^2} \frac{1}{\lr{\cosh(\pi b s_1)+(\cosh(\pi b s_2)}^2} \frac{1}{\cosh(\pi b s_1) \cosh(\pi b s_2)} \nn\\
&=\frac{1}{4} \frac{1}{\sqrt{E_0+\mu_1} \sqrt{E_0+\mu_2}} \frac{1}{\lr{\sqrt{E_0+\mu_1}+\sqrt{E_0+\mu_2}}^2}\,.
\end{align}
In the second line we went to boundary c.c. variables using $\mu_i = E_0 \cosh(2\pi b s_i)$. As discussed earlier the energy variable in the matrix integral is $E=-\mu_B$ so we get the universal answer for the one matrix integral in the case that the spectrum has edge at $E_0$, see for example \cite{Saad:2019lba}.

\paragraph{FZZT cylinder: DMS.}
The unmarked and marked cylinders in the $(p,p')$ DMS with $p'>p$ are given by, respectively, 
\begin{align}
Z(s_1,s_2) &=2\int_0^\infty \d P\, 
\frac{\cos(4\pi s_1 P)\cos(4\pi s_2 P)}{P\sinh(2\pi p' b P) } \Biggr( \underbrace{\frac{\sinh(2\pi \frac{P}{b} (p-1))}{\sinh(2\pi \frac{P}{b})}}_{\t{identical to A-series}} +1 \Biggr) ~,\nn \\
Z^{\t{M}}(s_1,s_2) &=\frac{8}{ (b E_0)^2}\int_0^\infty \d P\, 
\frac{P\sin(4\pi s_1 P)\sin(4\pi s_2 P)}{\sinh(2\pi p' b P) \sinh(2\pi b s_1 ) \sinh(2\pi b s_2 )} \nn\\ 
&\quad\times\Biggr( \underbrace{\frac{\sinh(2\pi \frac{P}{b} (p-1))}{\sinh(2\pi \frac{P}{b})}}_{\t{identical to A-series}} +1 \Biggr) \,,
\end{align}
Note that the first term is equal to the $(p,p')$ AMS amplitude, while the second is the $(2,p')$ AMS amplitude. This shows the relation \eqref{eqn:DMS=AMScyls} between cylinder amplitudes. Explicitly,
\begin{align}
&Z^M(s_1,s_2)=\frac{1}{(2\pi b  E_0)^2 }\csch\lr{2\pi b s_1} \csch\lr{2\pi b s_2} \nn \\
&\times  \partial_{s_1}\partial_{s_2}\log \left[2 \frac{\cosh\lr{2\pi \frac{b}{p} s_1}-\cosh\lr{2\pi \frac{b}{p} s_2}}{\cosh\lr{2\pi b s_1}-\cosh\lr{2\pi b s_2}} \lr{\cosh\lr{2\pi \tfrac{b}{p} s_1}+\cosh\lr{2\pi \tfrac{b}{p} s_2}}^{-1} \right]\,,
\end{align}
we have used $b/p=(b p')^{-1}$. This answer deviates from the universal two-matrix model result for the cylinder, and confirms that the DMS has a more intricate matrix model dual. We comment on the deviation of this answer below.

\paragraph{Fixed length cylinder: AMS.} To go to fixed length basis we take the inverse Laplace transform of the marked cylinder
\begin{gather} \label{eqn:fixed_length_transform}
Z(\beta_1,\beta_2) = \frac{1}{(2\pi i)^2}\int_{-i\infty}^{i \infty} \d \mu_{B_1} \d \mu_{B_2} \e^{\mu_{B_1}\beta_1+\mu_{B_2}\beta_2} \partial_{\mu_{B_1}} \partial_{\mu_{B_2}}Z(s_1, s_2)\,,
\end{gather}
where we denote it as a derivative of the unmarked cylinder for convenience. There is a convenient trick given by deforming the contour and evaluating the discontinuity across $\mu_B \in (-\infty,-E_0)$ using \cite{mtms} 
\begin{gather}
\operatorname{Disc} \partial_{\mu_B}[\cos 4 \pi P s]=\frac{4 i P}{b E_0}  \sinh\lr{\frac{2 \pi P}{b}} \frac{\cos(4 \pi P s)}{\sinh(2 \pi b s)},\\
\int_0^{\infty} \d s\, \e^{-\beta E_0 \cosh 2 \pi b s} \cos 4 \pi P s=\frac{1}{2 \pi b} K_{\frac{2 i P}{b}  }(\beta E_0)\,.    
\end{gather}
For the AMS this gives
\begin{align} \label{eqn:AMS_length_cylinder}
Z(&\beta_1,\beta_2)=\frac{8}{\pi^2 b^2} \int_0^\infty \d P K_{\frac{2 i P}{b}  }(\beta_1 E_0) K_{\frac{2 i P}{b}  }(\beta_2 E_0) \frac{P \sinh(2\pi \frac{P}{b}) \sinh(2\pi  \frac{P}{b} (p-1))}{\sinh(2\pi p' b P)} \nn\\
&= \frac{8}{\pi^2 b^2} \int_0^\infty \d P K_{\frac{2 i P}{b}  }(\beta_1 E_0) K_{\frac{2 i P}{b}  }(\beta_2 E_0) P \sinh\lr{2\pi \tfrac{P}{b}} \sum_{n=1}^{p-1} \frac{\frac{1}{p}\sin^2(\frac{\pi n}{p})}{\cosh\lr{2\pi \frac{P}{b}}-\cos(\frac{\pi n}{p})}\,,
\end{align}
In the second line we have given an equivalent form that commonly appears \cite{Martinec:2003ka,Moore:1991ir,Kutasov:2004fg}. For the $(2,p')$ model we can perform the integral
\begin{align} \label{eqn:double_trumpet_(2,p)}
Z(\beta_1,\beta_2)&=\frac{4}{\pi^2 b^2}\int_0^\infty \d P P \tanh \lr{2\pi\frac{P}{b}} K_{\frac{2 i P}{b}  }(\beta_1 E_0) K_{\frac{2 i P}{b}  }(\beta_2 E_0)\nn \\
&=\frac{\sqrt{\b_1 \b_2}}{2 \pi (\beta_1+\beta_2)} e^{-E_0(\b_1+\b_2)}\,.
\end{align}
This is the expected answer in the one-matrix model, see for example \cite{Moore:1991ir,Saad:2019lba}. The one-matrix model has a universal ramp in the cylinder given by analytically continuing $\beta_1 \to \beta + i T, \beta_2 \to \beta - i T$ and taking large $T$ which goes as $T/(4\pi \beta)$.

Note that the two-matrix model dual to the $(p,p')$ AMS \eqref{eqn:AMS_length_cylinder} has the same large $T$ behavior for the ramp as the $(2,p)$ \eqref{eqn:double_trumpet_(2,p)}. The only difference between the integrals are the trig functions in \eqref{eqn:AMS_length_cylinder}, which quickly asymptote to the behavior of the $(2,p)$ integrand \eqref{eqn:double_trumpet_(2,p)} with only different behavior near $P=0$. At large $T$ the integral is dominated by contributions away from $P=0$ so this difference does not matter for calculating the late time ramp.

\paragraph{Fixed length cylinder: DMS.} 
For the $(p,p')$ DMS, the fixed length cylinder is given by the same procedure as for the AMS. The first term is identical to the $(p,p')$ AMS cylinder
\begin{align} \label{eqn:DMS_doubletrumpet}
Z_{(p,p')}^{\t{DMS}}(\beta_1,\beta_2) &= Z_{(p,p')}^{\t{AMS}}(\beta_1,\beta_2) + \frac{4}{\pi^2 b^2}\int_0^\infty \d P P \tanh \lr{2\pi\frac{P}{b}} K_{\frac{2 i P}{b}  }(\beta_1 E_0) K_{\frac{2 i P}{b}  }(\beta_2 E_0) \nn\\
&=Z_{(p,p')}^{\t{AMS}}(\beta_1,\beta_2)+ \frac{\sqrt{\b_1 \b_2}}{2 \pi (\beta_1+\beta_2)} e^{-E_0(\b_1+\b_2)}\,.
\end{align}
Interestingly, the extra piece in the D-series is the universal one-matrix model answer. If we analytically continue $\beta_1 \to \beta + i T, \beta_2 \to \beta - i T$ to compute the ramp we would find at late times a behavior $2 \times T/(4\pi \beta)$, which is a ramp with twice the slope of the one-matrix model answer. This would be the expected slope in a one-matrix GSE ensemble, which is not the boundary dual of this model.

\paragraph{$|\text{FZZT}(s)\rangle\times|\frac{p}{2},1\rangle$ string cylinder diagram.} In the DMS we have an additional boundary state to consider $|\text{FZZT}(s)\rb \otimes |\frac{p}{2},1 \rb$, which we denote by subscript $n$. Cylinder amplitudes with this state are significantly more involved and explained in appendix \ref{app:FZZT-FZZT}. The unmarked cylinder with this state on both boundaries is
\begin{align} \label{eqn:n|n_double_trumpet}
Z_{n|n}&(s_1, s_2) \nn\\
&= \int_0^\infty \d t \int_\Gamma \d P \frac{\cos \lr{4\pi s_1 P} \cos \lr{4\pi s_2 P}}{\sqrt{2}\sinh \lr{2\pi b P} \sinh \lr{2\pi \frac{P}{b}}} \chi_P^{(b)}(q) \eta(q)^2 \nn \sum_{r = 1}^{p-1}  \delta_{r \t{\,mod\,} 4 =1} \chi^{(i b)}_{r,1}(q') \nn \\
&= \sum_{r = 1}^{p-1} \delta_{r \t{\,mod\,} 4 =1} Z^{\t{AMS}}(s_1,r,1;s_2,1,1) =\sum_{r = 1}^{p-1} \sum_{u\overset{2}{=}1-r}^{r-1} \delta_{r \t{\,mod\,} 4 =1} Z^{\t{AMS}}(s_1+\frac{i}{2} u b^{-1},s_2) \nn \\
&=2\sum_{r = 1}^{p-1} \sum_{u\overset{2}{=}1-r}^{r-1} \delta_{r \t{\,mod\,} 4 =1}\int_0^\infty \d P
\frac{\cos(4\pi P (s_1+\frac{i}{2} u b^{-1}))\cos(4\pi s_2 P)\sinh(2\pi \frac{P}{b} (p-1)) }{P \sinh(2\pi p' b P)\sinh(2\pi \frac{P}{b})} \nn\\
&=2\sum_{r = 1}^{p-1}  \delta_{r \t{\,mod\,} 4 =1}\int_0^\infty \d P
\frac{\cos(4\pi P s_1)\cos(4\pi s_2 P)\sinh(2\pi \frac{P}{b} (p-1)) }{P \sinh(2\pi p' b P)\sinh(2\pi \frac{P}{b})} \frac{\sinh(2\pi \frac{P}{b} r)}{\sinh(2\pi\frac{P}{b})} ~.
\end{align}
In the first line we have written the full integral representation. In the second line we have rewritten the DMS cylinder as an AMS cylinder with matter boundary states $|r,1\rb$ on one side and $|1,1\rb$ on the other which give a matter amplitude $\chi^{(i b)}_{r,1}(q')$ reproducing the first line. In the same line we used Seiberg-Shih equivalence \eqref{eqn:Seiberg-Shih-equivalence} to rewrite the AMS cylinder with non-trivial matter states in terms of the identity matter state. Finally, we wrote out the AMS amplitude and performed a partial resummation over $u$ in the last line.

In \eqref{eqn:Zn|n_fixedlength} we compute the fixed length amplitude which turns out to be naively divergent
\begin{align} \label{eqn:DMS_doubletrumpet_extrabrane}
&Z^{\t{DMS}}_{n|n}(\beta_1,\beta_2) = \frac{8}{\pi^2 b^2}\sum_{r = 1}^{p-1} \delta_{r \t{\,mod\,} 4 =1} \nn\\ &\int_0^\infty \d P K_{\frac{2 i P}{b}  }(\beta_1 E_0) K_{\frac{2 i P}{b}  }(\beta_2 E_0) \frac{P \sinh(2\pi \frac{P}{b}) \sinh(2\pi  \frac{P}{b} (p-1))}{\sinh(2\pi p' b P)} \frac{\sinh(2\pi b^{-1} P r)}{\sinh(2\pi b^{-1} P)} \nn \\
& =  \infty\,,   
\end{align}
One possibility is the brane boundary conditions identified are not cleanly mapped to reasonable operator insertions on the matrix model side, and should not be interpreted as computing $\lb \lr{\Tr e^{-\beta_1 M}} \lr{\Tr e^{-\beta_1 M}} \rb_c$ of some matrix $M$. In \eqref{eqn:AMS_Divergence} we point out a similar divergence appears in the AMS when cylinder partition functions are evaluated involving general matter states $|a,b\rb \neq |1,1\rb$. The divergence is absent for the $|1,1\rb$ boundary condition which has a well established matrix integral interpretation. 

The other amplitude has the identity matter state on one side and $|\frac{p}{2},1\rb$ on the other giving
\begin{align} \label{eqn:cyl_n|i}
Z_{n|i}^{\t{DMS}}(s_1,s_2)&= \int_0^\infty \d t \int_\Gamma \d P \frac{\cos \lr{4\pi s_1 P} \cos \lr{4\pi s_2 P}}{\sqrt{2}\sinh \lr{2\pi b P} \sinh \lr{2\pi \frac{P}{b}}} \chi_P^{(b)}(q) \eta(q)^2 \nn \times \chi^{(i b)}_{\frac{p}{2},1}(q') \\
&=\frac{1}{2} \sum_{u\overset{2}{=} 1-\frac{p}{2}}^{\frac{p}{2}-1} Z^{\rm AMS} (s_1+\frac{i}{2} u b^{-1},s_2)\,,
\end{align}
where we have given the final answer after rewriting the D-series matter cylinder in terms of A-series matter cylinders, and then using the Seiberg-Shih equivalence as in the previous calculation. The integral representation can be easily read off from \eqref{eqn:cyl_n|i}. The factor of $\frac{1}{2}$ comes from the seiberg-shih equivalence. We also compute the fixed length variant of this amplitude in appendix \ref{app:cylinder_computations}, and it is also divergent. 

The interpretation of the marked version of this amplitude should presumably be of a connected resolvent between two distinct matrices in the dual theory
\be
\left\lb \Tr \frac{1}{z_1 - M_1} \Tr \frac{1}{z_2 - M_2}  \right\rb_{g=0\t{, conn.}}\,
\ee
where $z_i$ should be related to the energies labelled by $s_1$. This should in principle include information on eigenvalue repulsion between different matrices. However, since the fixed length version of the amplitude is divergent it is unclear if the matter state $|\frac{p}{2},1\rb$ maps to this type of observable on the matrix side.

\paragraph{$|\text{ZZ}_{m,n}\rangle\times|r,s \rangle$ instantons.}
Up to now we have considered boundaries which must be specified by external boundary conditions. We now discuss ZZ instanton amplitudes which must be automatically included in the worldsheet path integral. 
These types of conformal boundary conditions have a very different interpretation: rather than introducing boundaries of a specific kind (e.g. asymptotic or fixed length), ZZ-instanton boundaries instead compute non-perturbative corrections of order $\e^{-g_s^{-1}}$ to the string scattering amplitudes $\mathsf{N}^{(b)}_{n}(P_{r_j,s_j})$.\footnote{For recent work on ZZ-instantons and non-perturbative effects see for example \cite{Balthazar:2019rnh,Balthazar:2019ypi,Sen:2019qqg,Sen:2020eck,Eniceicu:2022nay,Eniceicu:2022dru,Eniceicu:2022xvk,Collier:2024mlg,Alexandrov:2025pzs,Kaushik:2025neu,Marino:2022rpz,Schiappa:2023ned,Eynard:2023qdr,Johnson:2019eik,Johnson:2020heh,Johnson:2020exp}.}
In this paper, we will not undertake a full analysis of the non-perturbative sector of the D-series minimal string, as the precise relation between an independent set of such boundary conditions and the singularities of the corresponding dual spectral curve is not presently understood.
Instead, we restrict our attention to the computation of certain cylinder diagrams, which determine normalization factors in the transseries expansions of the non-perturbative sectors.
From these, we find evidence that the structure emerging in the D-series lies beyond that of conventional two-matrix models.

We first consider identity branes for the matter boundaries, both for the A and D series, while the Liouville branes will be indexed by general ZZ boundary conditions $(m,n)$ and $(m',n')$. Adopting the notation of \cite{Eniceicu:2022dru}, we define the ZZ-instanton normalization factors, which are given by exponentiated cylinder diagrams, as follows:\footnote{For identical boundary conditions on both sides the measure factor is instead $\frac{1}{2 t}$ due to the extra $\mathbb{Z}_2$ symmetry of the cylinder.
This factor comes from the enhanced CKV volume that must be divided by.} 
\begin{align}
C_{(m,n),(m',n')} &= \exp \lr{ \int_0^\infty \frac{\d t}{t} Z^M_{1|1}(q) Z^L_{(m,n),(m',n')}(q) Z_{\mathrm{b.c}}(q)}\,, \nn \\ 
B_{(m,n)} &= \exp \lr{ \int_0^\infty \frac{\d t}{2 t} Z^M_{1|1}(q) Z^L_{(m,n),(m,n)}(q) Z_{\mathrm{b.c}}(q)}\,,
\end{align}
where $B$ is the exponentiated cylinder with identical boundary conditions on both ends. The Liouville cylinder is given in \eqref{eqn:ZZ_liouville} while the matter is given in \eqref{eqn:matter_cylinder} for the A-series and \eqref{eqn:cylinder_Dseries_identity} for the D-series.

\paragraph{A-series ZZ-ZZ.}  We review the calculation in the A-series in appendix \ref{app:cylinder_computations}
\cite{Martinec:2003ka,Kutasov:2004fg,Eniceicu:2022dru,Eniceicu:2022nay,Eniceicu:2022xvk,Chakrabhavi:2024szk}. The final answer in a convenient representation is \cite{Eniceicu:2022dru}
\begin{align} \label{eqn:ZZ_AMS}
C^{\t{AMS}}_{(m,n),(m',n')}&=\frac{\cos \left(\frac{m \pi}{p}+\frac{n \pi}{p'}\right)-\cos \left(\frac{m' \pi}{p}+\frac{n' \pi}{p'}\right)}{\cos \left(\frac{m \pi}{p}+\frac{n \pi}{p'}\right)-\cos \left(\frac{m' \pi}{p}-\frac{n' \pi}{p'}\right)} \frac{\cos \left(\frac{m \pi}{p}-\frac{n \pi}{p'}\right)-\cos \left(\frac{m' \pi}{p}-\frac{n' \pi}{p'}\right)}{\cos \left(\frac{m \pi}{p}-\frac{n \pi}{p'}\right)-\cos \left(\frac{m' \pi}{p}+\frac{n' \pi}{p'}\right)}\,, \nn
\\
B^{\t{AMS}}_{(m,n)} &= i \frac{1}{\sqrt{8 \pi T_{m,n}}} \lr{\frac{\cot^2 \frac{n \pi}{p'}-\cot^2 \frac{ m \pi}{p}}{p'^2-p^2} }^{\frac{1}{2}}\,.
\end{align}
This representation nicely connects to the known two-matrix integral dual to the A-series. See \cite{Eniceicu:2022dru} for an explanation. Without detail, the double scaled matrix integral expression for the above is given by
\be \label{eqn:MI_ZZ-ZZ}
C^{\t{MI}}_{(m,n),(m',n')} = \frac{(z^+_{m,n}-z^+_{m',n'})}{(z^+_{m,n}-z^-_{m',n'})} \times \frac{(z^-_{m,n}-z^-_{m',n'})}{(z^-_{m,n}-z^+_{m',n'})}\,,
\ee
where $z^\pm_{m,n}$ are nodal singularities of what is known as the spectral curve.
This nicely matches the form of the bulk computation \eqref{eqn:ZZ_AMS} from which the location of the nodal singularities can be read off. A similar expression exists for $B_{(m,n)}$.

\paragraph{D-series ZZ-ZZ.} We leave the calculation to appendix \ref{app:cylinder_computations}. The final answer can be expressed as a product of the A-series answer
\begin{align} \label{eqn:ZZ-ZZ_instantons}
C^{\t{DMS}}_{(m,n),(m',n')}&=C^{\t{AMS}}_{(m,n),(m',n')}C^{\t{AMS}}_{(p-m,n),(m',n')}\equiv C^{\t{AMS}}_{(m,n),(m',n')}C^{\t{AMS}}_{(m,n),(p-m',n')}\,,\\
B^{\t{DMS}}_{(m,n)} &=\frac{i}{ \sqrt{8 \pi T_{m,n}}} \lr{ \frac{\cot^2(\frac{n \pi}{p'})-\cot^2(\frac{m \pi}{p})}{p'^2-p^2} }^{\frac{1}{2}}  \lr{\frac{2 \cos(\frac{m \pi}{p})^2 \cos(\frac{n \pi}{p'})^2  }{\cos \lr{\frac{2\pi m}{p}} + \cos \lr{\frac{2\pi n}{p'}} }}^{\frac{1}{2}}
\end{align}
We can also consider ZZ-ZZ instantons with the additional brane boundaries discussed around \eqref{eqn:ZZ_bdy_states}, we compute these amplitudes in appendix \ref{app:ZZ_Extra_branes_DMS}.

The form of these ZZ amplitudes makes it clear the theory is not dual to a two-matrix model \eqref{eqn:MI_ZZ-ZZ}. However, the structure of the cylinder amplitudes is surprisingly close to the two-matrix answers, and one may hope that the matrix dual has enough structure to be exactly solvable.

\section{Matrix integral duals}
\label{sec:mm duals}

\subsection{Two-matrix models}
The $(p,p')$ AMS is famously dual to a double-scaled matrix integral with multiple matrices coupled by a chain interaction term \cite{Kazakov:1986hu,Brezin:1989db,Douglas:1989dd,Douglas1991,Crnkovic:1989tn,Seiberg:2003nm,Seiberg:2004at}. We will be brief and focus on the main results, for detailed reviews see \cite{Ginsparg:1993is,DiFrancesco:1993cyw,Klebanov:1991qa,Eynard:2015aea,Anninos:2020ccj,Collier:2024lys}. It is known that the same critical point can be reached by studying the two matrix model. The two-matrix model dual is given by the double scaled limit of an integral over two $N\times N$ Hermitian matrices
\be
Z= \int \d M_1 \d M_2 \, e^{-N \Tr (V(M_1)+V(M_2)- h M_1 M_2)}\,,
\ee
with the Haar measure and a pair-wise coupling. The potentials take the form $V(H_1) = \sum_{i=1}^p a_n M_1^n$ and $V(H_2) = \sum_{i=1}^{p'} b_n M_2^n$. The explicit forms of the matrix potentials can be found for certain models in \cite{Daul:1993bg}. A common observable is products of the resolvent of a chosen matrix
\begin{gather}
R^{(1)}(x)=\Tr \lr{ \frac{1}{x-M_1}} \,, \qq R^{(1)}(x_1,\ldots, x_n) \equiv R^{(1)}(x_1)\ldots R^{(1)}(x_n)\\
\left\langle R^{(1)}\left(x_1, \ldots, x_n\right)\right\rangle_{\mathrm{c}}=\sum_{g=0}^{\infty} R^{(1)}_{g, n}\left(x_1, \ldots, x_n\right) N^{2-2 g-n}\,,
\end{gather}
where the connected component of the resolvents has the standard genus expansion in the large $N$ limit.\footnote{There is an independent resolvent for each matrix, and products of resolvents involving all matrices can be computed. In this work we will not be interested in these correlators.} At large $N$ the saddlepoint eigenvalue distribution for both matrices becomes continuous and described by a density of states. It is supported along some cut of the real line. Along the cut, the resolvent can be used to calculate the leading density of states 
\be
\rho^{(1)}(x)=-\frac{1}{2 \pi i}\lim_{\epsilon \to 0}\left(R^{(1)}_{0,1}(x+i \varepsilon)-R^{(1)}_{0,1}(x-i \varepsilon)\right)\,.
\ee
The density can be computed for both matrices $M_1$ and $M_2$ using their respective resolvents. 

Observables in the matrix integral can be solved recursively in $\frac{1}{N}$ using the loop equations \cite{Eynard:2002kg}, which are equations arising from differentiating everything inside the integral $\int d M_1 d M_2 \partial_{M_1}(\ldots)=0$. The loop equations give a way to obtain higher resolvent data $R_{g,n}$ from lower resolvents. One output of the loop equations is an object known as the spectral curve, which allows one to bypass the loop equations and directly determine all resolvents using data of the curve. 

One intuitive way to understand the spectral curve is the following \cite{Eynard:2015aea,Collier:2024lys}. At large $N$ the resolvent $R^{(1)}(x)$ is defined on the complex plane away from the cuts that support the eigenvalue density. If we go through the cut we go onto a second sheet of the complex plane. However, one can imagine that we take the two sheets of the complex plane, slightly open up the cuts, and smoothly glue them together into a Riemann surface $\Sigma$ on which the resolvent is single-valued. The spectral curve can then be extracted from the leading order term of a particular large $N$ loop equation, giving a defining equation
\be \label{eqn:spectral_curve_basic}
P_0(x(z),y(z))=0,
\ee
where $z$ is a uniformization parameter of the Riemann surface $\Sigma$, and $x(z),y(z)\in \mathbb{C}$. For a point $z\in \Sigma$, we can view the above equation as specifying a value $x(z)$ in the complex plane, with the dual parameter
\be \label{eqn:y(z)}
y(z) = V'_1(x(z))-R^{(1)}_{0,1}(x(z))\,,
\ee
encoding information on the large $N$ resolvent of the first matrix at that point. The spectral curve has a symmetric interpretation where $y(z)$ is a associated to the second matrix and the value
\be \label{eqn:x(z)}
x(z) = V_2'(y(z))-R^{(2)}_{0,1}(y(z))\,,
\ee
tells us information about the resolvent of the second matrix.
\paragraph{Cylinder diagram.}
Another immediate application of the loop equations is that $R_{0,2}$ has a universal answer for two-matrix models \cite{Daul:1993bg}, see for example \cite{Collier:2024lys} for a recent review. Knowing the spectral curve, the answer is
\begin{align} \label{eqn:R_{0,2}}
&R_{0,2}\left(\mathrm{x}(z), \mathrm{x}\left(z^{\prime}\right)\right) \mathrm{dx}(z) \mathrm{dx}\left(z^{\prime}\right)=\frac{\mathrm{d} z \mathrm{~d} z^{\prime}}{\left(z-z^{\prime}\right)^2}-\frac{\mathrm{dx}(z) \mathrm{dx}\left(z^{\prime}\right)}{\left(\mathrm{x}(z)-\mathrm{x}\left(z^{\prime}\right)\right)^2}\,, \nn \\
&R_{0,2}\left(\mathrm{x}(z), \mathrm{x}\left(z^{\prime}\right)\right)=\partial_{x(z)} \partial_{x(z')} \log \frac{z-z'}{x(z)-x(z')}
\end{align}
Connected amplitudes with more boundaries can be easily obtained using recursion relations, but we will not study them in this work.

\subsection{AMS spectral curve}
The spectral curve for the $(p,p')$ AMS is given by a genus zero Riemann surface with $\frac{1}{2}(p-1)(p'-1)$ pinch singularities \cite{Seiberg:2003nm,Seiberg:2004at}.\footnote{The pinch singularities are not interpreted as generating a higher genus surface.} The uniformization parameter for the surface is $z\in \mathbb{C}$. The spectral curve and it's uniformization is
\begin{gather}
T_{p'}(x(z))-T_{p}(y(z))=0\\ \nn
x(z)= T_{p}(z), \qq y(z)= T_{p'}(z), 
\end{gather}
where $T_p$ are Chebyshev polynomials.\footnote{For additional recent work on matching bulk and boundary calculations in the AMS using the spectral curve see \cite{Belavin:2013nba,Artemev:2024rck,Artemev:2025pvk,Artemev:2023bqj}.} We can solve for the variables in terms of each other which will be useful later
\be
y(x)=\cos \lr{\frac{1}{b^2} \arccos(x)}, \qq x(y)=\cos \lr{b^2 \arccos(y)}\,,
\ee
with $b=\sqrt{\frac{p}{p'}}$. The uniformization parameter is commonly written as $z=\cosh \frac{2\pi s}{\sqrt{p p'}}$ where $s$ is the FZZT variable from earlier.

\paragraph{Density of states.} The density of states for matrices $M_1$ and $M_2$ can be extracted from the spectral curve using \eqref{eqn:x(z)} and \eqref{eqn:y(z)}. We have
\be
\underbrace{\cos \lr{\frac{1}{b^2} \arccos(x)}}_{y(x)}=-R^{(1)}_{0,1}(x) + V_1'(x), \qq \underbrace{\cos \lr{b^2 \arccos(y)}}_{x(y)} = -R^{(2)}_{0,1}(y) + V_2'(y)\,.
\ee
Since the potentials $V_1, V_2$ are smooth polynomials they have no branch cuts and we can ignore it while calculating the discontinuity of the resolvent. From the above formula $R_{0,1}(x)$ has a branch cut $x\in(-\infty,-1)$ which is the location of the spectrum. The energy is identified according to $E=-x$ and the density of states for both matrices is\footnote{We use $\t{Disc} \cos \lr{a \arccos(x)} = 2 i \sin(\pi a) \sinh\lr{a ~\arccosh|x|}, \qq x<-1\,.$}
\begin{align}
&\rho^{(1)}_0(E)=\frac{1}{2\pi i} \text{Disc\,}R_{0,1}(x)=\frac{1}{\pi} \sin\lr{\frac{\pi}{b^2}} \sinh \(\frac{1}{b^2} \arccosh E\)\,, \\  
&\rho^{(2)}_0(E)=\frac{1}{2\pi i} \text{Disc\,}R_{0,1}(x)=\frac{1}{\pi} \sin\lr{\pi b^2} \sinh \lr{b^2 \arccosh E}\,.
\end{align}
We now explain how these are related to the bulk calculation from earlier.

\paragraph{FZZT and dual FZZT branes.} In the above we computed two densities of states for the two matrices $M_1$ and $M_2$. We see the first density exactly matches our bulk computation \eqref{eqn:bulk_dos}, what about the other one? As explained in \cite{Seiberg:2003nm}, the other density/resolvent can be computed using dual FZZT branes as follows. Liouville CFT data is invariant under the transformation $b\to b^{-1}$ and $\mu \to \tilde{\mu}$ where $\tilde{\mu}=\mu^{1/b^2}$ is a dual cosmological constant related to $\mu$ with a particular rescaling. Similarly, the dual FZZT parameter is given by $\tilde{\mu}_B=\tilde{\mu}^{1/2} \cosh(\frac{2\pi}{b} s)$. All Liouville observables without boundaries are invariant under $b\to b^{-1}$ if the cosmological constants are sent to their duals. The density of states computed using the dual brane matches what we found above. More concisely, the FZZT brane is the resolvent of the $x$ variable while the dual FZZT is the resolvent of the $y$ variable. 

In principle, to calculate a product of resolvents of both matrices on the string theory side we would need to consider worldsheets with both FZZT and dual FZZT boundaries. Of course with standard worldsheet technology it is not possible to simultaneously consider FZZT and dual FZZT boundaries, so such observables remain unreachable from a bulk calculation. Also note that when we restrict to the $(2,p)$ model where one of the matrices, say $M_2$, can be integrated out, it is still reasonable from the bulk to calculate resolvents with respect to $M_2$. 

\paragraph{Cylinder diagram.} In the two matrix integral the connected correlator for the resolvent is universal result given in \eqref{eqn:R_{0,2}}. This gives us
\begin{align}
Z^{\t{M}}(s_1,s_2) &= \partial_{x_1} \partial_{x_2}\log \lr{ \frac{\cosh\lr{2\pi \frac{b}{p} s_1}-\cosh\lr{2\pi \frac{b}{p} s_2}}{\cosh\lr{2\pi b s_1}-\cosh\lr{2\pi b s_2}}}\,,\\
&= \partial_{x_1} \partial_{x_2} \log \frac{z(x_1)-z(x_2)}{x_1-x_2}
\end{align}
where $z=\cosh(2\pi \frac{b}{p}s)$ was introduced above and $x(z)=\cosh(2\pi b s)$. This matches the bulk result \eqref{eqn:Z^M_Cylinder_AMS}. 



\subsection{Matrix integral: DMS}
\paragraph{Height models.}
We first explain how the matrix integral for the DMS/EMS was conjectured. This requires a detour through the lattice regularization of $(p,p+1)$ unitary ADE minimal model CFTs, known as height models \cite{Pasquier:1987xj,Warnaar:1993ka,Saleur:1988zx,DiFrancesco:1991st,Kostov:1991cg}. 

Consider a fixed 2d lattice of given topology. Height models are defined by putting heights $h_i$ on the vertices of the lattice, with the heights taking values in $h_i \in \{ \t{nodes of } A/D/E \t{ Dynkin diagram} \}$. The rules are that adjacent lattice sites must have either identical or adjacent heights on the chosen Dynkin diagram. The heights are dynamical and path integrated over, with a particular weight assigned to each height function on the lattice. In the continuum as the lattice spacing goes to zero these models flow to the $(p,p+1)$ ADE minimal models. For example, if the heights take values in $D_4$ the model flows to the $(5,6)$ D-series three-state Potts model, while if the heights take values in $A_5$ we get the $(3,4)$ A-series Ising model. 

\paragraph{Matrix integrals.} To get a theory of quantum gravity we must make the lattice dynamical by integrating over all discretizations along with heights. Remarkably, height models on dynamical lattices can be re-written as integrals over random matrices \cite{DiFrancesco:1991st,Kostov:1995xw}, where each height in the Dynkin diagram is mapped to a random Hermitian matrix
\be \label{eqn:Matrix_Model_ADE}
\mathcal{Z} = \int  \prod_i \d M_i \hspace{.03cm} e^{-N \Tr ( V_i(M_i)-
h_{i j} M_i M_j)}\,,
\ee
and we have as many Hermitian matrices $M_i$ as nodes in the Dynkin diagram, and
\be
h_{i j} = \begin{cases}
    h \qq \t{nodes $i, j$ adjacent on Dynkin diagram}\,,\\
    0 \qq \t{otherwise}\,,
\end{cases}
\ee
with $h$ some constant. For $A$ type Dynkin diagrams this reproduces the matrix chain potential standard for the AMS. For the $D_4$ diagram we get the three-state Potts model \cite{DiFrancesco:1990jd,DiFrancesco:1990mc}
\be
\mathcal{Z} = \int \prod_{i=1}^4 \d M_i \hspace{.03cm} e^{-N \Tr ( V_i(M_i)-h M_1(M_2+M_3+M_4))}\,,
\ee
where the matrix $M_1$ is the central one in the Dynkin diagram. This entire discussion has been for unitary minimal models which have a height model realization, but it was suspected that the form of the matrix model potential \eqref{eqn:Matrix_Model_ADE} holds more generally for $(p,p')$ ADE minimal string theories, although to date we are not aware of much progress to confirm this beyond \cite{DiFrancesco:1990jd,DiFrancesco:1990mc}.

\subsection{DMS: Matrix bootstrap} \label{sec:5.4_matrixbootstrap}

In this section we present some preliminary positivity bootstrap results \cite{Lin:2020mme,Kazakov:2021lel}. We will first begin with the simpler case of the 3-state Potts model given by the DMS$_{5,6}$. This model has been extensively studied on the matrix side, see \cite{Kazakov:1987qg,Daul:1994qy,Eynard:1999gp,Zinn-Justin:1999qww,Kulanthaivelu:2019zia,Kulanthaivelu:2019atg,Atkin:2015ksy}. The continuum string theory can be found by tuning the couplings so that the free energy $F=-\frac{1}{N^2} \log Z(g_i)$ develops a non-analytic term $(g-g_c)^\#$ in its series expansion. While this is difficult to detect directly via bootstrap methods, the bootstrap will bound the space of couplings, thus bounding the location of the critical point. 

We first study the 3-matrix model conjecturally dual to the $(5,6)$ DMS theory \cite{DiFrancesco:1990jd}. While generic DMS models are expected to correspond to four-matrix integrals, in this case one of the matrices can be integrated out, yielding an effective three-matrix model. The potential we take is
\be\label{eq:3mm boot}
V(M_1,M_2,M_3)=\sum_{i=1}^3\big (  \frac{1}{2} M_i^2+\frac{g}{4} M_i^4\big )+ h \big( M_1 M_2+M_2 M_3+M_3 M_1 \big)
\ee
The symmetries of the model include: Hermiticity, which ensures that $\langle \tr (M_{i_1} M_{i_2} \allowbreak \cdots \allowbreak M_{i_k})\rangle \allowbreak =  \langle \tr(M_{i_1} M_{i_2}\cdots M_{i_k})\rangle^* \allowbreak= \langle \tr(M_{i_k} M_{i_{k-1}}\cdots M_{i_1})\rangle \in \mb{R}$; the vanishing of traces over odd numbers of matrices; permutation symmetry $\mc{S}_3$ among the matrices; and the cyclicity of the trace.\footnote{We use notation $\tr = \frac{1}{N} \Tr$ such that $\tr 1 = 1$.} The loop equations are given by
\be
\int \prod_{i=1}^3 d M_i \tr \frac{\partial}{\partial (M_n)}\lr{ \mathcal{O}_i e^{-N \Tr V(M_i)} } = 0\,,
\ee
where $\mathcal{O}_i\in\{ \mathbf{1}, M_1, M_2, M_3, M_1^2, M_1 M_2, M_2 M_1, \ldots \}$ is an arbitrary word in the matrices. We also impose large-$N$ factorization on the resulting relations among matrix expectation values. 
An example is
\be
&\int \prod_{i=1}^3 d M_i \pd_{M_1} (M_1^4M_2 e^{-N \Tr V}) =0 \nn\\
\Rightarrow \, &\lb M_1^5M_2\rb + g\lb M_1^7 M_2 \rb + h\(\lb M_1^4M_2^2\rb+\lb M_1^4M_2M_3\rb\)=\lb M_1^2\rb \lb M_1M_2\rb +\lb M_1^3M_2\rb\,,
\ee
where it is understood that each expectation value denotes a traced quantity $\tr$. 
The existence of nontrivial bounds in coupling and moment space arises from imposing the loop equations together with the following positivity condition:
\bal
\langle \tr \mc{O}^\dagger \mc{O} \rangle \geq 0 \,
\eal
where $\mc{O}=\sum_i \alpha_i \mathcal{O}_i$ is the sum over all words without symmetries imposed, and $\alpha_i\in\mathbb{R}$ are arbitrary coefficients. This positivity condition is equivalent to $\mathbf{a}^T \mathbf{M} \,\mathbf{a} \geq 0$ with $\mathbf{M}_{ij}= \langle \mc{O}^\dagger_i \mc{O}_j \rangle$ for all real vectors $\mathbf{a}$. This is simply the statement that the moment matrix is positive semidefinite,
\bal
\mathbf{M} \succeq 0\,.
\eal
Note that the large-$N$ loop equations have quadratic terms in the expectation values, yielding a nonlinear optimization problem. A resolution is the relaxation bootstrap method \cite{Kazakov:2021lel}, which relaxes the quadratic loop equations to linear inequalities, yielding a standard semidefinite programming (SDP) problem.\footnote{The standard primal form of SDP is
\bal
\text{minimize}~~~\sum_{i=1}^m c_i x_i ~~~\text{w.r.t} \, \{ x_1,x_2,\ldots,x_m\} \in \mb{R},~~~\text{subject to}~~~\sum_{i=1}^m \mathbf{F}_i x_i - \mathbf{F}_0 \succeq 0,~~~\mathbf{F}_i \in \mc{S}^n \,.
\eal} 
To achieve this, we introduce new variables $\mathbf{X}_{ij}=\langle \tr \mc{O}_i\rangle \langle \tr \mc{O}_j\rangle$, which replace the quadratic terms in the loop equations and render the system linear. 
Thus, the resulting SDP problem is to extremize a cost function, such as a chosen single-trace expectation value, subject to:\footnote{
Additional constraints may also be imposed \cite{Lin:2025srf}, such as
\bal
\langle  \tr \(\mc{O}_i - \langle \tr \mc{O}_i \rangle\)^\dagger \(\mc{O}_j - \langle \tr \mc{O}_j \rangle\) \rangle \, \succeq \, 0 \Rightarrow \mathbf{M}-\mathbf{X} \succeq 0
\eal
However, we found that such constraints do not significantly improve the bounds at low truncation order, and therefore we did not impose them in this work.}
\begin{gather} \label{eqn:SDP_problem}
\mathbf{M} \succeq 0 \,, \\
\mathbf{X}_{ij}  -\langle \tr \mc{O}_i\rangle \langle \tr \mc{O}_j\rangle \succeq 0 \nn\\
\t{(linearized) loop equations}=0\,. \nn
\end{gather}

The practical implementation proceeds as follows: 1) generate all words up to a fixed length $n$; 2) construct the positive semidefinite (PSD) matrices $\mathbf{M}_{i j}$ and the relaxation matrix appearing in \eqref{eqn:SDP_problem}; 3) impose symmetries on the matrix entries at this stage, for example $\lb \tr M_1^2 \rb \equiv \lb \tr M_2^2 \rb \equiv \lb \tr M_3^2 \rb$, to reduce the number of free variables; 4) generate the loop equations for words up to a length sufficient to relate all free variables; and 5) feed the PSD constraints and loop equations from \eqref{eqn:SDP_problem} into an SDP solver. For our results, we simply used Mathematica's \texttt{SemidefiniteOptimization} function with MOSEK's commercial SDP solver.

\begin{figure}
    \centering
    {{\includegraphics[width=0.42\linewidth]{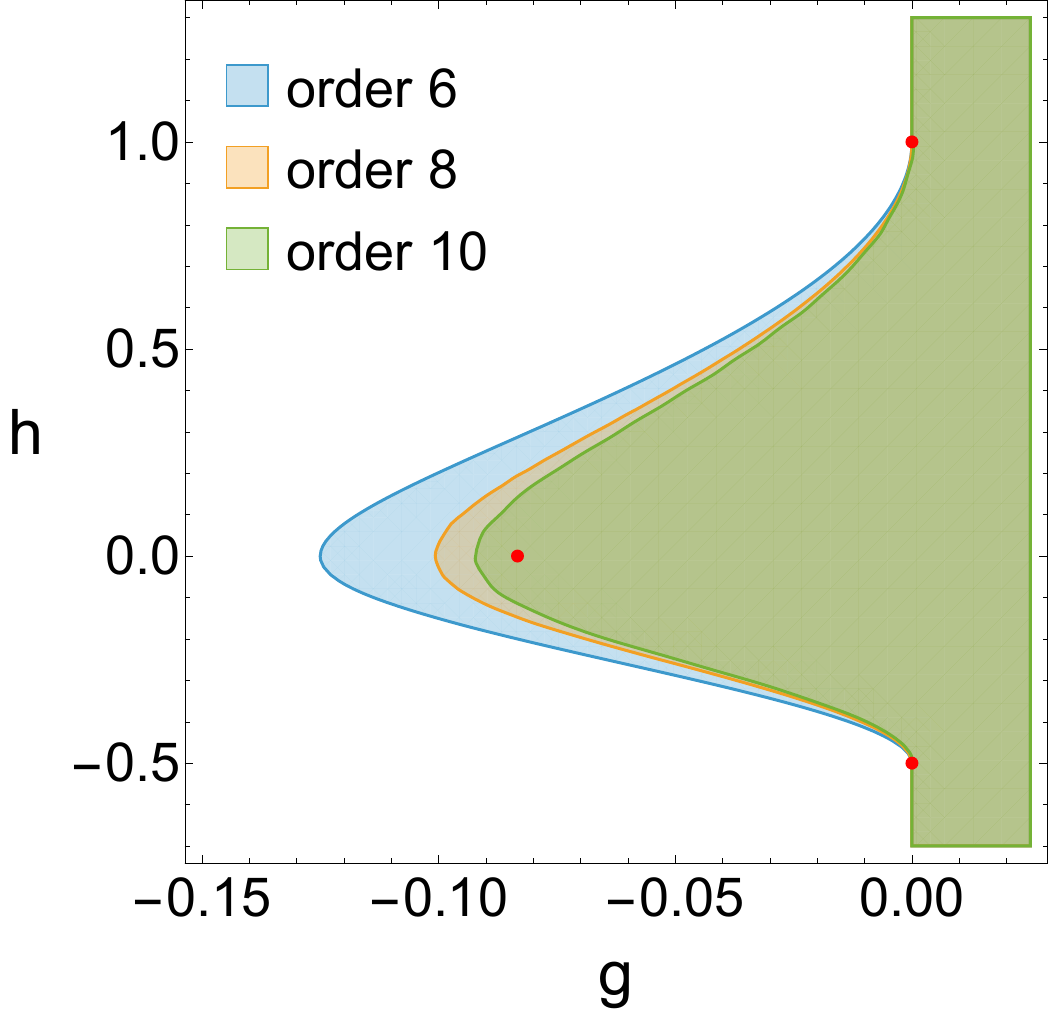}}}
    \quad
    {{\includegraphics[width=0.465\linewidth]{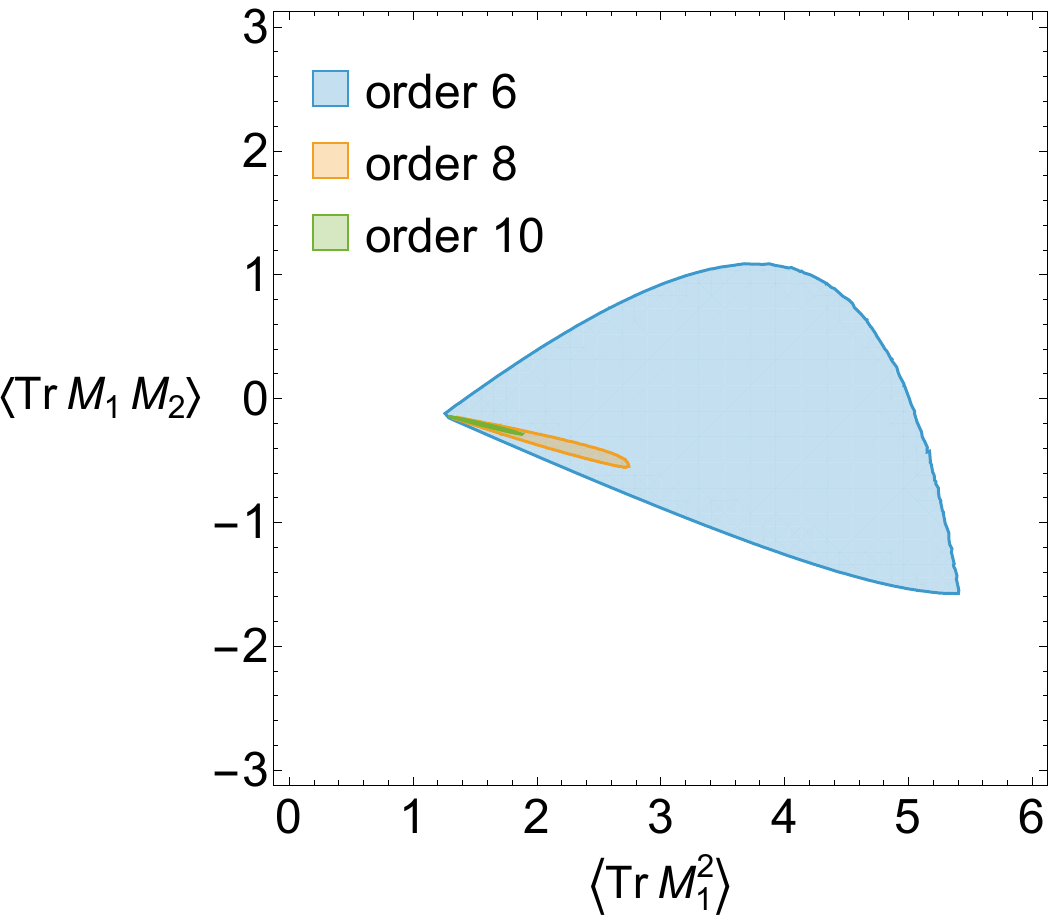} }}
    \caption{\textbf{Left:} The allowed shaded region of the coupling space $(g,h)$ for the 3-matrix model \eqref{eq:3mm boot} at large-$N$ with positivity matrices composed of words up to orders 6, 8, and 10. The three red dots mark solvable critical points, which are expected to lie on the boundary of the true island (infinite-order limit). \textbf{Right:} The large-$N$ allowed region for $\langle \tr M_1^2\rangle$, $\langle \tr M_1M_2\rangle$ at orders 6, 8, and 10, for the case $g=-0.07$, $h=0.1$.}
    \label{fig:PottsBootstrap}
\end{figure}

We now explain how the plots are generated.\footnote{We thank Xiaoyi Liu for valuable discussions on improving the efficiency of our plotting.} For a fixed numerical value of the couplings $(g,h)$, we use SDP to determine whether the moment $\langle \tr M_1^2 \rangle$ admits a minimum consistent with the imposed positivity constraints. If the SDP solver finds a solution, the point is considered valid and lies within the allowed region (the island). See left figure \ref{fig:PottsBootstrap}.
For the second plot, we again fix values $(g,h)$, along with trial values for $\lb \tr M_1^2\rb$ and $\lb \tr M_1 M_2\rb$. We then use the SDP solver to minimize another moment, such as $\lb \tr M_1^8 \rb$. If a solution is found, the chosen trial expectation values are considered consistent with the constraints and are plotted as part of the allowed region. See right figure \ref{fig:PottsBootstrap}.

We note this model has two solvable special cases. For $h=0$ it becomes three de-coupled single matrix integrals and can be solved for $g\in (-1/12,\infty)$, see for example \cite{Lin:2020mme}. Second, for $g=0$ the integral is gaussian and has no negative modes between $h \in (-1/2,1)$ and can be directly integrated.

\begin{figure}
    \centering
    \includegraphics[width=0.42\linewidth]{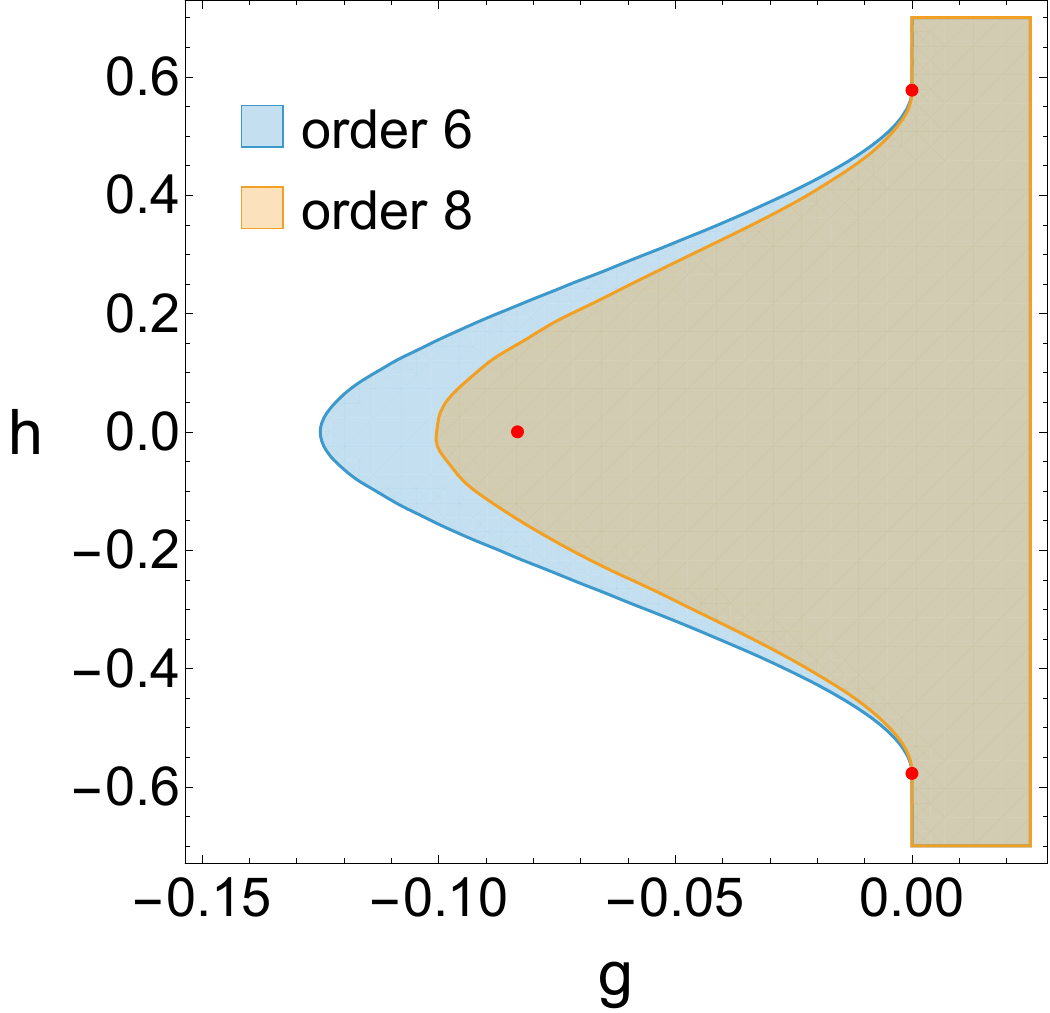}
    {{\includegraphics[width=0.465\linewidth]{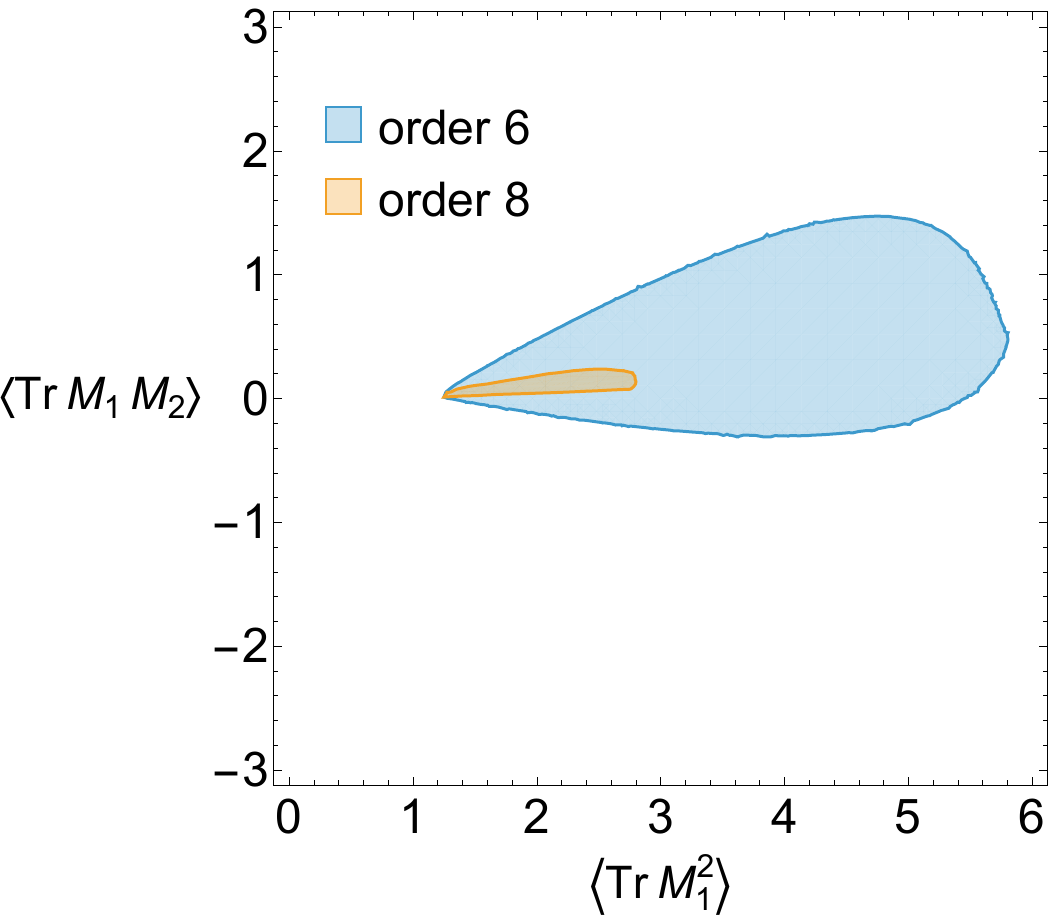} }}
    
    \caption{\textbf{Left:} The allowed shaded region of the coupling space $(g,h)$ for the 4-matrix model \eqref{eq:4mm boot} at large-$N$ with positivity matrices composed of words up to orders 6 and 8. The three red dots mark solvable critical points, which are expected to lie on the boundary of the true island (infinite-order limit). \textbf{Right:} The large-$N$ allowed region for $\langle \tr M_1^2\rangle$, $\langle \tr M_1M_2\rangle$ at orders 6 and 8, for the case $g=-0.07$, $h=0.1$.
    }
    \label{fig:Mercedesbootstrap}
\end{figure}

\paragraph{Four-matrix model.} 
We now bootstrap the more generic four-matrix model given by
\be\label{eq:4mm boot}
V(M_1,M_2,M_3,M_4)=\sum_{i=1}^4\big (  \frac{1}{2} M_i^2+\frac{g}{4} M_i^4\big )+ h M_1 \big( M_2+M_3 +M_4 \big)\,.
\ee
An example loop equation is
\be
&\int \prod_{i=1}^4 d M_i \pd_{M_1} (M_1^4M_2 e^{-N \Tr V}) =0 \nn\\
\Rightarrow \, &\lb M_1^5M_2\rb + g\lb M_1^7 M_2 \rb + h\(\lb M_1^4M_2^2\rb+2\lb M_1^4M_2M_3\rb\)=\lb M_1^2\rb \lb M_1M_2\rb +\lb M_1^3M_2\rb\,,
\ee
where again the traces $\tr$ are implicit. We follow the same procedure as the three-matrix model. The results are in figure \ref{fig:Mercedesbootstrap}. 

While we are unable to identify the string theory critical point from this preliminary analysis, we believe one may exist for an intermediate value of $h$ inside the island. As explained earlier, the critical point can be found by searching for a point in coupling space where the free energy is developing a non-analytic piece as a function of the couplings.\footnote{For example, the one-matrix model with $g=-\frac{1}{12}$ is the $(2,3)$ AMS of the identity CFT coupled to Liouville. For two-matrix potentials dual to AMS models see \cite{Daul:1993bg}. } Similar remarks should apply for traces such as $\lb \tr M_1^2 \rb$. We are unable to search for such points since we did not precisely pin down the island. In the multi-matrix case there are many more free variables entering the positivity matrix than in the single matrix case where the island very quickly converges \cite{Lin:2020mme,Kazakov:2021lel}. We believe imposing positivity at higher order will give much sharper islands and may allow for the identification of critical points. 

\section{Discussion}
We end with some open questions. 

\paragraph{CFT questions.} There are a number of open questions involving minimal model CFTs. The structure constants for the E-series minimal model are not known very explicitly apart from the $(12,p)$ series, see \cite{Nivesvivat:2025odb} for recent work. It would be useful to have them in a more explicit form for the evaluation of numerical amplitudes in for example the E-series minimal string. Furthermore, the boundary states and structure constants for the E-series models have not been worked out, although see \cite{Behrend:1998fd,Behrend:1998mu}. When the boundary states are worked out, the inequivalent branes in the EMS can be analyzed. We suspect there will be four inequivalent branes just as in the DMS case due to the conjectured matrix potential. 

Conformal blocks for multiply degenerate Virasoro representations are not known in closed form nor through recursion relations although a well-checked conjecture exists for blocks on the plane \cite{Ribault:2018jdv}, and see \cite{Gaberdiel:2008ma} for some torus one-point functions. It would be interesting to derive a recursion relation for general degenerate blocks that allows for a numerical implementation.  The main difficulty appears to lie in correctly subtracting the contributions of new null descendants as one approaches the degenerate limit from generic (non-degenerate) blocks. 
Nevertheless, for the purposes of integrating correlation functions over moduli space to compute string amplitudes, this turned out not be necessary as evidenced by the very precise numerical results we obtained in this paper. 

\paragraph{Further landscape of minimal string theories.}
There are several related two-dimensional CFTs that one may consider as the matter sector of the string worldsheet theory \eqref{eq:minimal string}. One example is two copies of the Ising model, which may be dual to a four-matrix model due to height model intuition. Other interesting theories are generalized minimal models, non-analytic Liouville theory, loop model CFTs.
Given the intricate connections between these theories and Liouville CFT \cite{Ribault:2024rvk}, it is reasonable to expect that their corresponding minimal string theories are also ultimately related. In particular, it would be very interesting to explore whether these theories can be recovered from the complex Liouville string. 

\paragraph{Generalizations of topological recursion and bootstrap.}
The conjectured matrix model duals to the D- and E-series minimal strings lie beyond the class of matrix models that admit standard topological recursion.
A natural and important question is whether the large $N$ loop equations of these models can be leveraged to carry out an analytic bootstrap following the approach developed by Eynard \cite{Eynard:2007kz}, potentially leading to generalized forms of topological recursion.
Complementarily, it would also be interesting to explore whether the numerical bootstrap based on positivity constraints can be efficiently extended to double-scaled matrix integrals.
Relatedly, it would be very interesting to investigate whether these models admit an interpretation in terms of 3d gravity, analogous to what occurs in the VMS and CLS.

\paragraph{Target space interpretation.} In \cite{Seiberg:2003nm} a target space interpretation of the AMS was given. The target space is a Riemann surface described by the spectral curve, and the nodal singularities of the Riemann surface indicate locations where ZZ instantons are localized. It would be interesting to develop a similar understanding of the DMS target space; suggestively, it may be a Riemann surface that does not admit a standard representation in terms of a spectral curve. 

\paragraph{Matrix integral interpretation.} It would be useful to directly identify asymptotic boundary insertions in the DMS to matrix integral insertions involving the four matrices. Two of the four branes appear to have an obvious interpretation of inserting $\Tr \frac{1}{x-M_i}$, while the other two branes have mysterious properties such as the double trumpet diverging \eqref{eqn:DMS_doubletrumpet_extrabrane}, which is a property of non-identity branes in the AMS as well. It is possible the diverging branes do not insert a simple trace of a matrix. 

\paragraph{Gluing discrete trumpets.}
In this paper, we have only begun to explore the role of conformal boundary conditions and the open string sector in minimal string theory. Our motivation was to gather evidence that the D-series minimal string exhibits a structure suggestive of a multi-matrix model dual involving more than two matrices.
Recently, a remarkably simple gluing formula applicable to a broad class of conformal boundary conditions was introduced in \cite{Collier:2024mlg}, providing a string-theoretic derivation of the operation of gluing trumpets in JT gravity. While this formula holds in various string theories, such as the VMS and CLS, it remains an open question whether an analogous formula exists for discrete minimal string theories like the ADE models studied here.
Deriving such a formula, for example in the case of resolvent trumpets, would offer a more direct dictionary between string amplitudes and resolvents in the dual matrix model.

\section*{Acknowledgements}

We would like to thank David Berenstein, Scott Collier, David Simmons-Duffin, Lorenz Eberhardt, Alexander Frenkel, Clifford V. Johnson, Douglas Stanford, Joaquin Turiaci for useful discussions.
We acknowledge the use of ChatGPT Astra to find the conjectural formulas \eqref{eq:conjecture torus 1pt AMS}, \eqref{eq:conjecture torus 1pt DMS}, and \eqref{eq:conjecture torus 1pt EMS} from our numerical results. 
VAR is supported by the University of California President’s Postdoctoral Fellowship. This work was performed in part at the Aspen Center for Physics, which is supported by a grant from the Simons Foundation (1161654, Troyer). 
MU was supported in part by grant NSF PHY-2309135 to the Kavli Institute for Theoretical Physics (KITP), and by grants from the Simons Foundation (Grant Number 994312, DG), (216179, LB). 
ZW is supported by Heising-Simons Foundation grants \#2024-5307, and by funds from the University of California. ZW gives special thanks to Steven Giddings for support.

\appendix
\section{$(12,p')$ E-series minimal model CFT data} \label{app:E-series}
In this appendix, we collect some details on the E-series minimal model CFT data following closely the notation of \cite{Nivesvivat:2025odb}.
Although the fusion rules are known for the $p=12,18,$ and $30$ models, explicit expressions for the structure constants are available only in the $p=12$ case. Thus, our analysis in this paper will be restricted to the $p=12$ models.

\paragraph{Fusion rules.}

Let us begin from the \emph{chiral} fusion rules: 
\begin{align}\label{eq:chiral fusion}
\mathcal{R}_{r_1,s_1} \times \mathcal{R}_{r_2,s_2} &= \sum_{r\overset{2}{=}|r_1-r_2|+1}^{\min(r_1+r_2-1,2p-r_1-r_2-1)} \sum_{s\overset{2}{=}|s_1-s_2|+1}^{\min(s_1+s_2-1,2p'-s_1-s_2-1)} \mathcal{R}_{r,s} ~,
\end{align}
and similarly for the anti-chiral fusion rule. 
The procedure to obtain the full (non-chiral) fusion rules of two E-series vertex operators in \eqref{eqn:E_series_fields} is as follows.
First, take the chiral (and antichiral) fusion of the given two vertex operators, using \eqref{eq:chiral fusion}. 
Second, take the outer product of the resulting chiral and anti-chiral sums, in order to obtain a list of full vertex operators (with both chiral and anti-chiral parts).
Third, discard any resulting vertex operators that are \emph{not} part of the spectrum listed in \eqref{eqn:E_series_fields}. 

This procedure produces the list of vertex operators that may appear in the fusion. However, it does not determine the multiplicity (either zero or one) of each term on the RHS. 
Determining these multiplicities constitutes a bootstrap problem. This analysis has been carried out in \cite{Nivesvivat:2025odb}, and we present the result below, following their notation. The fusion rules take the explicit form:
\begin{subequations}\label{eq:E_series_fusion list}
\begin{align}
D_1 &\times X =X\\
D_4 &\times D_4=D_1+{}^e D_7+N_3+\widetilde{N}_3\\
D_4 &\times D_7={}^e D_4 + {}^e N_2\\
D_7 &\times D_7=D_1+D_7+N_3+\widetilde{N}_3\\
D_4 &\times N_2={}^e D_7 +N_3+\widetilde{N}_3\\
D_4 &\times N_3=D_4+N_2\\
D_7 &\times N_2={}^e D_4 + {}^e N_2\\
D_7 &\times N_3=D_7 + \widetilde{N}_3\\
N_2 &\times N_2=D_1 + {}^e D_7 + N_3 + \widetilde{N}_3\\
N_2 &\times N_3=D_4 + N_2\\
N_3 &\times N_3=D_1+N_3\\
N_3 &\times \widetilde{N}_3=D_7
\end{align}
\end{subequations}
Here, the notation is as in \eqref{eqn:E_series_fields}, where the subscript is related to the first index of the pair $(r,s)$, but not the second. Further, the superscript ${}^e$ is used when the second index runs over both even and odd values, whereas it is omitted when the parity is fixed (either even or odd). Further fusion rules related by ``conjugation", such as $D_7 \times \widetilde{N}_3 = D_7 + N_3$ are not shown in \eqref{eq:E_series_fusion list}. 

The fourth step is therefore, to discard any vertex operators are \emph{not} allowed in the precise fusion rules \eqref{eq:E_series_fusion list}. 
We have checked crossing symmetry of several four-point functions and torus one-point functions using this procedure together the fusion rules \eqref{eq:E_series_fusion list}. 

For example, in the $E$-series $(12,5)$ model, the fusion of 
$\widehat{V}^{D_4}_{4,1}$ with itself is  
\bal
\widehat{V}^{D_4}_{4,1} \times \widehat{V}^{D_4}_{4,1} 
  &= \widehat{V}^{D_1}_{1,1} 
   + \widehat{V}^{D_7}_{7,1} 
   + \widehat{V}^{D_7}_{7,4} 
   + \widehat{V}^{N_3}_{1,1} 
   + \widehat{V}^{\widetilde{N}_3}_{7,1},
\eal
where $\widehat{V}^{D_7}_{7,4}$ belongs to the ${}^e D_7$ type, since it does not 
preserve parity. In contrast,
\bal
\widehat{V}^{D_7}_{7,1} \times \widehat{V}^{D_7}_{7,1} 
  &= \widehat{V}^{D_1}_{1,1} 
   + \widehat{V}^{D_7}_{7,1} 
   + \widehat{V}^{N_3}_{1,1} 
   + \widehat{V}^{\widetilde{N}_3}_{7,1},
\eal
because ${}^e D_7$ with the superscript ${}^e$ does not arise in the fusion rules 
\eqref{eq:E_series_fusion list}.

\paragraph{Structure constants.} The structure constants for the $(12,p')$ E-series minimal model CFT are given by (not including the indicator function $f$ enforcing the fusion rule \eqref{eq:E_series_fusion list}) \cite{Nivesvivat:2025odb}
\begin{align} \label{eqn:E_series_C_appendix}
C^{\t{E}}_{r_1,s_1;r_2,s_2;r_3,s_3}=c_{r_1,s_1;r_2,s_2;r_3,s_3} C^{\rm ref}_{r_1,s_1;r_2,s_2;r_3,s_3} ~,
\end{align}
where the $c$ are numerical constants specified below, and $C^{\rm ref}$ is the following ``reference" structure constant of DOZZ type:
\begin{align}\label{eq:Eseries Cref normalized}
C^{\rm ref}_{r_1,s_1;r_2,s_2;r_3,s_3} &= \frac{\sqrt{C^{\text{ref, un-norm}}_{1,1;1,1;1,1}}C^{\text{ref, un-norm}}_{r_1,s_1;r_2,s_2;r_3,s_3}}{\prod_{i=1}^3 \sqrt{C^{\text{ref, un-norm}}_{1,1;r_i,s_i;r_i,s_i}}} ~, \\
c_{r_1,s_1,r_2,s_2,r_3,s_3} &= \frac{\sqrt{c^{\text{un-norm}}_{1,1;1,1;1,1}}c^{\text{un-norm}}_{r_1,s_1;r_2,s_2;r_3,s_3}}{\prod_{i=1}^3 \sqrt{c^{\text{un-norm}}_{1,1;r_i,s_i;r_i,s_i}}} ~,
\end{align}
where, in turn, the un-normalized $C^{\text{ref, un-norm}}$ is given by,
\begin{align}
C^{\text{ref, un-norm}}_{r_1,s_1,r_2,s_2,r_3,s_3}=\frac{1}{\prod_{\epi_1,\epi_2,\epi_3=\pm} \Gamma_\b \big (\frac{\b+\b^{-1}}{2}+ \frac{\b^{-1}}{2} |\sum_i \epi_i u(r_i,s_i)|+\frac{\b}{2} \sum_i \epi_i v(r_i,s_i) \big )} 
\end{align}
with 
\begin{align}
u(r,s)=\begin{cases}
0 &\text{ for }D_{1,4,7}\\
2 &\text{ for }N_2\\
3 &\text{ for }N_3, \tilde{N}_3
\end{cases} ~,
\quad\quad
v(r,s)=\begin{cases}
r\beta^{-2}-s &\text{ for }D_{1,4,7}\\
\frac{p'}{2}-s &\text{ for }N_2\\
\frac{p'}{3}-s &\text{ for }N_3\\
-\frac{p'}{3}+s &\text{ for }\tilde{N}_3 
\end{cases} ~.
\end{align}
Lastly, the constants $c^{\text{un-norm}}_{r_1,s_1;r_2,s_2;r_3,s_3}$ are
\begin{align}
&c^{\text{un-norm}}_{111}=c^{\rm un-norm}_{144}=c^{\rm un-norm}_{177}=1\\
&c^{\text{un-norm}}_{122}=-1\\
&c^{\text{un-norm}}_{133}=(-1)^{3s_3} \nu \sqrt{6}\\
&c^{\text{un-norm}}_{247}=
\begin{cases}
-\sqrt{\frac{3}{2}} n^{-1} &\text{ when } \sum s_i  \in 2\mb{Z}+1\\
\sqrt{\frac{3}{2}} n^{-1} &\text{ when } \sum s_i  \in 2\mb{Z}
\end{cases}\\
&c^{\text{un-norm}}_{447}=
\begin{cases}
\nu \sqrt{\frac{1}{2}} &\text{ when } \sum s_i  \in 2\mb{Z}+1\\
\mu \sqrt{\frac{3}{2}} &\text{ when } \sum s_i  \in 2\mb{Z}
\end{cases}\\
&c^{\text{un-norm}}_{227}=
\begin{cases}
-\mu\nu \frac{1}{2}n^{-3} &\text{ when } \sum s_i  \in 2\mb{Z}+1\\
-\sqrt{\frac{3}{4}}n^{-1} &\text{ when } \sum s_i  \in 2\mb{Z}
\end{cases}\\
&c^{\text{un-norm}}_{777}=\nu \sqrt{2}\\
&c^{\text{un-norm}}_{443}=(-1)^{s_1} \sqrt{2} n^{-1}\\
&c^{\text{un-norm}}_{773}=(-1)^{s_1} \mu \sqrt{12} n^{-1}\\
&c^{\text{un-norm}}_{73\tilde{3}}=-(-1)^{s_3}\mu\sqrt{6}n^{-1}\\
&c^{\text{un-norm}}_{333}=(-1)^{\frac{p'}{2}-\frac{1}{2}\sum s_i}\mu\nu\sqrt{6} n^{-2}\\
&c^{\text{un-norm}}_{223}=(-1)^{\frac{p'}{2}-\frac{1}{2}(s_1-s_2-3s_3)}\sqrt{\frac{1}{2}} n^{-3}\\
&c^{\text{un-norm}}_{243}=(-1)^{s_1} \mu\nu \sqrt{3} n^{-1}
\end{align}
in which 
\begin{align}
n=-\frac{\mu+\sqrt{3}\nu}{\sqrt{2}}~, \quad
\quad
\mu&=\begin{cases}
+1 \text { for } p'=1,7, 17, 23 ~\,\text{ mod }24\\
-1 \text { for } p'=5,11,13,19 \text{ mod }24\\
\end{cases} ~, \\
\nu&=\begin{cases}
+1 \text { for } p'=1,5, 19, 23 ~\,\text{ mod }24\\
-1 \text { for } p'=7,11,13,17 \text{ mod }24\\
\end{cases} ~.
\end{align}
Here, we have used the short-hand notation employed in \cite{Nivesvivat:2025odb}, where we drop the second index $s_j$, $c^{\text{un-norm}}_{r_1r_2r_3}\equiv c^{\text{un-norm}}_{r_1,s_1;r_2,s_2;r_3,s_3}$.\footnote{We have specified the numbers $c_{r_1,s_1;r_2,s_2;r_3,s_3}$ in the given particular ordering of vertex operators. If we need the structure constants in a different permutation ordering, the sign might flip depending on the (non-)diagonality of the vertex operators. 
Such signs are explicitly given in \cite{Nivesvivat:2025odb}.
}

For diagonal fields the structure constant simplifies to be proportional to the timelike DOZZ formula
\begin{align} \label{eqn:E_series_C_diagonal_appendix}
\t{For diagonal fields:} \qq C^{\t{E}}_{r_1,s_1;r_2,s_2;r_3,s_3}=c_{r_1,s_1;r_2,s_2;r_3,s_3} \widehat{C}_{\beta}(P_{r_1,s_1},P_{r_2,s_2}P_{r_3,s_3}) ~.
\end{align}

\section{Minimal model CFT conventions}
\label{sec:Minimal model CFT conventions}

In this appendix, we briefly explain the convention used for vertex operators in minimal model CFTs that lead to the structure constants \eqref{eq:C_structureconstants_AMM}, \eqref{eq:CDseries DDD}, \eqref{eq:CDseries DNN}, as well as those of the E-series in appendix \ref{app:E-series}, in comparison to the more standard convention where the two-point function of bulk vertex operators is normalized to unity.

\paragraph{A-series}
We begin by considering the A-series minimal model CFTs. Let us denote by $V^{\rm m}_{r,s}$, with m for minimal, the vertex operators in the spectrum \eqref{eq:A_spectrum} with unit normalized two-point function,
\begin{align}
\lb V'^{\rm m}_{r_1,s_1}(\infty)  V^{\rm m}_{r_2,s_2}(0) \rb = \delta_{r_1,r_2} \delta_{s_1,s_2} ~.
\end{align}
The structure constants for these operators in the A-series minimal model CFT take the form \cite{Ribault:2024rvk}
\begin{align}\label{eq:C_structureconstants_AMM m}
&\lb V'^{\rm m}_{r_1,s_1}(\infty)  V^{\rm m}_{r_2,s_2}(0) V^{\rm m}_{r_3,s_3}(1) \rb \equiv C^{\mathrm{A},\,\rm m}_{r_1,s_1;r_2,s_2;r_3,s_3} \,f_{r_1,s_1;r_2,s_2;r_3,s_3} ~, \quad \t{with } \\
&C^{\mathrm{A},\,\rm m}_{r_1,s_1;r_2,s_2;r_3,s_3} = \left[ \frac{\prod_{j=1}^3\gbeta (\beta^{\pm 1} \pm 2 P_{r_j,s_j})}{\gbeta(\beta^{\pm 1} \pm 2P_{1,1} )^3} \right]^{\frac{1}{2}}\frac{\gbeta (\frac{\beta+\beta^{-1}}{2}\pm P_{1,1}\pm P_{1,1}\pm P_{1,1})}{\gbeta (\frac{\beta+\beta^{-1}}{2}\pm P_{r_1,s_1}\pm P_{r_2,s_2}\pm P_{r_3,s_3})}
 ~.
\end{align}
This has the property that $C^{\mathrm{A},\,\rm m}_{1,1;r_2,s_2;r_2,s_2}=1$. 
As an example, the Ising model is given by $(p,p')=(4,3)$ with three primary fields $\id =V^{\rm m}_{1,1}, \sigma=V^{\rm m}_{2,2}, \epsilon=V^{\rm m}_{1,2}$. Using \eqref{eq:C_structureconstants_AMM m} we find the only non-trivial structure constant of the Ising model $C_{\sigma \sigma \epsilon}=-\frac{1}{2}$.\footnote{This OPE is usually quoted as $C_{\sigma \sigma \epsilon}=\frac{1}{2}$, but all fields can be independently redefined $V^{\rm m}_{r,s}\to - V^{\rm m}_{r,s}$ without modifying the two-point function while flipping the sign of various OPE coefficients. The OPE we have chosen makes an implicit choice of these signs, which gives $C_{\sigma \sigma \epsilon}=-\frac{1}{2}$.}

In order to obtain our three-point function \eqref{eq:C_structureconstants_AMM}, we rescale the vertex operators by a normalization constant $N_{r,s}$,
\begin{gather}\label{eq:Nrs A 1}
\widehat{V}_{r,s} \equiv N_{r,s} \, V^{\rm m}_{r,s} ~,\\
N_{r,s} = 
d(\beta) \times 2^{\frac{5}{4}}
\left[ \frac{1}{\gbeta(\beta^{\pm 1} \pm 2  P_{r,s})} \right]^{\frac{1}{2}} \gbeta(\beta+\beta^{-1} \pm 2  P_{r,s})~,\\
d(\beta) = 2^{-\frac{5}{4}} 
\gbeta(\beta+\beta^{-1}) 
\left[ \frac{\sqrt{2}}{\gbeta(2(\beta+\beta^{-1}))} \right]^{\frac{1}{3}}
\frac{\gbeta(\beta^{\pm 1}\pm 2P_{1,1})^{\frac{1}{2}}}{\gbeta(\frac{\beta+\beta^{-1}}{2}\pm P_{1,1}\pm P_{1,1}\pm P_{1,1})^{\frac{1}{3}}} ~.
\end{gather}
Indeed, one can verify that $(\prod_{j=1}^3 N_{r_j,s_j})C^{\mathrm{A},\,\rm m}_{r_1,s_1;r_2,s_2;r_3,s_3} = \widehat{C}_\beta(P_{r_1,s_1},P_{r_2,s_2},P_{r_3,s_3})$.
The first factor $d(\beta)$ in \eqref{eq:Nrs A 1} is a constant that is independent of the momenta $P_{r,s}$. 
We note that it can be simplified using the identity
\begin{align}\label{eq:dbeta6simplify}
\left( d(\beta)\right)^6 = -\frac{\pi^2}{64\sqrt{2}} \frac{(\beta^{-2}-\beta^2)^2}{\sin(\pi \beta^2)\sin(\pi \beta^{-2})} ~.
\end{align}
The second factor in \eqref{eq:Nrs A 1} depends on the momenta $P_{r,s}$. We have the relations
\begin{align}
\lr{N_{r,s}}^2 &= d(\beta)^2 2^{\frac{5}{2}}\frac{\gbeta(\beta+\beta^{-1} \pm 2  P_{r,s})^2}{\gbeta(\beta^{\pm 1} \pm 2  P_{r,s})} 
= d(\beta)^2\frac{4\sqrt{2}\sin(2\pi\beta P_{r,s})\sin(2\pi\beta^{-1} P_{r,s})}{P_{r,s}^2}
\label{eq:measuresimplify 0} \\ 
&= d(\beta)^2\frac{\rho_\beta(i P_{r,s})}{(i P_{r,s})^2} ~, \nn \\
N^2_{1,1} &= \frac{\pi^2 Q^2}{4 d(\beta)^2} ~. \label{eq:measuresimplify}
\end{align}
Moreover, when performing a conformal block decomposition of a correlation function via the OPE, one must divide by this nontrivial two-point function, resulting in a nontrivial integration measure over intermediate states. For instance, for the torus one-point function,
\begin{align} \label{eq:torus_1pt_MM 0}
\lb \widehat{V}_{r_1,s_1}(0) \rb_{T^2(\tau)} &= d(\beta)^{-2} \sum_{(r,s) \in \mathcal{S}^{\t{A}}} \frac{(iP_{r,s})^2}{\rho_\beta(iP_{r,s})}\, C^{\t{A}}_{r_1,s_1;r,s;r,s} \, f_{r_1,s_1;r,s;r,s} \nn \\
&\quad\quad\quad\quad\quad \times \mF^{(i\beta),\text{deg}}_{1,1}(P_{r_1,s_1};P_{r,s}|\tau) \mF^{(i\beta),\text{deg}}_{1,1}(P_{r_1,s_1};P_{r,s}|\ol\tau) ~,
\end{align}
and similarly for higher genus and higher point functions.
Note that the RHS of \eqref{eq:dbeta6simplify} is real-valued but not sign-definite. For an expression such as \eqref{eq:torus_1pt_MM 0} one may choose the real branch of $d(\beta)$, but as explained at the end of section \ref{sec:stringbkg}, this overall constant factor is physically irrelevant for our purposes.  
Finally, let us note that for the identity operator we have 
\be
\lb \widehat{V}_{1,1}(0) \rb_{T^2(\tau)} = N_{1,1}  \lb V^{\rm m}_{1,1}(0) \rb_{T^2(\tau)} = N_{1,1} Z^{\t{MM}}_{p,p'}(\tau)\,,
\ee
where the partition function of the minimal model is given by degenerate Virasoro characters as in \cite{DiFrancesco:1997nk}. 

We note that the numerical results reported in section \ref{sec:String Perturbation Theory} have also been computed using the more standard conventions of minimal model CFTs (e.g \eqref{eq:C_structureconstants_AMM m}). 
In the main text, however, we have adopted conventions more closely aligned with those of \cite{Collier:2023cyw,Collier:2024kmo} with the aim of making the relationship between these string theories more transparent. 
A detailed exploration of this connection is left for future work.

\paragraph{D-series}
We can follow the same procedure for the D-series minimal model CFT. In this case, the two nontrivial structure constants of type $DDD$ and $DNN$ in the convention in which two-point functions of vertex operators are unit normalized take the form \cite{Ribault:2024rvk}
\begin{align}
\lb V'^{\t{m},\,D}_{r_1,s_1}(\infty)  V^{\t{m},\,D}_{r_2,s_2}(0) V^{\t{m},\,D}_{r_3,s_3}(1) \rb \equiv C^{\t{D},\,\t{m},\,DDD}_{r_1,s_1;r_2,s_2;r_3,s_3} \,f_{r_1,s_1;r_2,s_2;r_3,s_3} ~, \\
\lb V'^{\t{m},\,D}_{r_1,s_1}(\infty)  V^{\t{m},\,N}_{r_2,s_2}(0) V^{\t{m},\,N}_{r_3,s_3}(1) \rb \equiv C^{\t{D},\,\t{m},\,DNN}_{r_1,s_1;r_2,s_2;r_3,s_3} \,f_{r_1,s_1;r_2,s_2;r_3,s_3} ~, 
\end{align}
with
\begin{align}
C^{\t{D},\,\t{m},\,DDD}_{r_1,s_1;r_2,s_2;r_3,s_3} &= (-1)^{\frac{1}{2}(r_1+r_2+r_3+1)}C^{\mathrm{A},\,\rm m}_{r_1,s_1;r_2,s_2;r_3,s_3} ~, \\
C^{\t{D},\,\t{m},\,DNN}_{r_1,s_1;r_2,s_2;r_3,s_3} 
&= (-1)^{\frac{r_3}{2}-\frac{p}{4}} \times \gbeta (\frac{\beta+\beta^{-1}}{2}\pm P_{1,1}\pm P_{1,1}\pm P_{1,1})\nn \\ 
&\mkern-18mu\!\!\times\left [\frac{\gbeta(\beta^{\pm 1} \pm 2 P_{r_1,s_1}) \prod_{j=2}^3 (-1)^{\frac{r_j}{2}-\frac{p}{4}} \gbeta(\beta \pm 2 P_{r_j,s_j}) \gbeta(\beta^{-1} \pm 2 \widetilde{P}_{r_j,s_j})}{\gbeta(\beta^{\pm 1} \pm 2 P_{1,1})^3} \right ]^{\frac{1}{2}} \nn \\ 
&\mkern-18mu\!\!\times \frac{1}{\gbeta(\frac{\beta+\beta^{-1}}{2} +P_{r_1,s_1}\pm P_{r_2,s_2}\pm P_{r_3,s_3})\gbeta(\frac{\beta+\beta^{-1}}{2} - P_{r_1,s_1}\pm \widetilde{P}_{r_2,s_2}\pm \widetilde{P}_{r_3,s_3})} ~,
\end{align}
where, as a reminder, $\widetilde{P}_{r,s}=P_{p-r,s}$.
Next, we rescale the vertex operators by normalization constants,
\begin{align}
\widehat{V}^{D}_{r,s} \equiv N^{D}_{r,s} \, V^{\t{m},\,D}_{r,s} ~,
\quad\quad
\widehat{V}^{N}_{r,s} \equiv N^{N}_{r,s} \, V^{\t{m},\,N}_{r,s} ~.
\end{align}
For operators in the diagonal sector,
\begin{align}
N^{D}_{r,s} &= (-1)^{\frac{r-1}{2}} N_{r,s} ~,\label{eq:Nrs D D}
\end{align}
is equal to that of the A-series \eqref{eq:Nrs A 1}, up to a sign. 
For operators in the non-diagonal sector,
\begin{align}
N^{N}_{r,s} 
&=d(\beta) \times 2^{\frac{5}{4}} \left[ \frac{\gbeta(\beta+\beta^{-1} \pm 2  P_{r,s})\gbeta(\beta+\beta^{-1} \pm 2 \widetilde{P}_{r,s})}{(-1)^{\frac{r}{2}-\frac{p}{4}}\gbeta(\beta \pm 2 P_{r,s}) \gbeta(\beta^{-1} \pm 2 \widetilde{P}_{r,s})} \right]^{\frac{1}{2}} \nn\\
&= d(\beta) \left[ (-1)^{\frac{r}{2}-\frac{p}{4}} \frac{4\sqrt{2}\sin(2\pi bP_{r,s})\sin(2\pi b^{-1}\widetilde{P}_{r,s})}{P_{r,s}\widetilde{P}_{r,s}} \right]^{\frac{1}{2}} ~. \label{eq:Nrs D N}
\end{align}
With these normalization factors, the structure constants take the form shown in \eqref{eq:CDseries DDD} and \eqref{eq:CDseries DNN}.\footnote{Note that for an OPE of type $DDD$, the extra sign factor $(-1)^{\frac{r-1}{2}}$ would contribute an overall sign $(-1)^{r-1}$ to the OPE measure in the sum over intermediate states. However, since $r$ is odd in the diagonal D-series spectrum, this factor is always equal to 1. This explains the absence of any additional sign in \eqref{eq:sphere_4pt_DMM} and \eqref{eqn:DseriesMM torus1pt}.}

\paragraph{E-series} Lastly, a nicer convention for the $(12,p')$ E-series minimal model CFT can be obtained by following normalization factors:
\begin{align}\label{eq:Eseries MM norms}
N^{D_r}_{r,s} &= d(\beta)  \left[ \frac{\rho_\beta(i P_{r_j,s_j})}{(i P_{r_j,s_j})^2} \right]^{\frac{1}{2}} ~, \nn\\
N^{N_2}_{r,s} &= d(\beta) \left[  \frac{\gbeta(\beta+\beta^{-1} \pm 2  P_{4,s})\gbeta(\beta+\beta^{-1} \pm 2  P_{8,s})}{\gbeta(\beta^{\pm 1} - 2  P_{4,s})\gbeta(\beta^{\pm 1} + 2  P_{8,s})} \right]^{\frac{1}{2}} ~, \nn\\
N^{N_3}_{r,s} &= d(\beta) \left[  \frac{\gbeta(\beta+\beta^{-1} \pm 2  P_{1,s})\gbeta(\beta+\beta^{-1} \pm 2  P_{7,s})}{\gbeta(\beta^{\pm 1} - 2  P_{1,s})\gbeta(\beta^{\pm 1} + 2  P_{7,s})} \right]^{\frac{1}{2}} ~, \nn\\
N^{\widetilde{N}_3}_{r,s} &= d(\beta) \left[  \frac{\gbeta(\beta+\beta^{-1} \pm 2  P_{1,s})\gbeta(\beta+\beta^{-1} \pm 2  P_{7,s})}{\gbeta(\beta^{\pm 1} - 2  P_{1,s})\gbeta(\beta^{\pm 1} + 2  P_{7,s})} \right]^{\frac{1}{2}} ~. 
\end{align}
With these normalization factors \eqref{eq:Eseries MM norms}, the structure constants become of timelike DOZZ form and with an OPE measure dictated appropriately by the inverse of the term inside the square-roots in \eqref{eq:Eseries MM norms}. 

\paragraph{Explicit torus one-point amplitude in the EMS.} In explicit detail, the torus one-point amplitude in the EMS takes the form:
\begin{align}\label{eq:torus1ptdiagramexplicit E}
\mathsf{N}_{1,1}^{(b)} (P^D_{r_1,s_1}) 
&=C_{T^2}\cN^{(b)}  \frac{(2\pi)^2 }{2} \int_{F_0} \d^2 \tau \nn\\
&\quad\times
\int_{\mathcal{C}} \d P \rho_b(P) C_b(iP_{r_1,s_1},P,P) |q|^{2P^2} \mathcal{H}^{(b)}_{1,1}(iP_{r_1,s_1};P|q) \mathcal{H}^{(b)}_{1,1}(iP_{r_1,s_1};P|\bar{q}) \nonumber\\
&\quad\times \sum_{X} \sum_{(r,s) \in \mathcal{S}^{\t{E},\,X}} \frac{1}{(N^{X}_{r,s})^2}\, C^{\t{E},\,DXX}_{r_1,s_1;r,s;r,s} \, f_{r_1,s_1;r,s;r,s}
\nn \\
&\quad\times 
q^{P_{r,s}} \overline{q}^{\widetilde{P}_{r,s}} \mathcal{H}^{(ib)}_{1,1}(P_{r_1,s_1};P_{r,s}|q) \mathcal{H}^{(ib)}_{1,1}(P_{r_1,s_1};\widetilde{P}_{r,s}|\ol q) \,,
\end{align}
where the first sum is over all possible labels ${X\in \{D,N_2,N_3,\widetilde{N}_3\}}$ of the distinct sectors with spectra $\mathcal{S}^{\t{E},\,X}$ of the E-series listed in \eqref{eqn:E_series_fields}, $f_{r_1,s_1;r,s;r,s}$ is the indicator function enforcing the fusion rules listed in \eqref{eq:E_series_fusion list}, and the factors $N^{X}_{r,s}$ are given in \eqref{eq:Eseries MM norms}.

\paragraph{Explicit sphere four-point amplitude in DMS and EMS.} Similarly, the (semi-)explicit form of the four-point amplitude is
\begin{align}\label{eq:sphere4pt amp explicit DE}
\mathsf{N}_{0,4}^{(b)}&(P^{Y_1}_{r_1,s_1},P^{Y_2}_{r_2,s_2},P^{Y_3}_{r_3,s_3},P^{Y_4}_{r_4,s_4}) \nn\\
&= \pi^2 C^{(b)}_{S^2} (\mathcal{N}^{(b)})^4
\int_{F_0} \!\d t
\int_{\mathcal{C}} \d P \rho_b(P) C_b(iP_{r_1,s_1},iP_{r_2,s_2},P) C_b(iP_{r_3,s_3},iP_{r_4,s_4},P) 
\nn\\
&\quad
\times|16q|^{2P^2} \mathcal{H}^{(b)}_{0,4}(iP_{r_j,s_j};P|q) \mathcal{H}^{(b)}_{0,4}(iP_{r_j,s_j};P|\bar{q}) \nonumber\\
&\quad\times \sum_{X} \sum_{(r,s) \in \mathcal{S}^{\t{D/E},\,X}} \frac{1}{(N^{X}_{r,s})^2}\, C^{\t{D/E},\,Y_1Y_2X}_{r_1,s_1;r_2,s_2;r,s} \, f_{r_1,s_1;r_2,s_2;r,s} \,C^{\t{D/E},\,Y_3Y_4X}_{r_3,s_3;r_4,s_4;r,s} \, f_{r_3,s_3;r_4,s_4;r,s} 
\nn\\
&\quad
\times 16^2q^{P_{r,s}} \overline{q}^{\widetilde{P}_{r,s}}  \mathcal{H}^{(ib)}_{0,4}(P_{r_j,s_j};P_{r,s}|q) \mathcal{H}^{(ib)}_{0,4}(P_{r_j,s_j};\widetilde{P}_{r,s}|\ol q) 
\nn\\
&\quad + \big( \text{5 perms of } \{1,2,3\} \big) ~.
\end{align}
Here, the sum over $X$ runs over all possible sectors of the D- or E-series minimal model CFT listed in \eqref{eq:Dseries vertexops} and \eqref{eqn:E_series_fields}, respectively.
Note that the remaining five permutation terms in \eqref{eq:sphere4pt amp explicit DE} correspond to different OPE channels, so the sum over intermediate states will generally differ for each term.

\paragraph{Connection to timelike Liouville CFT.}
We comment on the connection of the two-point function for minimal model compared to timelike Liouville. The two-point function for minimal models \eqref{eqn:two_pt_A_series} differs from standard conventions for timelike Liouville, while the structure constants are identical \cite{Collier:2023cyw}. This difference arises because for timelike Liouville with general central charge the continuation $P_3 \to \frac{i Q}{2}$ does not insert the identity operator which has a null descendant at level one $L_{-1}|h=0\rb = 0$. The continuation inserts another non-degenerate operator in the spectrum with $h=0$ \cite{Ribault:2015sxa}. Another way to see this is that the timelike structure constant does not reduce to a delta function in this continuation. The two-point function is thus typically defined by demanding consistency with crossing, which defines it up to a normalization factor. However, minimal models have a discrete spectrum and the operator with $P_{r_3,s_3}=P_{1,1}=\frac{i Q}{2}$ is the genuine identity operator, giving us the two-point function unambiguously used in the main text.

\section{Torus one-point diagram of the $(1,1)$  tachyon}
\label{app:torus_1pt}

We have the following useful integrals \cite{diFrancesco:1987ses,Bershadsky:1990xb} for minimal model torus partition functions integrated over the fundamental domain of the torus, where $\tau_2=\t{Im}(\tau)$ 
\begin{gather} \label{app:identity_Z_MM_int}
\int_{F_0} \frac{d^2 \tau}{\tau_2^{3/2}} |\eta(q)|^2 Z^{\t{AMM}}_{p,p'}(\tau) = \frac{\pi (p-1)(p'-1)}{6 \sqrt{p p'}}\,,\\  
\int_{F_0} \frac{d^2 \tau}{\tau_2^{3/2}} |\eta(q)|^2 Z^{\t{DMM}}_{p,p'}(\tau) = \frac{\pi (p-1)(p'+2)}{12 \sqrt{p p'}}\,,
\end{gather}
where for the $D$-series $p'$ is even. Here $Z_{p,p'}(\tau)$ is the torus partition function of the minimal model given as a sum over degenerate Virasoro characters as given in \cite{DiFrancesco:1997nk}. A similar formula applies for the E-series. The trick to evaluate the integral is to rewrite the minimal model partition function in terms of compact free boson partition functions, perform a Poisson resummation, and unwrap the integral over $F_0$ into a half infinite strip at which point the integral can be performed \cite{Bershadsky:1990xb}.

We use the above formula to prove \eqref{eqn:torus_1pt_analytics}. In main text conventions the one-point function of the identity in the minimal model is \eqref{eq:measuresimplify}
\be
\lb \widehat{V}_{1,1} \rb_{T^2(\tau)} = \frac{\pi Q}{2 d(\beta)} Z_{p,p'}^{\t{MM}}(\tau)\,,
\ee
The one-point function of $\cT_{1,1}$ for a general ADE minimal string is then
\begin{align} 
\lb \cT_{1,1} \rb_{T^2} = C_{T^2} \frac{(2\pi)^2 \cN^{(b)}}{2 d(\beta)} \times \frac{\pi Q}{2} \int_{F_0} d^2 \tau |\eta(\tau)|^4 \Big\lb V_{i P_{1,1}} (0) \Big\rb_{T^2(\tau)}^{\t{L}} Z^{\t{MM}}_{p,p'}(\tau)\,, \\
\langle V_{i P_{1,1}}(0) \rangle^{\t{L}}_{T^2(\tau)} = \int_{0}^\infty \!\!\d P \rho_b(P) C_b(i P_{1,1},P,P) \mathcal{F}^{(b)}_{1,1}(i P_{1,1};P;\tau) \mathcal{F}^{(b)}_{1,1}(i P_{1,1};P;\bar{\tau})\,.
\end{align}
Since $i P_{1,1}=i\frac{Q}{2}$ the external operator is $h_{\t{ext}}=0$, and the structure constant at this special value and the Liouville block have exact forms
\be
C_b(i \frac{Q}{2},P,P)= \frac{2 P^2}{\pi Q \rho_b(P)}\,, \qq \mF_{1,1}^{(b)}(h_{\t{ext}}=0,P|q)= \frac{q^{P^2}}{\eta(q)}\,,
\ee
see for instance \cite{Collier:2023cyw}. Using these, we can integrate over the momentum exactly
\be \label{eqn:app_identity1}
\int_0^\infty d P \rho_b(P) C_b(i \frac{Q}{2},P,P) |\eta(\tau)|^2  |\mF_{1,1}^{(b)}(i P_{1,1},P;q)|^2 = \frac{1}{16  \pi^2 Q } \times \frac{1}{\tau_2^{3/2}}\,.
\ee
Combining \eqref{eqn:app_identity1} with \eqref{app:identity_Z_MM_int} we get
\begin{gather}
\t{AMS:} \qq\lb \cT_{1,1} \rb_{T^2} = C_{T^2} \frac{(2\pi)^2 \cN^{(b)} \pi Q}{4 d(\beta)} \times \frac{1}{16 \pi^2 Q} \times \frac{\pi (p-1)(p'-1)}{6 \sqrt{p p'}}\,,\\
\t{DMS:} \qq \lb \cT_{1,1} \rb_{T^2} = C_{T^2} \frac{(2\pi)^2 \cN^{(b)} \pi Q}{4 d(\beta)} \times \frac{1}{16 \pi^2 Q} \times \frac{\pi (p-1)(p'+2)}{12 \sqrt{p p'}}\,,
\end{gather}
The numerator in the last term is given by $(\#\text{ nodes of } A_{p-1}  ) \times (\# \text{ nodes of } A, D, E)$ of the dynkin diagrams that classify the minimal model. The E-series is classified by $E_6, E_7, E_8$ so the identity one point function of E-series minimal string for $(12,p')$, classified by $E_6$, is given by 
\be
(12,p') \q \t{EMS:} \qq\lb \cT_{1,1} \rb_{T^2} = C_{T^2} \frac{(2\pi)^2 \cN^{(b)} \pi Q}{4 d(\beta)} \times \frac{1}{16 \pi^2 Q} \times \frac{\pi (p'-1)}{\sqrt{6 p'}}\,.
\ee
These give \eqref{eqn:torus_1pt_analytics}.

\section{Seiberg-Shih equivalence for D-series} \label{app:seiberg_shih_D_series}
\paragraph{AMS review.} We explain the Seiberg-Shih equivalence for the A-series and then give a similar argument for the D-series. In \cite{Seiberg:2003nm} it was noticed that the FZZT boundary states appear to satisfy the identity
\be
|\sigma\rb \otimes |r,s\rb =\sum_{u\overset{2}{=}1-r}^{r-1}\sum_{v\overset{2}{=}1-s}^{s-1} \, |\sigma + i (u b^{-1} +v b)\rangle \otimes |1,1\rb  + (\t{BRST exact})
\ee
with $r,s$ taking values in the Kac table, and we have introduced more compact notation to label the FZZT brane for this section $\sigma = 2s$ which makes the manipulations easier. This identity is only true in the full string theory. This identity can be motivated by noticing that tachyon one-point functions on the disk satisfy the above relation, for example for $\mc{T}_{r_1,s_1}$ we can use equation (3.6) of \cite{Seiberg:2003nm}
\begin{align}
&\lb \mc{T}_{r_1,s_1}\rb_{\sigma;r,s}=(-1)^{r s_1+r_1 s} \sin \frac{r r_1 p' \pi}{p}  \sin \frac{s s_1 p \pi}{p'} \cos \Big(i\pi \si (r_1b^{-1}-s_1b) \Big )\\ \nn
&=\sum_{u\overset{2}{=}1-r}^{r-1}\sum_{v\overset{2}{=}1-s}^{s-1} (-1)^{r_1+s_1} \sin \frac{r_1 p' \pi}{p}  \sin \frac{s_1 p \pi}{p'} \cos \Big(\pi \big (\si+i(u b^{-1}+ v b)\big) i (r_1b^{-1}-s_1b) \Big )\,.
\end{align}
A similar relation exists for ZZ branes which we quote \cite{Seiberg:2003nm}
\bal
|m,n\rangle_{\t{L}} \otimes |r,s\rangle =\sum_{u\overset{2}{=}1-r}^{r-1}\sum_{v\overset{2}{=}1-s}^{s-1} |m+u,n+v\rangle_{\t{L}}\otimes |1,1\rangle\,.
\eal

\paragraph{DMS brane state equivalence.} We will now discuss the analogue of Seiberg-Shih for the DMS. For FZZT boundaries, we find evidence that there exist two distinct FZZT states to which all other states can be reduced to
\bal
\Big \{\left| \si \right\rb \otimes |1,1 \rb\,,~~~\left| \si  \right\rb \otimes  |\frac{p}{2},1 \rb  \Big \}\,.
\eal
The evidence for this statement is that a cylinder amplitude with states $Z_{a|b}$ can be equivalently given by the $|a\rb$ state rewritten as a linear combination of the above states.

We compute DMS cylinders as in the main text to check the equivalence. We define the amplitude according to
\bal
A^{\rm DMS}_{\si,r,s;\si',r',s'}
=\int \frac{\d t}{t} \, \eta(it)^2   \langle \si|_{\t{L}} \otimes \langle r,s | ~ e^{-\frac{2\pi}{t} (L_0+\bar{L}_0-\frac{c}{12})} ~ |\si'\rangle_{\t{L}}\otimes | r',s'\rangle\,.
\eal
For an i-type boundary on the left, general i- or n-type boundary on the right the cylinder amplitude can be written as
\bal
A^{\rm DMS}_{\si, r,s;\si',r',s'} = \sum_{u\overset{2}{=}1-r}^{r-1}\sum_{v\overset{2}{=}1-s}^{s-1} A^{\rm DMS}_{\si + i (u b^{-1} + v b ) , 1,1;\si',r',s'}\,.
\eal
The above relation can be argued for as follows. On the LHS the matter cylinder amplitude can be computed using \eqref{eqn:cylinder_Dseries}, and the characters can be rewritten as cylinder amplitudes in a different theory, the corresponding A-series minimal model with an identity boundary state $|1,1\rb$ on one side and a general boundary state on the other,
\bal
Z^{\rm DMM}_{r,s|r',s'}(q)=\sum_{r_1,s_1} f_{r,s;r',s';r_1,s_1} \chi^{(ib)}_{r_1,s_1}(q) = \sum_{r_1,s_1 }f_{r,s;r',s';r_1,s_1} Z^{\rm AMM}_{r_1,s_1|1,1}(q)\,,
\eal
where the values that are included are described by the fusion rules in \eqref{eqn:cylinder_Dseries}. Therefore the string amplitude is 
\bal
A^{\rm DMS}_{\si,r,s;\si',r',s'}=\sum_{r_1,s_1}  f_{r,s;r',s';r_1,s_1} A^{\rm AMS}_{\si,r_1,s_1;\si',1,1} \, .
\eal
A-series Seiberg-Shih can then be applied to the non-trivial matter boundary state to rewrite everything in terms of identity boundary states. This gives some answer for the LHS DMS amplitude in terms of AMS amplitudes between identity matter states.
\bal \label{eqn:FZZT_Cylinder_SS}
A^{\rm DMS}_{\si,r,s;\si',r',s'} = \sum_{r_1,s_1} \sum_{u\overset{2}{=}1-r_1}^{r_1-1} \sum_{v\overset{2}{=}1-s_1}^{s_1-1} f_{r,s,r',s';r_1,s_1} A^{\rm AMS}_{\si+i(u b^{-1} + v b),1,1;\si',1,1} \, .
\eal
The same procedure can be applied to the RHS, to get the RHS as a cylinder amplitude in the AMS with identity matter states. The end result is the amplitudes turn out to be equivalent, which gives evidence for the DMS boundary conditions to be equivalent as stated earlier.

For n-type on the left, and i-type boundary on the other side, we see
\bal
A^{\rm DMS}_{\si,\frac{p}{2},s;\si',r',s'}=A^{\rm DMS}_{\si,\frac{p}{2}+1,s;\si',r',s'}=\frac{1}{2}\sum_{u\overset{2}{=}1-\frac{p}{2}}^{\frac{p}{2}-1}\sum_{v\overset{2}{=}1-s}^{s-1} A^{\rm DMS}_{\si + i (u b^{-1} + v b ),1,1;\si',r',s'} \, .
\eal
For n-type on the left, n-type boundary on the other side,
\bal
A^{\rm DMS}_{\si, \frac{p}{2},s;\si',r',s'} +A^{\rm DMS}_{\si, \frac{p}{2}+1,s;\si',r',s'}= \sum_{u\overset{2}{=}1-\frac{p}{2}}^{\frac{p}{2}-1}\sum_{v\overset{2}{=}1-s}^{s-1} A^{\rm DMS}_{\si + i (u b^{-1} + v b ) , 1,1;\si',r',s'} \, .
\eal
The above two relations imply that the boundary state $|\sigma\rb \otimes(|\frac{p}{2},s\rb+|\frac{p}{2}+1,s\rb)$ can also be written as a superposition of shifted FZZTs with the matter identity. The final set of states we must consider are either $|\sigma\rb \otimes |\frac{p}{2},s\rb$ or $|\sigma\rb \otimes |\frac{p}{2}+1,s\rb$. Let us consider the first set since with linear combinations we can go to the second set. We have the additional identity
\bal
A^{\rm DMS}_{\si, \frac{p}{2},s;\si',r',s'}=\sum_{v\overset{2}{=} 1-s}^{s-1}  A^{\rm DMS}_{\si+ i v b,  \frac{p}{2},1;\si',r',s'}\,.
\eal
This completes the identification of boundary states.

We now consider ZZ boundary states. We argue that a basis for the ZZ boundary states is
\begin{align}
\Big \{ |m,n\rangle_{\t{L}}\otimes |1,1\rangle ~ \Big | ~ m=1,2,\ldots,\frac{p}{2},~n=1,3,\ldots,p'-2 \Big \}& \nn\\
\bigcup \Big \{ |1,n\rangle_{\t{L}} \otimes |\frac{p}{2},1\rangle \Big | ~ n=1,3,\ldots,p'-2 \Big \}&
\end{align}
We motivate this conjecture by looking at the ZZ cylinders. We define the amplitude by
\bal
A^{\rm DMS}_{m,n,r,s;m',n',r',s'}
=\int \frac{\d t}{t} \, \eta(it)^2   \langle m,n|_{\t{L}} \otimes \langle r,s | ~ e^{-\frac{2\pi}{t} (L_0+\bar{L}_0-\frac{c}{12})} ~ |m',n'\rangle_{\t{L}}\otimes | r',s'\rangle\,,
\eal
note this is not the string amplitude when both boundaries are identical, since it differs by a factor of $\frac{1}{2}$. We always define the amplitudes with open string measure $\frac{1}{t}$ in this section. Similar to the FZZT case, we can compute DMS ZZ cylinders by first using \eqref{eqn:cylinder_Dseries} to rewrite the cylinder in the D-series minimal model by a cylinder in the A-series minimal model
\bal
Z^{\rm DMM}_{r,s|r',s'}(q)=\sum_{r_1,s_1} f_{r,s;r',s';r_1,s_1} \chi^{(ib)}_{r_1,s_1}(q) = \sum_{r_1,s_1 }f_{r,s;r',s';r_1,s_1} Z^{\rm AMM}_{r_1,s_1|1,1}(q)\,,
\eal
and then we use the A-series Seiberg-Shih equivalence to get
\bal \label{eqn:ZZ_Cylinder_SS}
A^{\rm DMS}_{m,n,r,s;m',n',r',s'} = \sum_{r_1,s_1} \sum_{u\overset{2}{=}1-r_1}^{r_1-1} \sum_{v\overset{2}{=}1-s_1}^{s_1-1} f_{r,s;r',s';r_1,s_1} A^{\rm AMS}_{m+u,n+v,1,1;m',n',1,1} \, .
\eal
Using this method, everything can be written in terms of $A^{\rm AMS}_{m,n,1,1;m',n',1,1}$ and the following equalities between cylinder amplitudes can be found.

For i-type on the left, and general i- or n-type boundary on the other side we can derive
\bal
A^{\rm DMS}_{m,n,r,s;m',n',r',s'}=\sum_{u\overset{2}{=}1-r}^{r-1}\sum_{v\overset{2}{=}1-s}^{s-1} A^{\rm DMS}_{m+u,n+v,1,1;m',n',r',s'}\,,
\eal
where this can be seen by expanding both the left and right side using \eqref{eqn:ZZ_Cylinder_SS}. We also find
\bal
A^{\rm DMS}_{m,n,1,1;m',n',r',s'}=A^{\rm DMS}_{p-m,n,1,1;m',n',r',s'} \, .
\eal
which further halves the number of independent states, so we may take $|m,n\rangle_{\t{L}}\otimes |1,1\rangle$ with $m=1,2,\ldots,\frac{p}{2}$ as a convenient set of independent i-type boundaries.

Applying the same steps as before for n-type on the left, and i-type boundary on the other side, we see
\bal \label{eqn:D.19}
A^{\rm DMS}_{m,n,\frac{p}{2},s;m',n',r',s'}=A^{\rm DMS}_{m,n,\frac{p}{2}+1,s;m',n',r',s'}=\frac{1}{2}\sum_{u\overset{2}{=}1-\frac{p}{2}}^{\frac{p}{2}-1}\sum_{v\overset{2}{=}1-s}^{s-1} A^{\rm DMS}_{m+u,n+v,1,1;m',n',r',s'}\,.
\eal
The factor of $\frac{1}{2}$ is not obvious to see. When the RHS is expanded out using \eqref{eqn:ZZ_Cylinder_SS} every individual amplitude will be included twice, which cancels out the factor of $\frac{1}{2}$ to match with the expansion of the LHS.

For n-type boundaries on both sides we have
\bal 
A^{\rm DMS}_{m,n,\frac{p}{2},s;m',n',r',s'}+A^{\rm DMS}_{m,n,\frac{p}{2}+1,s;m',n',r',s'}=\sum_{u\overset{2}{=}1-\frac{p}{2}}^{\frac{p}{2}-1}\sum_{v\overset{2}{=}1-s}^{s-1} A^{\rm DMS}_{m+u,n+v,1,1;m',n',r',s'} \, .
\eal
The above two equalities imply that $|m,n\rb_{\t{L}} \otimes \(|\frac{p}{2},s\rb + |\frac{p}{2}+1,s\rb\)$ can be written as a superposition of shifted ZZs with the matter identity  $|m',n'\rb_{\rm L} \otimes |1,1\rb$. Now if we look at the difference between these two n-type states we find
\bal \label{eqn:D.21}
A&^{\rm DMS}_{m,n,\frac{p}{2},s;m',n',r',s'}-A^{\rm DMS}_{m,n,\frac{p}{2}+1,s;m',n',r',s'} \\\nn
& ~ =\begin{cases}
(-1)^{\frac{m-1}{2}} \sum_{v\overset{2}{=} 1-s}^{s-1}\(  A^{\rm DMS}_{1,n+v,\frac{p}{2},1;m',n',r',s'}-A^{\rm DMS}_{1,n+v,\frac{p}{2}+1,1;m',n',r',s'} \)& m \text{ odd}\\
0& m \text{ even}
\end{cases} \, .
\eal
The conclusion of \eqref{eqn:D.19} and \eqref{eqn:D.21} is that state $|m,n\rb_{\rm L} \otimes \(|\frac{p}{2},s\rb-|\frac{p}{2}+1,s\rb\)$ can be written as $|1,n\rb_{\rm L} \otimes \(|\frac{p}{2},1\rb-|\frac{p}{2}+1,1\rb\)$. Furthermore, the below identity \eqref{eqn:D.22} shows the states can be further reduced to $|1,n'\rb_{\rm L} \otimes \(|\frac{p}{2},1\rb-|\frac{p}{2}+1,1\rb\)$ with $n'$ odd. The identity is
\bal \label{eqn:D.22}
A^{\rm DMS}_{1,n,\frac{p}{2},1;m',n',r',s'}-&A^{\rm DMS}_{1,n,\frac{p}{2}+1,1;m',n',r',s'} \nn\\
&= (-1)^{\frac{p}{2}-1}\( A^{\rm DMS}_{1,p'-n,\frac{p}{2},1;m',n',r',s'}-A^{\rm DMS}_{1,p'-n,\frac{p}{2}+1,1;m',n',r',s'} \)\, .
\eal
Finally, combining the fact that $|m,n\rb_{\rm L}\otimes \(|\frac{p}{2},s\rb + |\frac{p}{2}+1,s\rb\)$ can be rewritten as a superposition of shifted ZZs with the matter identity $|m',n'\rb_{\rm L} \otimes |1,1\rb$, only the states $|1,n\rb_{\t{L}}\otimes |\frac{p}{2},1\rb$ with odd $n$ need to be included as independent representatives.

\section{Cylinder diagram computations} \label{app:cylinders}
\label{app:cylinder_computations}
We list some useful discontinuity formulas. We have the simple relation $\sqrt{x-1\pm i \epsilon}=\pm i \sqrt{1-x}$ for $|x|<1$. We have
\begin{align}
\t{Disc}~ \partial_x \cos(a~ \arccosh(x)) &= -\frac{2 i a  \cos\!\big(a \operatorname{arccosh}(x)\big)\, \sinh(a\pi)}{\sqrt{x^2-1}}, \qq x<-1\\
\t{Disc}~ \partial_x \sin(a~ \arccosh(x)) &= -\frac{2 i a  \sin\!\big(a \operatorname{arccosh}(x)\big)\, \sinh(a\pi)}{\sqrt{x^2-1}}, \qq x<-1\\
&= -\frac{2 i a \, \cos\!\big(a \,\operatorname{arccosh}(x)\big)\, \cosh(a\pi)}{\sqrt{1-x^2}}, \qq |x| < 1\,.
\end{align}

\subsection{FZZT-FZZT cylinders} \label{app:FZZT-FZZT}

We explain how to perform the following integrals from the main text for marked cylinder amplitudes. The two relevant hyperbolic functions in the denominator should be expanded as $(e^x(1-e^{-2x}))^{-1}$, and rewritten as a double infinite series. The integral reduces to exponential which can be performed, and one of the series can be re-summed to give the following
\begin{align}
&\int_0^\infty \!\!\d P
\frac{P \sin(4\pi s_1 P)\sin(4\pi s_2 P)}{\sinh(2\pi pb^{-1}P)  \sinh(2\pi b^{-1}P)} \frac{\sinh(2\pi b^{-1} (p-1)P)}{\sinh (2\pi b s_1) \sinh (2\pi b s_2)} \nn\\
&~~= \frac{b^2 }{32\pi^2 \sinh(2\pi b s_1) \sinh(2\pi b s_2)} \sum_{m=0}^\infty \sum_{\epsilon_{1,2}=\pm1} (-\epsilon_1 \epsilon_2) \Big[ \psi^{(1)}\big( 1+p m + i b \lr{\epsilon_1 s_1 + \epsilon_2 s_2 } \big) \nn\\
&\qq~~ -\psi^{(1)}\big( m + p m + i b \lr{\epsilon_1 s_1 + \epsilon_2 s_2 } \big) \Big] \,,
\end{align}
written as a sum over eight Polygamma functions with alternating signs and arguments. The remaining series can be re-summed using the following identity
\be
\sum_{m=0}^\infty \Big[ \psi^{(1)}\big( 1+p m + i x \big) -\psi^{(1)}\big( m + p m + i x \big) \Big] = \psi^{(1)}(1+i x) - \frac{1}{p^2} \psi^{(1)}(1+ \frac{i x}{p})\,,
\ee
which gives the simple relation using properties of Polygamma functions
\begin{align}
\sum_{m=0}^\infty \sum_{\epsilon_{1,2}=\pm1} &(-\epsilon_1 \epsilon_2)\Big[ \psi^{(1)}\big( 1+p m + i b \lr{\epsilon_1 s_1 + \epsilon_2 s_2 } \big) -\psi^{(1)}\big( m + p m + i b \lr{\epsilon_1 s_1 + \epsilon_2 s_2 } \big) \Big] \nn \\
&= \pi^2 \sum_{\epsilon=\pm 1} \epsilon \lr{ \csch(\pi b (s_1 + \epsilon s_2)) - \frac{1}{p^2} \csch(\frac{\pi}{p} b (s_1 + \epsilon s_2))}\, \nn \\
&= \Biggl[
\csch^{2}\lr{\frac{\pi b}{p} (s_{1}-s_{2})}
- \csch^{2}\lr{\frac{\pi b}{p} (s_{1}+s_{2})}
\nn\\ 
&\qq+ p^{2} \lr{
\csch^{2} \lr{\pi b (s_{1}+s_{2})}
- \csch^{2}\lr{\pi b (s_{1}-s_{2})}
}
\Biggr]
\end{align}
The final answer to the integral is
\begin{align}
&\int_0^\infty \!\!\d P
\frac{P \sin(4\pi s_1 P)\sin(4\pi s_2 P)}{\sinh(2\pi pb^{-1}P)  \sinh(2\pi b^{-1}P)} \frac{\sinh(2\pi b^{-1} (p-1)P)}{\sinh (2\pi b s_1) \sinh (2\pi b s_2)} \nn\\
&=
\frac{b^{2}}{32 p^{2}}
\csch\lr{2 \pi b s_{1}} \csch\lr{2 \pi b s_2}
\Biggl[
\csch^{2}\lr{\frac{\pi b}{p} (s_{1}-s_{2})}
- \csch^{2}\lr{\frac{\pi b}{p} (s_{1}+s_{2})}
\nn\\
&\qq+ p^{2} \lr{
\csch^{2} \lr{\pi b (s_{1}+s_{2})}
- \csch^{2}\lr{\pi b (s_{1}-s_{2})}
}
\Biggr] \nn \\
&= \frac{1}{32 \pi^{2}}
\csch\lr{2 \pi b s_{1}} \csch\lr{2 \pi b s_2} \partial_{s_1}\partial_{s_2}\log \Biggr[ \frac{\cosh\lr{2\pi \frac{b}{p} s_1}-\cosh\lr{2\pi \frac{b}{p} s_2}}{\cosh\lr{2\pi b s_1}-\cosh\lr{2\pi b s_2}}\Biggr] ~,
\end{align}
where we have given an alternative representation.

\paragraph{Unmarked cylinder.} The unmarked cylinder is given by integrating the previous formulas
\be
Z(s_1,s_2) = \log \frac{\cosh\lr{2\pi \frac{b}{p} s_1}-\cosh\lr{2\pi \frac{b}{p} s_2}}{\cosh\lr{2\pi b s_1}-\cosh\lr{2\pi b s_2}}
\ee
where are ignoring the multiplicative normalization factor in front of the log.

\paragraph{DMS extra cylinders fixed length.}
In the DMS  there is an extra boundary state. We compute the fixed length double trumpets involving this state in this appendix given for unmarked fixed energy boundaries by \eqref{eqn:n|n_double_trumpet}
\begin{align}
&Z_{n|n}(s_1, s_2) \nn\\
&=\sum_{r = 1}^{p-1} \sum_{u\overset{2}{=}1-r}^{r-1} \delta_{r \t{\,mod\,} 4 =1}\int_0^\infty \d P
\frac{\cos(4\pi P (s_1+\frac{i}{2} u b^{-1}))\cos(4\pi s_2 P)\sinh(2\pi \frac{P}{b} (p-1)) }{P \sinh(2\pi p' b P)\sinh(2\pi \frac{P}{b})}\nn \\
&=\sum_{r = 1}^{p-1} \delta_{r \t{\,mod\,} 4 =1} \int_0^\infty \d P
\frac{\cos(4\pi P s_1)\cos(4\pi s_2 P)\sinh(2\pi \frac{P}{b} (p-1)) }{P \sinh(2\pi p' b P)\sinh(2\pi \frac{P}{b})} \frac{\sinh(2\pi b^{-1} P r)}{\sinh(2\pi b^{-1} P)} ~,
\end{align}
where in the second line we performed the sum over $u$. The only difference between these boundaries and the AMS is the sum over $r$ and the extra factor in the integral at the end. We go to fixed length in the same way as for the AMS \eqref{eqn:fixed_length_transform}, where we notice that the extra factor doesn't impact the calculation of the discontinuity. We thus get the same integral as in the AMS with the extra prefactor in the integrand
\begin{align} \label{eqn:Zn|n_fixedlength}
&Z_{n|n}(\beta_1, \beta_2) = \frac{8}{\pi^2 b^2}\sum_{r = 1}^{p-1} \delta_{r \t{\,mod\,} 4 =1} \\ &\times \int_0^\infty \d P K_{\frac{2 i P}{b}  }(\beta_1 E_0) K_{\frac{2 i P}{b}  }(\beta_2 E_0) \frac{P \sinh(2\pi \frac{P}{b}) \sinh(2\pi  \frac{P}{b} (p-1))}{\sinh(2\pi p' b P)} \frac{\sinh(2\pi b^{-1} P r)}{\sinh(2\pi b^{-1} P)}\,. \nn 
\end{align}
This amplitudes appears to be divergent and should be regulated in some way, perhaps by slightly deforming the $P$ contour. Note that the fixed length cylinder between the $i$ and $n$ type boundary is the above with the sum over $r$ missing and $r\to \frac{p}{2}$ from \eqref{eqn:cyl_n|i}, along with an overall factor of half
\begin{align}
&Z_{n|i}(\beta_1, \beta_2) \nn\\
&= \frac{2}{\pi^2 b^2} \int_0^\infty \!\!\d P K_{\frac{2 i P}{b}  }(\beta_1 E_0) K_{\frac{2 i P}{b}  }(\beta_2 E_0) \frac{P \sinh(2\pi \frac{P}{b}) \sinh(2\pi  \frac{P}{b} (p-1))}{\sinh(2\pi p' b P)} \frac{\sinh(2\pi b^{-1} P \frac{p}{2})}{\sinh(2\pi b^{-1} P)}\,.
\end{align}
This integral also appears to be divergent, we are not sure of the physical significance of this divergence. 

\paragraph{Divergence in fixed length AMS amplitudes.}
We point out that in the AMS, cylinder's with non-trivial matter states when transformed to the fixed length basis are similarly divergent. This is obvious from the preceding calculation. Consider an AMS amplitude with identity $a=|1,1\rb$ on one side and $b=|r,1\rb$ on the other. It can be extracted from our DMS calculation with the final result
\begin{align}
Z_{a|b}&(s_1, s_2) = Z^{\t{AMS}}(s_1,r,1;s_2,1,1) =\sum_{r = 1}^{p-1} \sum_{u\overset{2}{=}1-r}^{r-1} \delta_{r \t{\,mod\,} 4 =1} Z^{\t{AMS}}(s_1+\frac{i}{2} u b^{-1},s_2) \nn \\
&=\sum_{r = 1}^{p-1}  \delta_{r \t{\,mod\,} 4 =1}\int_0^\infty\d P
\frac{\cos(4\pi P s_1)\cos(4\pi s_2 P)\sinh(2\pi \frac{P}{b} (p-1)) }{P \sinh(2\pi p' b P)\sinh(2\pi \frac{P}{b})} \frac{\sinh(2\pi \frac{P}{b} r)}{\sinh(2\pi\frac{P}{b})}\,.
\end{align}
Marking and inverse Laplace transforming we obtain 
\begin{align} \label{eqn:AMS_Divergence}
&Z_{a|b}(\beta_1, \beta_2) \nn\\
&= \frac{8}{\pi^2 b^2} \int_0^\infty \!\!\d P K_{\frac{2 i P}{b}  }(\beta_1 E_0) K_{\frac{2 i P}{b}  }(\beta_2 E_0) \frac{P \sinh(2\pi \frac{P}{b}) \sinh(2\pi  \frac{P}{b} (p-1))}{\sinh(2\pi p' b P)} \frac{\sinh(2\pi b^{-1} P r)}{\sinh(2\pi b^{-1} P)} \nn \\
&= \infty\,.
\end{align}
This amplitude should presumably be interpreted as the connected contribution to the inverse Laplace transformed resolvent of some matrix $R_{0,2}(E_1, E_2)$, and should naively be finite. The issue may be that the matter boundary condition chosen somehow does not have a reasonable interpretation on the matrix model side. One hint of this is that we know the matter identity states $|1,1\rb$ cleanly match resolvents of the Hermitian matrix. The insertion of other matter states appears mysterious from the matrix model side.

\subsection{ZZ-ZZ cylinders}
We compute ZZ-ZZ amplitudes closely following the approach of \cite{Eniceicu:2022dru}. The Liouville cylinder with different and identical ZZ boundaries on both ends is \eqref{eqn:ZZ_liouville}
\begin{gather}
Z_{(m, n)\left(m^{\prime}, n^{\prime}\right)}^{\text {L }}(q)=\sum_{k=\left|m-m^{\prime}\right|+1,2}^{m+m^{\prime}-1} \sum_{l=\left|n-n^{\prime}\right|+1,2}^{n+n^{\prime}-1} \chi^{(b)}_{k, l}(q), \qquad  Z_{(m, n)}^{\text {L }}(q) =\sum_{k=1}^m \sum_{l=1}^n \chi_{2 k-1,2 l-1}(q) \nn \\
\chi^{(b)}_{k, l}(q)  =\eta(q)^{-1}\left(q^{-k l}-1\right) q^{-\left(k p-l p^{\prime}\right)^2 / 4 p p^{\prime}}, \quad q:=e^{-2 \pi t}.
\end{gather}
The matter contribution will always be given by a sum over degenerate matter characters $Z_{a|b}^{\t{M}}(q)=\sum_{n,m} f_{n,m}^{a|b} \chi^{(i b)}_{n, m}(q)$ which depends on the boundary conditions for the matter through the indicator function. The characters are again given by
\be
\chi^{(i b)}_{r, s}(q)=\frac{1}{\eta(q)} \sum_{k \in \mathbb{Z}}\left(q^{a_{r, s}(k)}-q^{a_{r,-s}(k)}\right), \quad a_{r, s}(k)=\frac{\left(2 p' p k+p' r-p s\right)^2}{4 p' p}
\ee

The cylinder diagram takes the form $ \int_0^\infty \frac{\d t}{t} \eta(q)^2 Z_{a|b}^{\text {M}}(q) Z_{(m, n) (m',n')}^{\text {L}}(q) $ where we have included the ghost contribution.
\paragraph{Performing the integral.}
We have the integral
\be
\int_0^\infty \frac{\d t}{t} (\e^{-2\pi h_1 t} - \e^{-2\pi h_2 t})=\int_0^\infty \frac{\d t}{t} (q^{h_1} - q^{h_2})= \log \frac{h_2}{h_1}\,.
\ee
The basic building block for the full cylinder amplitude is
\begin{align}
&\int_0^\infty \frac{\d t}{t} \eta(q)^2 \chi_{k,l}^{(b)}(t) \chi_{r,s}^{(ib)}(t) \nn\\
&= \sum_{j=-\infty}^\infty \int_0^\infty  \frac{\d t}{t} \lr{ q^{a_{r,-s}(j)-d} - q^{a_{r,s}(j)-d} +  q^{a_{r,s}(j)-d-k l}-q^{a_{r,-s}(j)-d-k l}   } \nn \\
&=\sum_{j=-\infty}^\infty \log \lr{\frac{(2 p p' j + r p' + s p)^2 - (k p - l p')^2 }{(2 p p' j + r p' + s p)^2 - (k p + l p')^2} 
 \, \frac{(2 p p' j + r p' - s p)^2 - (k p + l p')^2 }{(2 p p' j + r p' - s p)^2 - (k p - l p')^2} }\,.
\end{align}
Where in the first line $d= \frac{(k p - l p')^2}{4 p p'}$. We can take the exponential of the above and factor it into the following form
\begin{align}
\exp& \lr{ \int_0^\infty \frac{\d t}{t} \eta(q)^2 \chi_{k,l}^{(b)}(t) \chi_{r,s}^{(ib)}(t) } \nn\\
&= \prod_{j=-\infty}^\infty \frac{(j-\frac{l-s}{2p'}+\frac{k+r}{2 p}) (j+\frac{l-s}{2p'}+\frac{k+r}{2 p}) (-j+\frac{l+s}{2p'}+\frac{k-r}{2 p}) (-j-\frac{l+s}{2p'}+\frac{k-r}{2 p})  }{(j+\frac{l+s}{2p'}+\frac{k+r}{2 p}) (j-\frac{l+s}{2p'}+\frac{k+r}{2 p}) (-j-\frac{l-s}{2p'}+\frac{k-r}{2 p}) (-j+\frac{l-s}{2p'}+\frac{k-r}{2 p})} \nn \\
M_{k,l;r,s}&\equiv \frac{\left(\sin ^2\left(\frac{\pi(l+s)}{2 p'}\right)-\sin ^2\left(\frac{\pi(k+r)}{2 p}\right)\right)\left(\sin ^2\left(\frac{\pi(l-s)}{2 p'}\right)-\sin ^2\left(\frac{\pi(k-r)}{2 p}\right)\right)}{\left(\sin ^2\left(\frac{\pi(l-s)}{2 p'}\right)-\sin ^2\left(\frac{\pi(k+r)}{2 p}\right)\right)\left(\sin ^2\left(\frac{\pi(l+s)}{2 p'}\right)-\sin ^2\left(\frac{\pi(k-r)}{2 p}\right)\right)} \nn \\
&= \frac{\lr{\cos(\frac{\pi (l+s)}{p'})-\cos(\frac{\pi (k+r)}{p})} \lr{\cos(\frac{\pi (l-s)}{p'})-\cos(\frac{\pi (k-r)}{p})} }{\lr{\cos(\frac{\pi (l+s)}{p'})-\cos(\frac{\pi (k-r)}{p})}    \lr{\cos(\frac{\pi (l-s)}{p'})-\cos(\frac{\pi (k+r)}{p})} }
\end{align}
If we consider the identity matter state on both ends for the DMS the matter contribution is $Z^{\t{M}}=\chi^{(i b)}_{1,1}(q)+\chi^{(i b)}_{p-1,1}(q)$. The Cylinder with different Liouville ZZ branes is
\begin{align}
C^{\t{DMS}}_{(m, n),\left(m^{\prime}, n^{\prime}\right)}&=\prod_{k=\left|m-m^{\prime}\right|+1,2}^{m+m^{\prime}-1} \prod_{l=\left|n-n^{\prime}\right|+1,2}^{n+n^{\prime}-1} \exp \lr{ \int_0^\infty \frac{\d t}{t} \eta(q)^2 \chi_{k,l}^{(b)}(t) (\chi_{1,1}^{(ib)}(t)+\chi_{p-1,1}^{(ib)}(t)) } \nn \\
 &=\prod_{k=\left|m-m^{\prime}\right|+1,2}^{m+m^{\prime}-1} \prod_{l=\left|n-n^{\prime}\right|+1,2}^{n+n^{\prime}-1} M_{k,l;1,1} M_{k,l;p-1,1}\,,
\end{align}
where for the AMS we would only have the first line above for $M_{k,l;1,1}$. A telescoping series satisfies $\prod_{i=m}^n \frac{a_{i+1}}{a_i} = \frac{a_{n+1}}{a_m}$ which for example gives
\be
\prod_{l=|n-n'|+1,2}^{n+n'-1} \frac{\lr{\cos(\frac{\pi (l+1)}{p'})-\cos(\frac{\pi (k+r)}{p})} }{    \lr{\cos(\frac{\pi (l-1)}{p'})-\cos(\frac{\pi (k+r)}{p})} } = \frac{\lr{\cos(\frac{\pi (n+n')}{p'})-\cos(\frac{\pi (k+r)}{p})} }{    \lr{\cos(\frac{\pi (|n-n'|)}{p'})-\cos(\frac{\pi (k+r)}{p})} }\,.
\ee
Applying the telescoping rule eight times we get
\begin{align}
C^{\t{DMS}}_{(m, n),\left(m^{\prime}, n^{\prime}\right)}&=\frac{\lr{\cos(\frac{\pi (n+n')}{p'})-\cos(\frac{\pi (m+m')}{p})} \lr{\cos(\frac{\pi |n-n'|}{p'})-\cos(\frac{\pi |m-m'|}{p})}}{    \lr{\cos(\frac{\pi (n+n')}{p'})-\cos(\frac{\pi |m-m'|)}{p})}    \lr{\cos(\frac{\pi |n-n'|}{p'})-\cos(\frac{\pi (m+m')}{p})}  } \nn \\
& \times \frac{ \lr{\cos(\frac{\pi (n+n')}{p'})+\cos(\frac{\pi |m-m'|}{p})}    \lr{\cos(\frac{\pi |n-n'|}{p'})+\cos(\frac{\pi (m+m')}{p})}    }{   \lr{\cos(\frac{\pi (n+n')}{p'})+\cos(\frac{\pi (m+m')}{p})}     \lr{\cos(\frac{\pi |n-n'|}{p'})+\cos(\frac{\pi |m-m'|}{p})}}\,.
\end{align}
This is precisely the main result claimed in the main text, it is the product of two AMS ZZ-ZZ amplitudes. The amplitude in the main text can be obtained by using trig identities. The AMS ZZ-ZZ amplitude is simply the first line of the above \cite{Eniceicu:2022dru}.

\subsection{ZZ-ZZ with identical boundaries}
We now consider the annulus with identical boundary conditions
\begin{align}
B^{\t{DMS}}_{(m, n)}&=\prod_{k=1,2}^{2m-1} \prod_{l=1,2}^{2 n-1} \exp \lr{ \int_0^\infty \frac{\d t}{2t} \eta(q)^2 \chi_{k,l}^{(b)}(t) (\chi_{1,1}^{(ib)}(t)+\chi_{p-1,1}^{(ib)}(t)) } \nn\\
&= \prod_{k=1,2}^{2m-1} \prod_{l=1,2}^{2 n-1} M_{k,l;1,1}^{\frac{1}{2}} M_{k,l;p-1,1}^{\frac{1}{2}}
\end{align}
We have a square root because identical boundary conditions introduce an extra factor of $1/2$ into the integration measure. The product over $M_{k,l;1,1}$ was performed in \cite{Eniceicu:2022nay,Eniceicu:2022dru}, the term $M_{1,1;1,1}$ is naively divergent and needs to be regulated using SFT. The answer is \cite{Eniceicu:2022dru}
\begin{align}
M_{1,1;1,1}^{1/2}
&=\exp \lr{ \int_0^\infty \frac{\d t}{2t} \eta(q)^2 \chi_{1,1}^{(b)}(q) \chi_{1,1}^{(i b)}(q)} \nn\\
&\to (T_{m,n})^{-1/2} \frac{i}{\sqrt{8 \pi}} \lr{\frac{\cot^2(\frac{\pi}{p'})-\cot^2(\frac{\pi}{p})}{p'^2-p^2}}^{1/2} ~,
\end{align}
where $T_{m,n}$ is interpreted as the tension of the ZZ brane. The product over the other values in the AMS can be computed from the above formulas or read off from \cite{Eniceicu:2022dru}
\be
B_{(m, n)}^{\t{AMS}}= \prod_{k=1,2}^{2m-1} \prod_{l=1,2}^{2 n-1} M_{k,l;1,1}^{\frac{1}{2}} = \frac{i}{ \sqrt{8 \pi T_{m,n}}} \lr{ \frac{\cot^2(\frac{\pi n}{p'})-\cot^2(\frac{\pi m}{p})}{p'^2-p^2} }^{1/2}\,.
\ee
The DMS amplitude is given by multiplying the above by the following convergent product
\be
\prod_{k=1,2}^{2m-1} \prod_{l=1,2}^{2 n-1} M_{k,l;p-1,1}^{\frac{1}{2}} = \lr{\frac{2 \cos(\frac{m \pi}{p})^2 \cos(\frac{n \pi}{p'})^2  }{\cos \lr{\frac{2\pi m}{p}} + \cos \lr{\frac{2\pi n}{p'}} }}^{1/2}
\ee
This gives
\be \label{eqn:ZZ_DMS}
B^{\t{DMS}}_{(m, n)}=\frac{i}{ \sqrt{8 \pi T_{m,n}}} \lr{ \frac{\cot^2(\frac{\pi n}{p'})-\cot^2(\frac{\pi m}{p})}{p'^2-p^2} }^{1/2} \times \lr{\frac{2 \cos(\frac{m \pi}{p})^2 \cos(\frac{n \pi}{p'})^2  }{\cos \lr{\frac{2\pi m}{p}} + \cos \lr{\frac{2\pi n}{p'}} }}^{1/2}\,.
\ee

\subsection{ZZ-ZZ amplitudes with extra brane boundary in DMS} \label{app:ZZ_Extra_branes_DMS}

For DMS, we have additional ZZ boundary states as explained around \eqref{eqn:ZZ_bdy_states}. One basis of these state is $|\text{ZZ}_{1,s}\rb\otimes |\frac{p}{2},1\rb$. Assuming $s_1\neq s_2$ the amplitude in the closed channel is
\bal
&\int_0^\infty \frac{\d t}{t} \, \eta(q)^2   \langle \text{ZZ}_{1,s_1}| \otimes \langle \frac{p}{2},1 | ~ e^{-\frac{2\pi}{t} (L_0+\bar{L}_0-\frac{c}{12})} ~ |\text{ZZ}_{1,s_2}\rangle\otimes | \frac{p}{2},1\rangle_{\t{DMM}} \\ \nn
&=\sum_{k\overset{4}{=}1}^{4 \left\lfloor \tfrac{p-1}{4} \right\rfloor + 1} \int_0^\infty \frac{\d t}{t} \, \eta(q)^2   \langle \text{ZZ}_{1,s_1}| \otimes \langle k,1 | ~ e^{-\frac{2\pi}{t} (L_0+\bar{L}_0-\frac{c}{12})} ~ |\text{ZZ}_{1,s_2}\rangle\otimes | 1,1\rangle_{\t{AMM}} \\ \nn
&=\sum_{k\overset{4}{=}1}^{4 \left\lfloor \tfrac{p-1}{4} \right\rfloor + 1} \sum_{u\overset{2}{=}1-k}^{k-1} \int_0^\infty \frac{\d t}{t} \, \eta(q)^2   \langle \text{ZZ}_{1+u,s_1}| \otimes \langle 1,1 | ~ e^{-\frac{2\pi}{t} (L_0+\bar{L}_0-\frac{c}{12})} ~ |\text{ZZ}_{1,s_2}\rangle\otimes | 1,1\rangle_{\t{AMM}}\,, \nn
\eal
The final line is $2\log B_{(1+u,s_1)}^{\t{AMS}}$ if the two ZZ boundary states are identical, and $\log C_{(1+u,s_1),(1,s_2)}^{\t{AMS}}$ if the two ZZ boundary states are not identical. Note “identical” includes various relations between AMS ZZ states, for example, $|\text{ZZ}_{m,n}\rb \otimes | 1,1 \rb = |\text{ZZ}_{p-m,p'-n}\rb \otimes | 1,1 \rb$. Amplitudes with $s_1=s_2$ should be divided by $\frac{1}{2}$. There is also an amplitude between the extra state and the states considered in the main text
\bal
&\int_0^\infty \frac{\d t}{t} \, \eta(it)^2   \langle \text{ZZ}_{1,s}| \otimes \langle \frac{p}{2},1 | ~ e^{-\frac{2\pi}{t} (L_0+\bar{L}_0-\frac{c}{12})} ~ |\text{ZZ}_{m,n}\rangle\otimes | 1,1\rangle_{\t{DMM}} \\ \nn
&=\sum_{u\overset{2}{=}1-\frac{p}{2}}^{\frac{p}{2}-1}  \int_0^\infty \frac{\d t}{t} \, \eta(it)^2   \langle \text{ZZ}_{1+u,s}| \otimes \langle 1,1 | ~ e^{-\frac{2\pi}{t} (L_0+\bar{L}_0-\frac{c}{12})} ~ |\text{ZZ}_{m,n}\rangle\otimes | 1,1\rangle_{\t{AMM}}
\eal
The integral in the second line is again given by $2\log B^{\t{AMS}}_{(1+u,s)}$ and $\log C^{\t{AMS}}_{(1+u,s),(m,n)}$ as above.

\bibliographystyle{JHEP}
\bibliography{references}

@article{Moore:1991zv,
    author = "Moore, Gregory W. and Plesser, M. Ronen and Ramgoolam, Sanjaye",
    title = "{Exact S matrix for 2-D string theory}",
    eprint = "hep-th/9111035",
    archivePrefix = "arXiv",
    reportNumber = "YCTP-P35-91",
    doi = "10.1016/0550-3213(92)90020-C",
    journal = "Nucl. Phys. B",
    volume = "377",
    pages = "143--190",
    year = "1992"
}

@article{Kazakov:2021lel,
    author = "Kazakov, Vladimir and Zheng, Zechuan",
    title = "{Analytic and numerical bootstrap for one-matrix model and {\textquotedblleft}unsolvable{\textquotedblright} two-matrix model}",
    eprint = "2108.04830",
    archivePrefix = "arXiv",
    primaryClass = "hep-th",
    doi = "10.1007/JHEP06(2022)030",
    journal = "JHEP",
    volume = "06",
    pages = "030",
    year = "2022"
}

@article{Lin:2020mme,
    author = "Lin, Henry W.",
    title = "{Bootstraps to strings: solving random matrix models with positivity}",
    eprint = "2002.08387",
    archivePrefix = "arXiv",
    primaryClass = "hep-th",
    doi = "10.1007/JHEP06(2020)090",
    journal = "JHEP",
    volume = "06",
    pages = "090",
    year = "2020"
}

@article{Dotsenko:1984ad,
    author = "Dotsenko, V. S. and Fateev, V. A.",
    title = "{Four Point Correlation Functions and the Operator Algebra in the Two-Dimensional Conformal Invariant Theories with the Central Charge c {\ensuremath{<}} 1}",
    reportNumber = "NORDITA-84/22",
    doi = "10.1016/S0550-3213(85)80004-3",
    journal = "Nucl. Phys. B",
    volume = "251",
    pages = "691--734",
    year = "1985"
}

@article{Goulian:1990qr,
    author = "Goulian, M. and Li, Miao",
    title = "{Correlation functions in Liouville theory}",
    reportNumber = "UCSB-TH-90-61",
    doi = "10.1103/PhysRevLett.66.2051",
    journal = "Phys. Rev. Lett.",
    volume = "66",
    pages = "2051--2055",
    year = "1991"
}

@article{Rodriguez:2023kkl,
    author = "Rodriguez, Victor A.",
    title = "{A two-dimensional string cosmology}",
    eprint = "2302.06625",
    archivePrefix = "arXiv",
    primaryClass = "hep-th",
    doi = "10.1007/JHEP06(2023)161",
    journal = "JHEP",
    volume = "06",
    pages = "161",
    year = "2023"
}

@article{Balthazar:2022atu,
    author = "Balthazar, Bruno and Rodriguez, Victor A. and Yin, Xi",
    title = "{The S-matrix of 2D type 0B string theory. Part I. Perturbation theory revisited}",
    eprint = "2201.05621",
    archivePrefix = "arXiv",
    primaryClass = "hep-th",
    doi = "10.1007/JHEP05(2023)234",
    journal = "JHEP",
    volume = "05",
    pages = "234",
    year = "2023"
}

@article{Chang:2014jta,
    author = "Chang, Chi-Ming and Lin, Ying-Hsuan and Shao, Shu-Heng and Wang, Yifan and Yin, Xi",
    title = "{Little String Amplitudes (and the Unreasonable Effectiveness of 6D SYM)}",
    eprint = "1407.7511",
    archivePrefix = "arXiv",
    primaryClass = "hep-th",
    doi = "10.1007/JHEP12(2014)176",
    journal = "JHEP",
    volume = "12",
    pages = "176",
    year = "2014"
}

@article{Witten:2013pra,
    author = "Witten, Edward",
    title = "{The Feynman $i \epsilon$ in String Theory}",
    eprint = "1307.5124",
    archivePrefix = "arXiv",
    primaryClass = "hep-th",
    doi = "10.1007/JHEP04(2015)055",
    journal = "JHEP",
    volume = "04",
    pages = "055",
    year = "2015"
}

@article{Berera:1992tm,
    author = "Berera, Arjun",
    title = "{Unitary string amplitudes}",
    reportNumber = "UCB-PTH-92-11, LBL-32248",
    doi = "10.1016/0550-3213(94)90057-4",
    journal = "Nucl. Phys. B",
    volume = "411",
    pages = "157--180",
    year = "1994"
}

@article{Eberhardt:2022zay,
    author = "Eberhardt, Lorenz and Mizera, Sebastian",
    title = "{Unitarity cuts of the worldsheet}",
    eprint = "2208.12233",
    archivePrefix = "arXiv",
    primaryClass = "hep-th",
    doi = "10.21468/SciPostPhys.14.2.015",
    journal = "SciPost Phys.",
    volume = "14",
    number = "2",
    pages = "015",
    year = "2023"
}

@article{Sen:2019jpm,
    author = "Sen, Ashoke",
    title = "{String Field Theory as World-sheet UV Regulator}",
    eprint = "1902.00263",
    archivePrefix = "arXiv",
    primaryClass = "hep-th",
    doi = "10.1007/JHEP10(2019)119",
    journal = "JHEP",
    volume = "10",
    pages = "119",
    year = "2019"
}

@article{Sen:2020cef,
    author = "Sen, Ashoke",
    title = "{D-instanton Perturbation Theory}",
    eprint = "2002.04043",
    archivePrefix = "arXiv",
    primaryClass = "hep-th",
    doi = "10.1007/JHEP08(2020)075",
    journal = "JHEP",
    volume = "08",
    pages = "075",
    year = "2020"
}

@article{Balthazar:2019rnh,
    author = "Balthazar, Bruno and Rodriguez, Victor A. and Yin, Xi",
    title = "{ZZ instantons and the non-perturbative dual of c = 1 string theory}",
    eprint = "1907.07688",
    archivePrefix = "arXiv",
    primaryClass = "hep-th",
    doi = "10.1007/JHEP05(2023)048",
    journal = "JHEP",
    volume = "05",
    pages = "048",
    year = "2023"
}

@article{Balthazar:2019ypi,
    author = "Balthazar, Bruno and Rodriguez, Victor A. and Yin, Xi",
    title = "{Multi-instanton calculus in c = 1 string theory}",
    eprint = "1912.07170",
    archivePrefix = "arXiv",
    primaryClass = "hep-th",
    doi = "10.1007/JHEP05(2023)050",
    journal = "JHEP",
    volume = "05",
    pages = "050",
    year = "2023"
}

@article{Sen:2019qqg,
    author = "Sen, Ashoke",
    title = "{Fixing an Ambiguity in Two Dimensional String Theory Using String Field Theory}",
    eprint = "1908.02782",
    archivePrefix = "arXiv",
    primaryClass = "hep-th",
    doi = "10.1007/JHEP03(2020)005",
    journal = "JHEP",
    volume = "03",
    pages = "005",
    year = "2020"
}

@article{Sen:2020eck,
    author = "Sen, Ashoke",
    title = "{D-instantons, string field theory and two dimensional string theory}",
    eprint = "2012.11624",
    archivePrefix = "arXiv",
    primaryClass = "hep-th",
    doi = "10.1007/JHEP11(2021)061",
    journal = "JHEP",
    volume = "11",
    pages = "061",
    year = "2021"
}

@article{Alexandrov:2025pzs,
    author = "Alexandrov, Sergei and Kaushik, Rishabh",
    title = "{Multi-instantons in 2d string theory}",
    eprint = "2509.03293",
    archivePrefix = "arXiv",
    primaryClass = "hep-th",
    month = "9",
    year = "2025"
}

@article{Kaushik:2025neu,
    author = "Kaushik, Rishabh",
    title = "{c = 1, R = 1 and N {\ensuremath{\gg}} 1: ZZ instantons in 2D string theory and matrix integrals}",
    eprint = "2501.14023",
    archivePrefix = "arXiv",
    primaryClass = "hep-th",
    doi = "10.1007/JHEP08(2025)177",
    journal = "JHEP",
    volume = "08",
    pages = "177",
    year = "2025"
}

@article{Marino:2022rpz,
    author = "Marino, Marcos and Schiappa, Ricardo and Schwick, Maximilian",
    title = "{New Instantons for Matrix Models}",
    eprint = "2210.13479",
    archivePrefix = "arXiv",
    primaryClass = "hep-th",
    month = "10",
    year = "2022"
}

@article{Schiappa:2023ned,
    author = "Schiappa, Ricardo and Schwick, Maximilian and Tamarin, Noam",
    title = "{All the D-Branes of Resurgence}",
    eprint = "2301.05214",
    archivePrefix = "arXiv",
    primaryClass = "hep-th",
    month = "1",
    year = "2023"
}

@article{Eynard:2023qdr,
    author = "Eynard, Bertrand and Garcia-Failde, Elba and Gregori, Paolo and Lewanski, Danilo and Schiappa, Ricardo",
    title = "{Resurgent Asymptotics of Jackiw{\textendash}Teitelboim Gravity and the Nonperturbative Topological Recursion}",
    eprint = "2305.16940",
    archivePrefix = "arXiv",
    primaryClass = "hep-th",
    reportNumber = "CERN-TH-2021-097",
    doi = "10.1007/s00023-023-01412-z",
    journal = "Annales Henri Poincare",
    volume = "25",
    number = "9",
    pages = "4121--4193",
    year = "2024"
}

@article{Ribault:2024rvk,
    author = "Ribault, Sylvain",
    title = "{Exactly solvable conformal field theories}",
    eprint = "2411.17262",
    archivePrefix = "arXiv",
    primaryClass = "hep-th",
    month = "11",
    year = "2024"
}

@article{Ribault:2015sxa,
    author = "Ribault, Sylvain and Santachiara, Raoul",
    title = "{Liouville theory with a central charge less than one}",
    eprint = "1503.02067",
    archivePrefix = "arXiv",
    primaryClass = "hep-th",
    doi = "10.1007/JHEP08(2015)109",
    journal = "JHEP",
    volume = "08",
    pages = "109",
    year = "2015"
}

@article{Hadasz:2009db,
    author = "Hadasz, Leszek and Jaskolski, Zbigniew and Suchanek, Paulina",
    title = "{Recursive representation of the torus 1-point conformal block}",
    eprint = "0911.2353",
    archivePrefix = "arXiv",
    primaryClass = "hep-th",
    doi = "10.1007/JHEP01(2010)063",
    journal = "JHEP",
    volume = "01",
    pages = "063",
    year = "2010"
}

@article{Cho:2017oxl,
    author = "Cho, Minjae and Collier, Scott and Yin, Xi",
    title = "{Recursive Representations of Arbitrary Virasoro Conformal Blocks}",
    eprint = "1703.09805",
    archivePrefix = "arXiv",
    primaryClass = "hep-th",
    doi = "10.1007/JHEP04(2019)018",
    journal = "JHEP",
    volume = "04",
    pages = "018",
    year = "2019"
}

@article{Zamolodchikov:1984eqp,
    author = "Zamolodchikov, A. B.",
    title = "{CONFORMAL SYMMETRY IN TWO-DIMENSIONS: AN EXPLICIT RECURRENCE FORMULA FOR THE CONFORMAL PARTIAL WAVE AMPLITUDE}",
    doi = "10.1007/BF01214585",
    journal = "Commun. Math. Phys.",
    volume = "96",
    pages = "419--422",
    year = "1984"
}

@article{Zamolodchikov:1987avt,
    author = "Zamolodchikov, Al. B.",
    title = "{Conformal symmetry in two-dimensional space: Recursion representation of conformal block}",
    doi = "10.1007/BF01022967",
    journal = "Theor. Math. Phys.",
    volume = "73",
    number = "1",
    pages = "1088--1093",
    year = "1987"
}

@article{Eberhardt:2023mrq,
    author = "Eberhardt, Lorenz",
    title = "{Notes on crossing transformations of Virasoro conformal blocks}",
    eprint = "2309.11540",
    archivePrefix = "arXiv",
    primaryClass = "hep-th",
    month = "9",
    year = "2023"
}

@article{Alexanian_2023,
	author = {Alexanian, S. and Kuznetsov, A.},
	doi = {10.1080/10652469.2023.2238115},
	issn = {1476-8291},
	journal = {Integral Transforms and Special Functions},
	month = jul,
	number = {12},
	pages = {891--914},
	publisher = {Informa UK Limited},
	title = "{On the Barnes double gamma function}",
	url = {http://dx.doi.org/10.1080/10652469.2023.2238115},
	volume = {34},
	year = {2023}}

@article{Collier:2024kmo,
    author = {Collier, Scott and Eberhardt, Lorenz and M\"uhlmann, Beatrix and Rodriguez, Victor A.},
    title = "{The complex Liouville string}",
    eprint = "2409.17246",
    archivePrefix = "arXiv",
    primaryClass = "hep-th",
    month = "9",
    year = "2024"
}

@article{Collier:2024kwt,
    author = {Collier, Scott and Eberhardt, Lorenz and M\"uhlmann, Beatrix and Rodriguez, Victor A.},
    title = "{The complex Liouville string: the worldsheet}",
    eprint = "2409.18759",
    archivePrefix = "arXiv",
    primaryClass = "hep-th",
    month = "9",
    year = "2024"
}

@article{Collier:2024lys,
    author = {Collier, Scott and Eberhardt, Lorenz and M\"uhlmann, Beatrix and Rodriguez, Victor A.},
    title = "{The complex Liouville string: the matrix integral}",
    eprint = "2410.07345",
    archivePrefix = "arXiv",
    primaryClass = "hep-th",
    doi = "10.21468/SciPostPhys.18.5.154",
    journal = "SciPost Phys.",
    volume = "18",
    pages = "154",
    year = "2025"
}

@article{Collier:2024mlg,
    author = {Collier, Scott and Eberhardt, Lorenz and M\"uhlmann, Beatrix and Rodriguez, Victor A.},
    title = "{The complex Liouville string: worldsheet boundaries and non-perturbative effects}",
    eprint = "2410.09179",
    archivePrefix = "arXiv",
    primaryClass = "hep-th",
    month = "10",
    year = "2024"
}

@article{Teschner:2001rv,
    author = "Teschner, J.",
    title = "{Liouville theory revisited}",
    eprint = "hep-th/0104158",
    archivePrefix = "arXiv",
    doi = "10.1088/0264-9381/18/23/201",
    journal = "Class. Quant. Grav.",
    volume = "18",
    pages = "R153--R222",
    year = "2001"
}

@article{Teschner:1995yf,
    author = "Teschner, Jorg",
    title = "{On the Liouville three point function}",
    eprint = "hep-th/9507109",
    archivePrefix = "arXiv",
    doi = "10.1016/0370-2693(95)01200-A",
    journal = "Phys. Lett. B",
    volume = "363",
    pages = "65--70",
    year = "1995"
}

@article{Collier:2023cyw,
    author = {Collier, Scott and Eberhardt, Lorenz and M\"uhlmann, Beatrix and Rodriguez, Victor A.},
    title = "{The Virasoro minimal string}",
    eprint = "2309.10846",
    archivePrefix = "arXiv",
    primaryClass = "hep-th",
    doi = "10.21468/SciPostPhys.16.2.057",
    journal = "SciPost Phys.",
    volume = "16",
    number = "2",
    pages = "057",
    year = "2024"
}

@article{Boulatov:1986sb,
    author = "Boulatov, D. V. and Kazakov, V. A.",
    title = "{The Ising Model on Random Planar Lattice: The Structure of Phase Transition and the Exact Critical Exponents}",
    reportNumber = "SPI-MOSCOW-1154",
    doi = "10.1016/0370-2693(87)90312-1",
    journal = "Phys. Lett. B",
    volume = "186",
    pages = "379",
    year = "1987"
}

@article{Douglas:2003um,
    author = "Douglas, Michael R.",
    title = "{The Statistics of string / M theory vacua}",
    eprint = "hep-th/0303194",
    archivePrefix = "arXiv",
    reportNumber = "RUNHETC-2003-09",
    doi = "10.1088/1126-6708/2003/05/046",
    journal = "JHEP",
    volume = "05",
    pages = "046",
    year = "2003"
}

@article{Denef:2008wq,
    author = "Denef, Frederik",
    editor = "Bachas, Costas and Baulieu, Laurent and Douglas, Michael and Kiritsis, Elias and Rabinovici, Eliezer and Vanhove, Pierre and Windey, Paul and Cugliandolo, Leticia F.",
    title = "{Lectures on constructing string vacua}",
    eprint = "0803.1194",
    archivePrefix = "arXiv",
    primaryClass = "hep-th",
    doi = "10.1016/S0924-8099(08)80029-7",
    journal = "Les Houches",
    volume = "87",
    pages = "483--610",
    year = "2008"
}

@article{mtms,
    author = "Mertens, Thomas G. and Turiaci, Gustavo J.",
    title = "{Liouville quantum gravity -- holography, JT and matrices}",
    eprint = "2006.07072",
    archivePrefix = "arXiv",
    primaryClass = "hep-th",
    month = "6",
    year = "2020"
}

@article{Brezin:1990rb,
    author = "Brezin, E. and Kazakov, V.A.",
    title = "{Exactly Solvable Field Theories of Closed Strings}",
    doi = "10.1016/0370-2693(90)90818-Q",
    journal = "Phys. Lett. B",
    volume = "236",
    pages = "144--150",
    year = "1990"
}

@article{Gross:1989vs,
    author = "Gross, David J. and Migdal, Alexander A.",
    editor = "Brezin, E. and Wadia, S.R.",
    title = "{Nonperturbative Two-Dimensional Quantum Gravity}",
    reportNumber = "PUPT-1148",
    doi = "10.1103/PhysRevLett.64.127",
    journal = "Phys. Rev. Lett.",
    volume = "64",
    pages = "127",
    year = "1990"
}

@article{Johnson:2019eik,
    author = "Johnson, Clifford V.",
    title = "{Non-Perturbative JT Gravity}",
    eprint = "1912.03637",
    archivePrefix = "arXiv",
    primaryClass = "hep-th",
    month = "12",
    year = "2019"
}

@article{Moore:1991ir,
    author = "Moore, Gregory W. and Seiberg, Nathan and Staudacher, Matthias",
    title = "{From loops to states in 2-D quantum gravity}",
    reportNumber = "RU-91-11, YCTP-P11-91",
    doi = "10.1016/0550-3213(91)90548-C",
    journal = "Nucl. Phys. B",
    volume = "362",
    pages = "665--709",
    year = "1991"
}

@article{Eynard:2007kz,
    author = "Eynard, Bertrand and Orantin, Nicolas",
    title = "{Invariants of algebraic curves and topological expansion}",
    eprint = "math-ph/0702045",
    archivePrefix = "arXiv",
    reportNumber = "SPHT-07-021",
    doi = "10.4310/CNTP.2007.v1.n2.a4",
    journal = "Commun. Num. Theor. Phys.",
    volume = "1",
    pages = "347--452",
    year = "2007"
}

@article{Saad:2019lba,
      author         = "Saad, Phil and Shenker, Stephen H. and Stanford, Douglas",
      title          = "{JT gravity as a matrix integral}",
      year           = "2019",
      eprint         = "1903.11115",
      archivePrefix  = "arXiv",
      primaryClass   = "hep-th",
      SLACcitation   = "%%CITATION = ARXIV:1903.11115;%%"
}

@article{Eynard:2015aea,
    author = "Eynard, Bertrand and Kimura, Taro and Ribault, Sylvain",
    title = "{Random matrices}",
    eprint = "1510.04430",
    archivePrefix = "arXiv",
    primaryClass = "math-ph",
    month = "10",
    year = "2015"
}

@book{DiFrancesco:1997nk,
    author = "Di Francesco, P. and Mathieu, P. and Senechal, D.",
    title = "{Conformal Field Theory}",
    doi = "10.1007/978-1-4612-2256-9",
    isbn = "978-0-387-94785-3, 978-1-4612-7475-9",
    publisher = "Springer-Verlag",
    address = "New York",
    series = "Graduate Texts in Contemporary Physics",
    year = "1997"
}

@article{Belavin:2008kv,
    author = "Belavin, A.A. and Zamolodchikov, A.B.",
    title = "{On Correlation Numbers in 2D Minimal Gravity and Matrix Models}",
    eprint = "0811.0450",
    archivePrefix = "arXiv",
    primaryClass = "hep-th",
    reportNumber = "RUNHETC-2008-24",
    doi = "10.1088/1751-8113/42/30/304004",
    journal = "J. Phys. A",
    volume = "42",
    pages = "304004",
    year = "2009"
}

@article{Kutasov:2004fg,
    author = "Kutasov, David and Okuyama, Kazumi and Park, Jong-won and Seiberg, Nathan and Shih, David",
    title = "{Annulus amplitudes and ZZ branes in minimal string theory}",
    eprint = "hep-th/0406030",
    archivePrefix = "arXiv",
    reportNumber = "EFI-04-19, PUPT-2121",
    doi = "10.1088/1126-6708/2004/08/026",
    journal = "JHEP",
    volume = "08",
    pages = "026",
    year = "2004"
}

@article{Johnson:2020heh,
      author         = "Johnson, Clifford V.",
      title          = "{JT Supergravity, Minimal Strings, and Matrix Models}",
      year           = "2020",
      eprint         = "2005.01893",
      archivePrefix  = "arXiv",
      primaryClass   = "hep-th",
      SLACcitation   = "%%CITATION = ARXIV:2005.01893;%%"
}

@article{Martinec:2003ka,
    author = "Martinec, Emil J.",
    archivePrefix = "arXiv",
    eprint = "hep-th/0305148",
    month = "5",
    reportNumber = "EFI-03-22",
    title = "{The Annular report on noncritical string theory}",
    year = "2003"
}

@article{Balthazar:2017mxh,
    author = "Balthazar, Bruno and Rodriguez, Victor A. and Yin, Xi",
    archivePrefix = "arXiv",
    doi = "10.1007/JHEP04(2019)145",
    eprint = "1705.07151",
    journal = "JHEP",
    pages = "145",
    primaryClass = "hep-th",
    title = "{The $c$ = 1 string theory S-matrix revisited}",
    volume = "04",
    year = "2019"
}

@article{Cardy:1989ir,
    author = "Cardy, John L.",
    doi = "10.1016/0550-3213(89)90521-X",
    journal = "Nucl. Phys. B",
    pages = "581--596",
    reportNumber = "UCSB-TH-89-06",
    title = "{Boundary Conditions, Fusion Rules and the Verlinde Formula}",
    volume = "324",
    year = "1989"
}

@article{Seiberg:2003nm,
    author = "Seiberg, Nathan and Shih, David",
    archivePrefix = "arXiv",
    doi = "10.1088/1126-6708/2004/02/021",
    eprint = "hep-th/0312170",
    journal = "JHEP",
    pages = "021",
    reportNumber = "PUPT-2102",
    title = "{Branes, rings and matrix models in minimal (super)string theory}",
    volume = "02",
    year = "2004"
}

@article{Zamolodchikov:2005fy,
    author = "Zamolodchikov, Alexei B.",
    archivePrefix = "arXiv",
    doi = "10.1007/s11232-005-0003-3",
    eprint = "hep-th/0505063",
    journal = "Theor. Math. Phys.",
    pages = "183--196",
    reportNumber = "LPM-04-10",
    title = "{Three-point function in the minimal Liouville gravity}",
    volume = "142",
    year = "2005"
}

@article{Kostov:2003uh,
    author = "Kostov, Ivan K. and Ponsot, Benedicte and Serban, Didina",
    archivePrefix = "arXiv",
    doi = "10.1016/j.nuclphysb.2004.02.009",
    eprint = "hep-th/0307189",
    journal = "Nucl. Phys. B",
    pages = "309--362",
    reportNumber = "SACLAY-SPHT-T03-102",
    title = "{Boundary Liouville theory and 2-D quantum gravity}",
    volume = "683",
    year = "2004"
}

@article{Fateev:2000ik,
    author = "Fateev, V. and Zamolodchikov, Alexander B. and Zamolodchikov, Alexei B.",
    archivePrefix = "arXiv",
    eprint = "hep-th/0001012",
    month = "1",
    reportNumber = "RUNHETC-2000-01",
    title = "{Boundary Liouville field theory. 1. Boundary state and boundary two point function}",
    year = "2000"
}

@article{Dorn:1994xn,
      author         = "Dorn, Harald and Otto, H. J.",
      title          = "{Two and three point functions in Liouville theory}",
      journal        = "Nucl. Phys.",
      volume         = "B429",
      year           = "1994",
      pages          = "375-388",
      doi            = "10.1016/0550-3213(94)00352-1",
      eprint         = "hep-th/9403141",
      archivePrefix  = "arXiv",
      primaryClass   = "hep-th",
      reportNumber   = "HU-BERLIN-IEP-94-2, DESY-94-066",
      SLACcitation   = "%%CITATION = HEP-TH/9403141;%%"
}

@article{Zamolodchikov:1995aa,
      author         = "Zamolodchikov, Alexander B. and Zamolodchikov, Alexei B.",
      title          = "{Structure constants and conformal bootstrap in Liouville
                        field theory}",
      journal        = "Nucl. Phys.",
      volume         = "B477",
      year           = "1996",
      pages          = "577-605",
      doi            = "10.1016/0550-3213(96)00351-3",
      eprint         = "hep-th/9506136",
      archivePrefix  = "arXiv",
      primaryClass   = "hep-th",
      reportNumber   = "LPM-95-24, RU-95-39",
      SLACcitation   = "%%CITATION = HEP-TH/9506136;%%"
}

@article{Ishibashi:1988kg,
    author = "Ishibashi, Nobuyuki",
    title = "{The Boundary and Crosscap States in Conformal Field Theories}",
    reportNumber = "UT-530-TOKYO",
    doi = "10.1142/S0217732389000320",
    journal = "Mod. Phys. Lett. A",
    volume = "4",
    pages = "251",
    year = "1989"
}

@article{Runkel:1998he,
      author         = "Runkel, Ingo",
      title          = "{Boundary structure constants for the A series Virasoro
                        minimal models}",
      journal        = "Nucl. Phys.",
      volume         = "B549",
      year           = "1999",
      pages          = "563-578",
      doi            = "10.1016/S0550-3213(99)00125-X",
      eprint         = "hep-th/9811178",
      archivePrefix  = "arXiv",
      primaryClass   = "hep-th",
      reportNumber   = "KCL-MTH-98-59",
      SLACcitation   = "%%CITATION = HEP-TH/9811178;%%"
}

@book{Polchinski:1998rq,
    author = "Polchinski, J.",
    title = "{String theory. Vol. 1: An introduction to the bosonic string}",
    doi = "10.1017/CBO9780511816079",
    isbn = "978-0-511-25227-3, 978-0-521-67227-6, 978-0-521-63303-1",
    publisher = "Cambridge University Press",
    series = "Cambridge Monographs on Mathematical Physics",
    month = "12",
    year = "2007"
}

@article{Belavin:2013nba,
    author = "Belavin, Alexander and Dubrovin, Boris and Mukhametzhanov, Baur",
    title = "{Minimal Liouville Gravity correlation numbers from Douglas string equation}",
    eprint = "1310.5659",
    archivePrefix = "arXiv",
    primaryClass = "hep-th",
    doi = "10.1007/JHEP01(2014)156",
    journal = "JHEP",
    volume = "01",
    pages = "156",
    year = "2014"
}

@article{Spodyneiko:2014lla,
    author = "Spodyneiko, Lev",
    title = "{Minimal Liouville Gravity on the Torus via Matrix Models}",
    eprint = "1407.3546",
    archivePrefix = "arXiv",
    primaryClass = "hep-th",
    month = "7",
    year = "2014"
}

@article{Artemev:2025pvk,
    author = "Artemev, Aleksandr",
    title = "{$x-y$ swap for $(2,2p+1)$ minimal string}",
    eprint = "2506.09222",
    archivePrefix = "arXiv",
    primaryClass = "hep-th",
    month = "6",
    year = "2025"
}

@article{Nivesvivat:2025odb,
    author = "Nivesvivat, Rongvoram and Ribault, Sylvain",
    title = "{Fusion rules and structure constants of E-series minimal models}",
    eprint = "2502.14295",
    archivePrefix = "arXiv",
    primaryClass = "hep-th",
    doi = "10.21468/SciPostPhys.18.5.163",
    journal = "SciPost Phys.",
    volume = "18",
    pages = "163",
    year = "2025"
}

@article{Belavin:2010sr,
    author = "Belavin, V.",
    title = "{Torus Amplitudes in Minimal Liouville Gravity and Matrix Models}",
    eprint = "1010.5508",
    archivePrefix = "arXiv",
    primaryClass = "hep-th",
    reportNumber = "ITEP-LAT-2010-09",
    doi = "10.1016/j.physletb.2011.03.003",
    journal = "Phys. Lett. B",
    volume = "698",
    pages = "86--90",
    year = "2011"
}

@article{Artemev:2022sfi,
    author = "Artemev, Aleksandr and Belavin, Vladimir",
    title = "{Torus one-point correlation numbers in minimal Liouville gravity}",
    eprint = "2210.14568",
    archivePrefix = "arXiv",
    primaryClass = "hep-th",
    doi = "10.1007/JHEP02(2023)116",
    journal = "JHEP",
    volume = "02",
    pages = "116",
    year = "2023"
}

@article{Bershadsky:1990xb,
    author = "Bershadsky, Michael and Klebanov, Igor R.",
    title = "{Genus one path integral in two-dimensional quantum gravity}",
    reportNumber = "PUPT-1197",
    doi = "10.1103/PhysRevLett.65.3088",
    journal = "Phys. Rev. Lett.",
    volume = "65",
    pages = "3088--3091",
    year = "1990"
}

@article{diFrancesco:1987ses,
    author = "di Francesco, P. and Saleur, H. and Zuber, J. B.",
    title = "{Relations between the Coulomb gas picture and conformal invariance of two-dimensional critical models}",
    doi = "10.1007/BF01009954",
    journal = "J. Statist. Phys.",
    volume = "49",
    number = "1",
    pages = "57--79",
    year = "1987"
}

@article{Belavin:2006ex,
    author = "Belavin, A. A. and Zamolodchikov, A. B.",
    title = "{Integrals over moduli spaces, ground ring, and four-point function in minimal Liouville gravity}",
    doi = "10.1007/s11232-006-0075-8",
    journal = "Theor. Math. Phys.",
    volume = "147",
    pages = "729--754",
    year = "2006"
}

@article{Artemev:2022nvl,
    author = "Artemev, A. A. and Belavin, A. A.",
    title = "{Five-Point Correlation Numbers in Minimal Liouville Gravity}",
    doi = "10.1134/S0021364022602135",
    journal = "JETP Lett.",
    volume = "116",
    number = "9",
    pages = "600--607",
    year = "2022"
}

@article{Tarnopolsky:2009ec,
    author = "Tarnopolsky, G.",
    title = "{Five-point Correlation Numbers in One-Matrix Model}",
    eprint = "0912.4971",
    archivePrefix = "arXiv",
    primaryClass = "hep-th",
    doi = "10.1088/1751-8113/44/32/325401",
    journal = "J. Phys. A",
    volume = "44",
    pages = "325401",
    year = "2011"
}

@article{Runkel:1999dz,
    author = "Runkel, Ingo",
    title = "{Structure constants for the D series Virasoro minimal models}",
    eprint = "hep-th/9908046",
    archivePrefix = "arXiv",
    doi = "10.1016/S0550-3213(99)00707-5",
    journal = "Nucl. Phys. B",
    volume = "579",
    pages = "561--589",
    year = "2000"
}

@article{Cardy:1984bb,
    author = "Cardy, John L.",
    title = "{Conformal Invariance and Surface Critical Behavior}",
    doi = "10.1016/0550-3213(84)90241-4",
    journal = "Nucl. Phys. B",
    volume = "240",
    pages = "514--532",
    year = "1984"
}

@article{Behrend:1998mu,
    author = "Behrend, Roger E. and Pearce, Paul A. and Zuber, Jean-Bernard",
    title = "{Integrable boundaries, conformal boundary conditions and A-D-E fusion rules}",
    eprint = "hep-th/9807142",
    archivePrefix = "arXiv",
    reportNumber = "SACLAY-SPH-T-98-076",
    doi = "10.1088/0305-4470/31/50/001",
    journal = "J. Phys. A",
    volume = "31",
    pages = "L763--L770",
    year = "1998"
}

@article{Behrend:1998fd,
    author = "Behrend, Roger E. and Pearce, Paul A. and Petkova, Valentina B. and Zuber, Jean-Bernard",
    title = "{On the classification of bulk and boundary conformal field theories}",
    eprint = "hep-th/9809097",
    archivePrefix = "arXiv",
    reportNumber = "SACLAY-SPH-T-98-080",
    doi = "10.1016/S0370-2693(98)01374-4",
    journal = "Phys. Lett. B",
    volume = "444",
    pages = "163--166",
    year = "1998"
}

@article{Eniceicu:2022nay,
    author = "Eniceicu, Dan Stefan and Mahajan, Raghu and Murdia, Chitraang and Sen, Ashoke",
    title = "{Normalization of ZZ instanton amplitudes in minimal string theory}",
    eprint = "2202.03448",
    archivePrefix = "arXiv",
    primaryClass = "hep-th",
    doi = "10.1007/JHEP07(2022)139",
    journal = "JHEP",
    volume = "07",
    pages = "139",
    year = "2022"
}

@article{Eniceicu:2022dru,
    author = "Eniceicu, Dan Stefan and Mahajan, Raghu and Murdia, Chitraang and Sen, Ashoke",
    title = "{Multi-instantons in minimal string theory and in matrix integrals}",
    eprint = "2206.13531",
    archivePrefix = "arXiv",
    primaryClass = "hep-th",
    doi = "10.1007/JHEP10(2022)065",
    journal = "JHEP",
    volume = "10",
    pages = "065",
    year = "2022"
}

@article{Eniceicu:2022xvk,
    author = "Eniceicu, Dan Stefan and Mahajan, Raghu and Maity, Pronobesh and Murdia, Chitraang and Sen, Ashoke",
    title = "{The ZZ annulus one-point function in non-critical string theory: A string field theory analysis}",
    eprint = "2210.11473",
    archivePrefix = "arXiv",
    primaryClass = "hep-th",
    doi = "10.1007/JHEP12(2022)151",
    journal = "JHEP",
    volume = "12",
    pages = "151",
    year = "2022"
}

@article{Chakrabhavi:2024szk,
    author = "Chakrabhavi, Vivek and Eniceicu, Dan Stefan and Mahajan, Raghu and Murdia, Chitraang",
    title = "{Normalization of ZZ instanton amplitudes in type 0B minimal superstring theory}",
    eprint = "2406.16867",
    archivePrefix = "arXiv",
    primaryClass = "hep-th",
    doi = "10.1007/JHEP09(2024)114",
    journal = "JHEP",
    volume = "09",
    pages = "114",
    year = "2024"
}

@article{Lewellen:1991tb,
    author = "Lewellen, David C.",
    title = "{Sewing constraints for conformal field theories on surfaces with boundaries}",
    reportNumber = "NSF-ITP-91-32",
    doi = "10.1016/0550-3213(92)90370-Q",
    journal = "Nucl. Phys. B",
    volume = "372",
    pages = "654--682",
    year = "1992"
}

@article{Eberhardt:2021ynh,
    author = "Eberhardt, Lorenz and Pal, Sridip",
    title = "{The disk partition function in string theory}",
    eprint = "2105.08726",
    archivePrefix = "arXiv",
    primaryClass = "hep-th",
    doi = "10.1007/JHEP08(2021)026",
    journal = "JHEP",
    volume = "08",
    pages = "026",
    year = "2021"
}

@article{Tseytlin:1987ww,
    author = "Tseytlin, Arkady A.",
    title = "{Renormalization of Mobius Infinities and Partition Function Representation for String Theory Effective Action}",
    reportNumber = "Print-88-0018 (LEBEDEV)",
    doi = "10.1016/0370-2693(88)90857-X",
    journal = "Phys. Lett. B",
    volume = "202",
    pages = "81--88",
    year = "1988"
}

@article{Liu:1987nz,
    author = "Liu, Jun and Polchinski, Joseph",
    title = "{Renormalization of the Mobius Volume}",
    reportNumber = "UTTG-26-87",
    doi = "10.1016/0370-2693(88)91566-3",
    journal = "Phys. Lett. B",
    volume = "203",
    pages = "39--43",
    year = "1988"
}

@article{Zamolodchikov:1982vx,
    author = "Zamolodchikov, A. B.",
    title = "{ON THE ENTROPY OF RANDOM SURFACES}",
    doi = "10.1016/0370-2693(82)90879-6",
    journal = "Phys. Lett. B",
    volume = "117",
    pages = "87--90",
    year = "1982"
}

@article{Alexandrov:2003nn,
    author = "Alexandrov, Sergei Yu. and Kazakov, Vladimir A. and Kutasov, David",
    title = "{Nonperturbative effects in matrix models and D-branes}",
    eprint = "hep-th/0306177",
    archivePrefix = "arXiv",
    reportNumber = "SACLAY-SPHT-T03-079, LPTENS-03-21, EFI-03-29",
    doi = "10.1088/1126-6708/2003/09/057",
    journal = "JHEP",
    volume = "09",
    pages = "057",
    year = "2003"
}

@article{Mahajan:2021nsd,
    author = "Mahajan, Raghu and Stanford, Douglas and Yan, Cynthia",
    title = "{Sphere and disk partition functions in Liouville and in matrix integrals}",
    eprint = "2107.01172",
    archivePrefix = "arXiv",
    primaryClass = "hep-th",
    doi = "10.1007/JHEP07(2022)132",
    journal = "JHEP",
    volume = "07",
    pages = "132",
    year = "2022"
}

@article{Behrend:1999bn,
    author = "Behrend, Roger E. and Pearce, Paul A. and Petkova, Valentina B. and Zuber, Jean-Bernard",
    title = "{Boundary conditions in rational conformal field theories}",
    eprint = "hep-th/9908036",
    archivePrefix = "arXiv",
    reportNumber = "SACLAY-SPH-T-99-085",
    doi = "10.1016/S0550-3213(99)00592-1",
    journal = "Nucl. Phys. B",
    volume = "570",
    pages = "525--589",
    year = "2000"
}

@article{Teschner:2000md,
    author = "Teschner, J.",
    editor = "Bernard, Denis and Bonora, Loriano and Mussardo, Giuseppe and Corrigan, Edward and Gomez, Cesar and Nahm, Werner and Julia, Bernard",
    title = "{Remarks on Liouville theory with boundary}",
    eprint = "hep-th/0009138",
    archivePrefix = "arXiv",
    doi = "10.22323/1.006.0041",
    journal = "PoS",
    volume = "tmr2000",
    pages = "041",
    year = "2000"
}

@article{Daul:1993bg,
    author = "Daul, J. M. and Kazakov, V. A. and Kostov, I. K.",
    title = "{Rational theories of 2-D gravity from the two matrix model}",
    eprint = "hep-th/9303093",
    archivePrefix = "arXiv",
    reportNumber = "CERN-TH-6834-93",
    doi = "10.1016/0550-3213(93)90582-A",
    journal = "Nucl. Phys. B",
    volume = "409",
    pages = "311--338",
    year = "1993"
}

@inproceedings{DiFrancesco:1990mc,
    author = "Di Francesco, P. and Kutasov, D.",
    title = "{Integrable models of two-dimensional quantum gravity}",
    booktitle = "{Cargese Study Institute: Random Surfaces, Quantum Gravity and Strings}",
    reportNumber = "PUPT-1206",
    month = "10",
    year = "1990"
}

@article{DiFrancesco:1990jd,
    author = "Di Francesco, P. and Kutasov, D.",
    title = "{UNITARY MINIMAL MODELS COUPLED TO 2-D QUANTUM GRAVITY}",
    reportNumber = "PUPT-90-1173",
    doi = "10.1016/0550-3213(90)90328-B",
    journal = "Nucl. Phys. B",
    volume = "342",
    pages = "589--624",
    year = "1990"
}

@article{Kazakov:1986hu,
    author = "Kazakov, V. A.",
    title = "{Ising model on a dynamical planar random lattice: Exact solution}",
    doi = "10.1016/0375-9601(86)90433-0",
    journal = "Phys. Lett. A",
    volume = "119",
    pages = "140--144",
    year = "1986"
}

@article{Brezin:1989db,
    author = "Brezin, Edouard and Douglas, Michael R. and Kazakov, Vladimir and Shenker, Stephen H.",
    title = "{The Ising Model Coupled to 2-$D$ Gravity: A Nonperturbative Analysis}",
    reportNumber = "RU-89-47",
    doi = "10.1016/0370-2693(90)90458-I",
    journal = "Phys. Lett. B",
    volume = "237",
    pages = "43--46",
    year = "1990"
}

@article{Douglas:1989dd,
    author = "Douglas, Michael R.",
    editor = "Brezin, E. and Wadia, S. R.",
    title = "{Strings in Less Than One-dimension and the Generalized $K^- D^- V$ Hierarchies}",
    reportNumber = "RU-89-51",
    doi = "10.1016/0370-2693(90)91716-O",
    journal = "Phys. Lett. B",
    volume = "238",
    pages = "176",
    year = "1990"
}

@Inbook{Douglas1991,
author="Douglas, Michael R.",
editor="Alvarez, Orlando
and Marinari, Enzo
and Windey, Paul",
title="The Two-Matrix Model",
bookTitle="Random Surfaces and Quantum Gravity",
year="1991",
publisher="Springer US",
address="Boston, MA",
pages="77--83",
isbn="978-1-4615-3772-4",
doi="10.1007/978-1-4615-3772-4\_6",
url="https://doi.org/10.1007/978-1-4615-3772-4_6"
}

@article{Crnkovic:1989tn,
    author = "Crnkovic, Cedomir and Ginsparg, Paul H. and Moore, Gregory W.",
    title = "{The Ising Model, the Yang-Lee Edge Singularity, and 2D Quantum Gravity}",
    reportNumber = "YCTP-P20-89, HUTP-89-A058",
    doi = "10.1016/0370-2693(90)91428-E",
    journal = "Phys. Lett. B",
    volume = "237",
    pages = "196--201",
    year = "1990"
}

@inproceedings{Ginsparg:1993is,
    author = "Ginsparg, Paul H. and Moore, Gregory W.",
    title = "{Lectures on 2-D gravity and 2-D string theory}",
    booktitle = "{Theoretical Advanced Study Institute (TASI 92): From Black Holes and Strings to Particles}",
    eprint = "hep-th/9304011",
    archivePrefix = "arXiv",
    reportNumber = "YCTP-P23-92, LA-UR-92-3479",
    pages = "277--469",
    month = "10",
    year = "1993"
}

@article{DiFrancesco:1993cyw,
    author = "Di Francesco, P. and Ginsparg, Paul H. and Zinn-Justin, Jean",
    title = "{2-D Gravity and random matrices}",
    eprint = "hep-th/9306153",
    archivePrefix = "arXiv",
    reportNumber = "LA-UR-93-1722, SACLAY-SPH-T-93-061",
    doi = "10.1016/0370-1573(94)00084-G",
    journal = "Phys. Rept.",
    volume = "254",
    pages = "1--133",
    year = "1995"
}

@article{Anninos:2020ccj,
    author = {Anninos, Dionysios and M{\"u}hlmann, Beatrix},
    title = "{Notes on matrix models (matrix musings)}",
    eprint = "2004.01171",
    archivePrefix = "arXiv",
    primaryClass = "hep-th",
    doi = "10.1088/1742-5468/aba499",
    journal = "J. Stat. Mech.",
    volume = "2008",
    pages = "083109",
    year = "2020"
}

@inproceedings{Klebanov:1991qa,
    author = "Klebanov, Igor R.",
    title = "{String theory in two-dimensions}",
    booktitle = "{Spring School on String Theory and Quantum Gravity (to be followed by Workshop)}",
    eprint = "hep-th/9108019",
    archivePrefix = "arXiv",
    reportNumber = "PUPT-1271",
    month = "7",
    year = "1991"
}

@article{Eynard:2002kg,
    author = "Eynard, B.",
    title = "{Large N expansion of the 2 matrix model}",
    eprint = "hep-th/0210047",
    archivePrefix = "arXiv",
    reportNumber = "SACLAY-SPHT-T02-128, CRM-2868",
    doi = "10.1088/1126-6708/2003/01/051",
    journal = "JHEP",
    volume = "01",
    pages = "051",
    year = "2003"
}

@article{Chekhov:2006vd,
    author = "Chekhov, Leonid and Eynard, Bertrand and Orantin, Nicolas",
    title = "{Free energy topological expansion for the 2-matrix model}",
    eprint = "math-ph/0603003",
    archivePrefix = "arXiv",
    reportNumber = "SPHT-T06-016, ITEP-TH-05-06",
    doi = "10.1088/1126-6708/2006/12/053",
    journal = "JHEP",
    volume = "12",
    pages = "053",
    year = "2006"
}

@article{Seiberg:2004at,
    author = "Seiberg, Nathan and Shih, David",
    title = "{Minimal string theory}",
    eprint = "hep-th/0409306",
    archivePrefix = "arXiv",
    doi = "10.1016/j.crhy.2004.12.007",
    journal = "Comptes Rendus Physique",
    volume = "6",
    pages = "165--174",
    year = "2005"
}

@article{Pasquier:1987xj,
    author = "Pasquier, Vincent",
    title = "{Operator Content of the Ade Lattice Models}",
    reportNumber = "SACLAY-PhT/87-014",
    doi = "10.1088/0305-4470/20/16/043",
    journal = "J. Phys. A",
    volume = "20",
    pages = "5707",
    year = "1987"
}

@article{Warnaar:1993ka,
    author = "Warnaar, S. Ole and Nienhuis, Bernard",
    title = "{Solvable lattice models labelled by Dynkin diagrams}",
    eprint = "hep-th/9301026",
    archivePrefix = "arXiv",
    reportNumber = "ITFA-93-01",
    doi = "10.1088/0305-4470/26/10/005",
    journal = "J. Phys. A",
    volume = "26",
    pages = "2301--2316",
    year = "1993"
}

@article{Saleur:1988zx,
    author = "Saleur, H. and Bauer, Michel",
    title = "{On Some Relations Between Local Height Probabilities and Conformal Invariance}",
    reportNumber = "SACLAY-SPHT-88-080",
    doi = "10.1016/0550-3213(89)90014-X",
    journal = "Nucl. Phys. B",
    volume = "320",
    pages = "591--624",
    year = "1989"
}

@article{DiFrancesco:1991st,
    author = "Di Francesco, P.",
    title = "{Integrable lattice models, graphs and modular invariant conformal field theories}",
    reportNumber = "PUPT-90-1243",
    doi = "10.1142/S0217751X92000223",
    journal = "Int. J. Mod. Phys. A",
    volume = "7",
    pages = "407--500",
    year = "1992"
}

@article{Kostov:1991cg,
    author = "Kostov, Ivan K.",
    title = "{Strings with discrete target space}",
    eprint = "hep-th/9112059",
    archivePrefix = "arXiv",
    reportNumber = "SACLAY-SPH-T-91-142",
    doi = "10.1016/0550-3213(92)90120-Z",
    journal = "Nucl. Phys. B",
    volume = "376",
    pages = "539--598",
    year = "1992"
}

@article{Kostov:1995xw,
    author = "Kostov, Ivan K.",
    editor = "Mussardo, G. and Randjbar-Daemi, S. and Saleur, H.",
    title = "{Solvable statistical models on a random lattice}",
    eprint = "hep-th/9509124",
    archivePrefix = "arXiv",
    doi = "10.1016/0920-5632(95)00611-7",
    journal = "Nucl. Phys. B Proc. Suppl.",
    volume = "45",
    pages = "13--28",
    year = "1996"
}

@article{Ribault:2018jdv,
    author = "Ribault, Sylvain",
    title = "{On 2d CFTs that interpolate between minimal models}",
    eprint = "1809.03722",
    archivePrefix = "arXiv",
    primaryClass = "hep-th",
    doi = "10.21468/SciPostPhys.6.6.075",
    journal = "SciPost Phys.",
    volume = "6",
    number = "6",
    pages = "075",
    year = "2019"
}

@article{Zamolodchikov:2003yb,
    author = "Zamolodchikov, A.",
    editor = "Belavin, A. and Corrigan, Edward",
    title = "{Higher equations of motion in Liouville field theory}",
    eprint = "hep-th/0312279",
    archivePrefix = "arXiv",
    reportNumber = "PM-03-34",
    doi = "10.1142/S0217751X04020592",
    journal = "Int. J. Mod. Phys. A",
    volume = "19S2",
    pages = "510--523",
    year = "2004"
}

@article{Gaberdiel:2008ma,
    author = "Gaberdiel, Matthias R. and Lang, Samuel",
    title = "{Modular differential equations for torus one-point functions}",
    eprint = "0810.0106",
    archivePrefix = "arXiv",
    primaryClass = "hep-th",
    doi = "10.1088/1751-8113/42/4/045405",
    journal = "J. Phys. A",
    volume = "42",
    pages = "045405",
    year = "2009"
}

@article{Lin:2025srf,
    author = "Lin, Henry W. and Zheng, Zechuan",
    title = "{High-precision bootstrap of multi-matrix quantum mechanics}",
    eprint = "2507.21007",
    archivePrefix = "arXiv",
    primaryClass = "hep-th",
    month = "7",
    year = "2025"
}

@article{Kazakov:1987qg,
    author = "Kazakov, V. A.",
    editor = "Billoire, A. and Lacaze, R. and Morel, A. and Napoly, O. and Zinn-Justin, Jean",
    title = "{Exactly solvable Potts models, bond- and tree-like percolation on dynamical (random) planar lattice}",
    doi = "10.1016/0920-5632(88)90089-8",
    journal = "Nucl. Phys. B Proc. Suppl.",
    volume = "4",
    pages = "93--97",
    year = "1988"
}

@article{Daul:1994qy,
    author = "Daul, Jean-Marc",
    title = "{Q states Potts model on a random planar lattice}",
    eprint = "hep-th/9502014",
    archivePrefix = "arXiv",
    reportNumber = "LPTENS-94-",
    month = "11",
    year = "1994"
}

@article{Eynard:1999gp,
    author = "Eynard, B. and Bonnet, G.",
    title = "{The Potts - q random matrix model: Loop equations, critical exponents, and rational case}",
    eprint = "hep-th/9906130",
    archivePrefix = "arXiv",
    reportNumber = "SACLAY-SPH-T-99-060",
    doi = "10.1016/S0370-2693(99)00925-9",
    journal = "Phys. Lett. B",
    volume = "463",
    pages = "273--279",
    year = "1999"
}

@article{Zinn-Justin:1999qww,
    author = "Zinn-Justin, Paul",
    title = "{The Dilute Potts model on random surfaces}",
    eprint = "cond-mat/9903385",
    archivePrefix = "arXiv",
    reportNumber = "RU-99-13",
    doi = "10.1023/A:1018626906256",
    journal = "J. Statist. Phys.",
    volume = "98",
    pages = "245--264",
    year = "2001"
}

@article{Kulanthaivelu:2019zia,
    author = "Kulanthaivelu, Aravinth",
    title = "{Free Variable Loop Equations for the 3-State Potts Model Coupled to 2D Gravity}",
    eprint = "1911.00032",
    archivePrefix = "arXiv",
    primaryClass = "math-ph",
    month = "10",
    year = "2019"
}

@article{Kulanthaivelu:2019atg,
    author = "Kulanthaivelu, Aravinth and Wheater, John F.",
    title = "{Boundary Conditions and the q-state Potts model on Random Planar Maps}",
    eprint = "1911.00266",
    archivePrefix = "arXiv",
    primaryClass = "math-ph",
    month = "11",
    year = "2019"
}

@article{Atkin:2015ksy,
    author = "Atkin, Max R. and Niedner, Benjamin and Wheater, John F.",
    title = "{Sums of Random Matrices and the Potts Model on Random Planar Maps}",
    eprint = "1511.03657",
    archivePrefix = "arXiv",
    primaryClass = "math-ph",
    doi = "10.1088/1751-8113/49/18/185201",
    journal = "J. Phys. A",
    volume = "49",
    number = "18",
    pages = "185201",
    year = "2016"
}

@article{Johnson:2020exp,
    author = "Johnson, Clifford V.",
    title = "{Explorations of nonperturbative Jackiw-Teitelboim gravity and supergravity}",
    eprint = "2006.10959",
    archivePrefix = "arXiv",
    primaryClass = "hep-th",
    doi = "10.1103/PhysRevD.103.046013",
    journal = "Phys. Rev. D",
    volume = "103",
    number = "4",
    pages = "046013",
    year = "2021"
}

@article{Artemev:2024rck,
    author = "Artemev, Aleksandr and Chaban, Igor",
    title = "{(2, 2p + 1) minimal string and intersection theory I}",
    eprint = "2403.02305",
    archivePrefix = "arXiv",
    primaryClass = "hep-th",
    doi = "10.1007/JHEP01(2025)151",
    journal = "JHEP",
    volume = "01",
    pages = "151",
    year = "2025"
}

@article{Artemev:2023bqj,
    author = "Artemev, Aleksandr",
    title = "{p {\textrightarrow} {\ensuremath{\infty}} limit of tachyon correlators in (2, 2p + 1) minimal Liouville gravity from classical Liouville theory}",
    eprint = "2305.08118",
    archivePrefix = "arXiv",
    primaryClass = "hep-th",
    doi = "10.1007/JHEP12(2023)155",
    journal = "JHEP",
    volume = "12",
    pages = "155",
    year = "2023"
}

\end{document}